\documentclass[aps,pra,twocolumn,superscriptaddress,showpacs,amsmath,amssymb,amsfonts,10pt]{revtex4-2}
\usepackage{graphicx}
\usepackage{dcolumn}
\usepackage{stmaryrd}
\usepackage{latexsym}
\usepackage{amssymb}
\usepackage{amsfonts}
\usepackage{amsmath}
\usepackage{mathtools}
\usepackage{verbatim}
\usepackage{fancybox}
\usepackage{color}
\usepackage{bigdelim}
\usepackage[unicode=true,pdfusetitle,bookmarks=true,bookmarksnumbered=true,bookmarksopen=false, breaklinks=false,pdfborder={0 0 0},pdfborderstyle={},backref=false,colorlinks=false, pdftitle={Momentum-dependent quasiparticle properties of the Fermi polaron from the functional renormalization group}]{hyperref}
\usepackage{float}
\usepackage[capitalize]{cleveref}
\usepackage{epstopdf}
\usepackage{braket}
\usepackage{placeins}
\usepackage{appendix}
\usepackage[dvipsnames]{xcolor}
\usepackage{tikz}
\usepackage{tikz-feynman}
\usepackage{tikz-feynhand}
\usetikzlibrary{shapes.misc}
\tikzset{cross/.style={cross out, draw=black, minimum size=2*(#1-\pgflinewidth), inner sep=0pt, outer sep=0pt},
cross/.default={3pt}}
\usepackage[straightquotes]{newtxtt}

\usepackage{lmodern}
\newcommand{\kv}{\mathbf{k}}
\newcommand{\kvp}{\mathbf{k'}}
\newcommand{\qv}{\mathbf{q}}
\newcommand{\rv}{\mathbf{r}}
\newcommand{\zerov}{\mathbf{0}}

\newcommand{\nablav}{\boldsymbol \nabla}

\newcommand{\nnl}{\nonumber \\}

\usepackage{cleveref}
\DeclareMathOperator{\STr}{STr}
\usepackage{physics}
\renewcommand{\pv}{\mathbf{p}}

\begin{document}
\title{Momentum-dependent quasiparticle properties of the Fermi polaron from the functional renormalization group}
\author{Jonas von Milczewski}
\email[Corresponding author, E-Mail: ]{jvmilczewski@mpq.mpg.de}
\affiliation{Institute for Theoretical Physics, Heidelberg University, Philosophenweg 16, 69120 Heidelberg, Germany}
\affiliation{Max-Planck-Institute of Quantum Optics, Hans-Kopfermann-Strasse 1, 85748 Garching, Germany}
\affiliation{Institute for Theoretical Physics, ETH Zurich, 8037 Zurich, Switzerland}
\author{Richard Schmidt}
\affiliation{Institute for Theoretical Physics, Heidelberg University, Philosophenweg 16, 69120 Heidelberg, Germany}

\date{\today}
\begin{abstract}

We study theoretically the lifetimes of attractive and repulsive Fermi polarons, as well as the molecule at finite momentum in three dimensions. To this end, we develop a new technique that allows for the computation of Green's functions in the whole complex frequency plane using exact analytical continuation within the functional renormalization group. The improved numerical stability and reduced computational cost of this method yield access to previously inaccessible momentum-dependent quasiparticle properties of low-lying excited states. While conventional approaches like the  non-selfconsistent $T$-matrix approximation method cannot determine these lifetimes, we are able to find the momentum-dependent lifetime at different interaction strengths of both the attractive and repulsive polaron as well as the molecule. At weak coupling our results confirm predictions made from effective Fermi liquid theory regarding the decay of the attractive polaron, and we demonstrate that Fermi liquid-like  behavior extends far into the strong-coupling regime where attractive polaron and molecule exhibit a $p^4$ momentum scaling in their decay widths. Our results offer an intriguing insight into the momentum-dependent quasiparticle properties of the Fermi polaron problem, which can be measured using techniques such as Raman transfer and Ramsey interferometry.

\end{abstract}

\maketitle

\section{Introduction}\label{sect:Intro}

In recent years, the polaron problem, a single particle interacting with a quantum medium, has attracted significant theoretical and experimental attention, due to its fundamental nature, its significance in understanding strongly coupled systems such as ultracold atoms and two-dimensional semiconductor heterostructures, and its widespread occurrence in a range of different experimental and natural systems, such as dilute mixtures of protons within neutron stars \cite{Gezerlis2008,Forbes2014} or electrons moving through a crystal lattice of atoms \cite{landau1933electron,Froehlich1954}. 

In two-dimensional semiconductor heterostructures, the Fermi- and Bose-polaron problems capture the physical properties of the interaction of electrons and excitons, along with the formation of trion states \cite{
Sidler2016,Goldstein2020,
Xiao2021,Liu2021,tan2022bose,Zipfel2022}. As such, these systems have been used to implement these limiting cases of extreme population imbalance. An understanding of these limits is an important step towards understanding the strong-coupling physics in such systems, which may render useful in exploring whether they might be used for practical applications such as inducing superconductivity \cite{Laussy2010,Cotlet2016,Kinnunen2018,Julku_2022,vonMilczewski2023,Zerba2023}. 

In  ultracold atom systems, the understanding of  polaron problems has helped characterize the phase diagram of both Fermi-Fermi and Bose-Fermi mixtures at strong coupling \cite{Ludwig2011,Bertaina2013,Guidini2015,vonMilczewski2022,Diessel2022,Duda2023,Ness2020,Fritsche2021}.
Experimental observations of the Fermi \cite{Schirotzek2009,Koschorreck2012,Ness2020,Fritsche2021} and the Bose polaron problem \cite{Hu2016,jorgensen2016,yan2020bose} have been flanked by
theoretical insights obtained from different methods such as variational \cite{Combescot2008,Mora2009,Parish2011,Parish2013,cui2020fermi,Parish2020,Chevy2006,Punk2009,Zoellner2011,Trefzger_2012}, diagrammatic \cite{Schmidt2012,Combescot2007,Combescot2009,Bruun2010,Massignan2011}, Monte Carlo \cite{Lobo2006,Prokofev2008,Prokofev2008a,Bertaina2012,Kroiss2014,Vlietinck2014} and functional renormalization group (fRG) approaches \cite{Nikolic2007,Gubbels2008,Schmidt2011,rath2013,Kamikado2017,Isaule2021,vonMilczewski2022}. These methods have been used to characterize properties such as the polaron-to-molecule transition/crossover and the competition with the formation of higher-order bound states. Furthermore, quasiparticle properties such as the energy, effective mass, and quasiparticle width have been extracted from these methods with great success.

The decay widths, or equivalently the quasiparticle lifetimes of the different collective excitations within the Fermi polaron problem, however, have largely remained elusive to a theoretical description. As the decay width may be determined from the self-energy of a quasiparticle, its self-energy needs to contain the correct low-energy states to decay into. As a result, at $T=0$ common non-self-consistent $T$-matrix approaches which contain bare propagators can yield qualitatively correct decay widths for the repulsive polaron \cite{Massignan2011,Tajima2018,Adlong2020} but not for the attractive polaron or the molecule state as these renormalized particles lie lower in energy than the bare particles contained in their self-energies. Of course, at $T>0$ these particles may decay via thermal excitations \cite{Hu2018,Hu2022,Hu2023}. Thus, at strong coupling, a description of the decay channels of polaron and molecule states needs to feature a form of self-consistency, requiring the use of renormalized Green's functions within the computation of the quasiparticle self-energies and decay widths. 

Such self-consistency is challenging to achieve within conventional methods using a wavefunction Ansatz or a non-selfconsistent $T$-matrix approach \cite{Chevy2006,Combescot2007,Punk2009,Schmidt2012,Trefzger_2012,Zoellner2011}. As a result, decay widths have been analyzed using Fermi liquid theory and Fermi's golden rule \cite{Baym1991}, in which the renormalization process is taken into account by using simplified Green's functions with modified quasiparticle weight, energy gap, and effective mass. This works well when the lower-lying particles are well described using Fermi liquid theory \cite{Fritsche2021} and may yield scaling laws for the decay rates in different decay channels \cite{Bruun2008,Bruun2010,Sadeghzadeh2011,Massignan2011,Ngampruetikorn2012,Trefzger2014}, however it is expected to break down at strong coupling \cite{Cetina2015} and thus the applicability of Fermi liquid theory can only really be tested by comparison to a fully self-consistent calculation.

Within fRG treatments \cite{Schmidt2011} this self-consistency is naturally included and thus the decay width may be computed without the need to rely on the validity of Fermi liquid theory. However, as the decay width of the zero-momentum ground state vanishes identically, at a fixed interaction parameter momentum-dependent decay widths of low-lying excited states can vary across several orders of magnitude within a small momentum range. Especially at smaller decay widths, this puts high requirements of numerical stability and precision on the used methods. Previous treatments using fRG \cite{Schmidt2011} lacked precisely this stability due to the need of a costly Matsubara integration and an analytic continuation of the resulting Green's function to real frequencies using numerical methods. 

In this paper, we present a novel, improved fRG treatment of the method used in Ref.~\cite{Schmidt2011}. By incorporating all information about the analytical structure of the Fermi polaron problem, we are able to carry out the Matsubara integration over imaginary frequencies \textit{exactly}. By  a subsequent mapping of  the fRG onto a horizontal line above the real frequency axis we perform an \textit{exact} analytical continuation of the problem onto the whole complex frequency plane. While this treatment is formally equivalent to the treatment used in  Ref.~\cite{Schmidt2011}, it provides greatly enhanced numerical stability and precision at a significantly lower computational cost.
These improvements are not only used to study previously inaccessible quasiparticle properties such as momentum-dependent decay widths of low-lying excited states but also allow to revisit previous results in the literature that implied a $9/2$-power law scaling of the decay of the excited polaron and molecule as function of the energy gap towards the respective ground state \cite{Bruun2010}.

This paper is structured as follows: In \cref{sect:Model} the model along with the  fRG are introduced. In \cref{sect:Exact_integration} the exact frequency integration and the exact analytical continuation onto an equivalent fRG operating on a horizontal line above the real frequency axis are performed. Next, in \cref{sect:Solution_fRG} the numerical solution of the resulting coupled flow equations is described along with the initial conditions of the flow and the parametrization of the renormalized Green's functions. In \cref{sect:Results} the quasiparticle properties of the two polaron states and the molecule are analyzed using this method, complemented by an analysis in terms of Fermi liquid theory. Finally, in \cref{sect:concl} we discuss possible experimental probes of quasiparticle properties such as the momentum-dependent decay width and we consider theoretical extensions of our work to finite impurity concentrations.

\section{Model}
\label{sect:Model}
We study the  three-dimensional Fermi polaron problem consisting of a mixture in which a bosonic or fermionic impurity $\phi$ is immersed in a fermionic bath $\psi$. This is a well-studied system whose microscopic action is given by    
\begin{align}
S &= \int_{x} \psi_x^\ast\left(\partial_\tau -\frac{\nablav^2}{2m_\psi} -\mu_\psi\right)  \psi^{\phantom{\ast}}_x\nnl 
&+ \int_x \phi_x^\ast\left(\partial_\tau - \frac{\nablav^2}{2m_\phi} -\mu_\phi\right)  \phi^{\phantom{\ast}}_x \nnl
&+g  \int_{x} \psi_x^\ast \phi_x^\ast\phi_x^{\phantom{\ast}} \psi_x^{\phantom{\ast}} \label{singlechannel}
\end{align}
where $x=(\rv,\tau)$ denotes the coordinate $\rv$ and imaginary time  $\tau \in [0,1/T]$ and $\int_x=\int_0^{1/T} d \tau \int d^d \rv$ with $d=3$ the dimension. In the following, we consider zero temperature, $T=0$,  and  assume that  impurity and bath particles have a bare dispersion described by the same  mass $m=m_\psi=m_\phi$. We work in units $\hbar=k_\mathrm{B}=1$, and set $2m=1$ unless indicated otherwise. The field $\psi$ is of fermionic Grassmann nature, while the  statistic of $\phi$ is irrelevant due to the single-impurity limit taken in this work. The fields $\phi$ and $\psi$ interact by means of an attractive contact potential of strength $g<0$, regularized in the ultraviolet (UV) by a momentum cutoff $\Lambda$. 

In the vacuum and single-impurity limit this system can host a bound state between a bath and an impurity particle,  both in 2D and 3D. Thus, in order to facilitate the description of this composite particle in a convenient way we consider an equivalent two-channel model \cite{Lurie1964,Nikolic2007} in which the interspecies interaction is mediated by a molecule field $t$ describing the composite particle of mass $2m$ \cite{Holland2001,Timmermans2001,Bruun2004,Bloch2008}
\begin{align}
S &= \int_{\pv,\omega}\bigg\{\sum_{\sigma=\psi,\phi} \sigma^\ast(\omega,\pv)\left(-i\omega +\pv^2 -\mu_\sigma\right)  \sigma(\omega,\pv)\nnl 
&+t^{\ast}(\omega,\pv) G_{t,\Lambda}^{-1}(\omega,\pv) t(\omega,\pv) \bigg\} 
\nnl&+h\int_{x}\left\{ \psi^{\ast}_x \phi^{\ast}_x t^{\phantom{\ast}}_{x}+t^{\ast}_{x}\phi^{\phantom{\ast}}_x\psi^{\phantom{\ast}}_x\right\}\ . \label{twochannel}
\end{align}
Here the momentum $\pv$ and the Matsubara frequency $\omega$ are the Fourier variables of $\rv$ and $\tau$ and $\int_{\pv, \omega}\equiv(2\pi)^{-d-1}\int d^d \pv d\omega$.
In this two-channel model a bath and an impurity particle can be converted into a molecule with a Yukawa coupling $h$  and  $G_{t,\Lambda}$ denotes the bare molecule propagator.  We operate in  the limit where $h\to\infty$  such that   $t$ becomes a purely auxiliary Hubbard-Stratonovich field with no dynamics, i.e. it can be integrated out to yield  back  the original action \cref{singlechannel} for  $h^2 G_{t,\Lambda}=-g$~\cite{Lurie1964,Nikolic2007}. 

To obtain access to the physical properties of this system, inscribed in the full Green's and vertex functions  we deploy a functional renormalization group approach similar to the constructions used in Refs.~\cite{Schmidt2011,Kamikado2017,vonMilczewski2022}. For a detailed explanation of the Fermi polaron problem we refer to Refs.~\cite{Zoellner2011,Parish2011,Schmidt2012,Bertaina2012,Parish2013,Kroiss2014,Vlietinck2014,vonMilczewski2022,Chevy2006,Lobo2006,Combescot2007,Nikolic2007,Prokofev2008,Prokofev2008a,Gubbels2008,Combescot2008,Punk2009,Mora2009,Combescot2009,Bruun2010,Schmidt2011,Parish2020}, for a detailed discussion of the fRG in general we refer to Refs.~\cite{Berges2002,Delamotte2012,Gies2012,Dupuis2020,Metzner2012}. In the following we provide a brief summary of the involved steps, see Refs.~\cite{Schmidt2011,Kamikado2017,vonMilczewski2022} for more detail.

\subsection{fRG equations}

The fRG accounts for the renormalization of Green's functions due to quantum fluctuations by providing  coupled differential equations linking the quantum effective action $\Gamma$ (the generating functional of all one-particle irreducible vertices) to the bare action $S$ using a flowing effective action $\Gamma_k$. This is achieved using the Wetterich equation \cite{Wetterich1993}
\begin{align}
\partial_k \Gamma_k = \frac{1}{2} \STr \Big[ \Big(\Gamma_k^{(2)}+ R_k\Big)^{-1}\partial_k R_k \Big]\ ,  \label{wetterich}
\end{align} 
where $\Gamma_k^{(2)}$ represents the matrix of second functional derivatives of $\Gamma_k$ in the fields and $R_k$ is a matrix containing so-called regulator functions which control the integration of quantum fluctuations. The supertrace $\STr$ denotes a summation over all momenta and frequencies, as well as  the different fields, including a minus sign for fermions. 

Provided that the regulator functions within $R_k$ fulfill certain conditions \cite{Delamotte2012,Gies2012}, in the \emph{ultraviolet} (UV) at $k=\Lambda$ the flowing effective action will be equivalent to the bare action  $\Gamma_{k=\Lambda}=S+const.$ while in the \emph{infrared} at  $k=0$ the quantum effective action is obtained as $\Gamma_{k=0}=\Gamma$. Having obtained this functional, all physical information can be extracted from it.

While the treatment of the problem so-far using \cref{wetterich} is exact; it is also impossible to solve as the effective quantum action contains infinitely-many vertices  yielding an infinite-dimensional set of coupled differential equations. It is thus customary to introduce an \emph{Ansatz} containing finitely-many terms representing the physically most relevant processes in a so-called vertex expansion. Following the treatment in Refs.~\cite{Schmidt2011,Kamikado2017,vonMilczewski2022} we thus choose  the following effective action truncation 
\begin{align}
\Gamma_k= {}&\int_{\pv,\omega} \bigg\{\sum_{\sigma=\psi, \phi} \sigma(\omega,\pv)^\ast G_{\sigma,k}^{-1}(\omega,\pv)  \sigma(\omega,\pv)\nonumber \\
&{}+ t^\ast (\omega,\pv) G_{t,k}^{-1}(\omega,\pv)  t(\omega,\pv) \bigg\} \nnl
&+h\int_{x}  (\psi^{\ast}_{x} \phi^{\ast}_{x} t_{x}^{\phantom{\ast}}+t^{\ast}_{x}\phi_{x}^{\phantom{\ast}} \psi_{x}^{\phantom{\ast}})\ . \label{truncation2}
\end{align}

\begin{figure}[t]
    \normalsize
\begin{align*}
 \begin{tikzpicture}[baseline=-\the\dimexpr\fontdimen22\textfont2\relax]
\begin{feynhand}
\vertex (a) at (-3.8,0){};
\vertex[squaredot](b) at (-2.5,0){};
\vertex[squaredot](c) at (-1,0){};
\vertex (e) at (.3,0){};
\vertex (i) at (-3.4,1.){$\mathbf{(a)}$};
\graph{(a)--[charged boson](b)--[fermion,looseness=1.6, half right,relative=false](c)--[charged scalar,looseness=1.6, half left,relative=false](b),(c)--[charged boson](e)};
\end{feynhand}
\end{tikzpicture}
 \begin{tikzpicture}[baseline=-\the\dimexpr\fontdimen22\textfont2\relax]
\begin{feynhand}
\vertex (a) at (-3.8,0){};
\vertex[squaredot](b) at (-2.5,0){};
\vertex[squaredot](c) at (-1,0){};
\vertex (e) at (.3,0){};
\vertex (i) at (-3.4,1.){$\mathbf{(b)}$};
\graph{(a)--[fermion](b)--[charged boson,looseness=1.6, half right,relative=false](c), (b)--[charged scalar,looseness=1.6, half left,relative=false](c),(c)--[fermion](e)};
\end{feynhand}
\end{tikzpicture}
\end{align*}
\caption{Diagrammatic representation of the fRG flow equations in \cref{Gphiflow,Gtflow}. The flows of the impurity Green's function $\partial_k G_{\phi,k}^{-1}$ \textbf{(a)} as well as the molecular Green's function $\partial_k G_{t,k}^{-1}$ \textbf{(b)} are shown, where wiggly and dashed lines denote impurity and bath propagators, while solid lines denote molecular propgators. The coupling vertex ${\sim} h_k \psi^* \phi^* t$ is denoted by square dots.}
\label{floweq}
\end{figure}
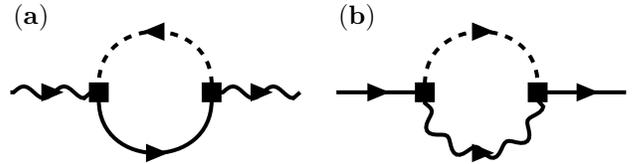

From this truncation one can obtain flow equations for its different constituents using  appropriate functional  derivatives of  \cref{wetterich}. Their diagrammatic representation is shown in \cref{floweq} and in terms of the flowing Green's functions they read \cite{Schmidt2011}
\begin{align}
\partial_k G^{-1}_{\phi,k}(\omega,\pv)&= h^2 \tilde{\partial}_k \int_{\qv,\nu} G^c_{t,k}(\omega+\nu,\pv+\qv) G^c_{\psi,k}(\qv, \nu) ,\label{Gphiflow}\\
\partial_k G^{-1}_{\psi,k}(\omega,\pv)&= -h^2 \tilde{\partial}_k \int_{\qv,\nu} G^c_{t,k}(\omega+\nu,\pv+\qv) G^c_{\phi,k}(\qv, \nu) ,\label{Gpsiflow}\\
\partial_k G^{-1}_{t,k}(\omega,\pv)&= -h^2 \tilde{\partial}_k \int_{\qv,\nu} G^c_{\phi,k}(\omega-\nu,\pv-\qv) G^c_{\psi,k}(\qv, \nu) . \label{Gtflow}
\end{align}
Here, the $G^c$ denote the regulated Green's functions given by 
\begin{equation}
(G_{\sigma,k}^c)^{-1} = (G_{\sigma,k})^{-1} + R_{\sigma,k},
\end{equation}where $R_{\sigma,k}$ is a regulator function contained within $R_k$, which will be defined in the following. 
In these expressions $\tilde{\partial}_k$ denotes  a derivative with respect to the $k$-dependence of the regulator only, i.e. $\tilde{\partial}_k=(\partial_k R_k) \partial_{R_k}$. As we will see in the following, $\partial_k G^{-1}_{\psi,k}=0$, and thus the bath Fermi energy $\epsilon_F$ is equivalent to its chemical potential $\mu_\psi=\epsilon_F$.

In the single-impurity limit, we expect the low-energy excitations of the impurity and the composite molecule particle to lie at low momenta while those of the bath lie around its Fermi surface where $p^2= \epsilon_F$. It is desirable for these fluctuations to be integrated out towards the end of the flow near $k=0$. To this end, we use sharp momentum regulators \cite{Metzner2012,Pawlowski2017} which yield regulated flowing propagators of the form  
\begin{align}
G^{c}_{\psi,k}(\omega,\pv)&= G_{\psi,k}(\omega,\pv)\Theta(|\pv^2 -\epsilon_F|-k^2) \ ,  \label{regulatorPsi}\\
G^{c}_{\phi,k}(\omega,\pv)&=G_{\phi,k}(\omega,\pv) \Theta(|\pv|-k) \ ,  \label{regulatorPhi}\\
G^{c}_{t,k}(\omega,\pv)&= G_{t,k}(\omega,\pv)\Theta(|\pv|-k) \ . \label{regulatorT}
\end{align}

\begin{figure}
 \normalsize
\begin{align*}&
\begin{tikzpicture}[baseline=-\the\dimexpr\fontdimen22\textfont2\relax]
\filldraw[color=Blue!50, fill= Blue!30] (-0.05,.-3) rectangle (0.05 , 2.8);
\filldraw[color=Red!50, fill= Red!30] (-2.9,.95) rectangle (2.9 ,1.05);
\draw[Red,fill=Red] (0,1) circle (2pt);
\draw (0,1) node[cross,Blue] {};
\draw (0,1.5) node[cross,Blue] {};
\draw (0,2) node[cross,Blue] {};
\draw (0,2.5) node[cross,Blue] {};
\draw (0,0.5) node[cross,Blue] {};
\draw (0,0) node[cross,Blue] {};
\draw[Red,fill=Red] (.5,1) circle (2pt);
\draw[Red,fill=Red] (1,1) circle (2pt);
\draw[Red,fill=Red] (1.5,1) circle (2pt);
\draw[Red,fill=Red] (2,1) circle (2pt);
\draw[Red,fill=Red] (2.5,1) circle (2pt);
\draw[Red,fill=Red] (-.5,1) circle (2pt);
\draw[Red,fill=Red] (-1,1) circle (2pt);
\draw[Red,fill=Red] (-1.5,1) circle (2pt);
\draw[Red,fill=Red] (-2,1) circle (2pt);
\draw[Red,fill=Red] (-2.5,1) circle (2pt);
	\begin{feynhand}
        \vertex (A) at (-0,-3){};
        \vertex (B) at (-3,-0){};
        \vertex (C1) at (-2.9,3.1){\( \textbf{(a)} \)};
        \vertex (C) at (-0,3){\( \phantom{i\omega} \)};
        \vertex (C1) at (-0,3.1 ){\( i\omega \)};
        \vertex (D) at (3,0){\( \phantom{\Omega} \)};
        \vertex (D11) at (3.05,0){\( \Omega \)};
        \vertex (R) at (-.1,2.2){};
        \vertex (S) at (1.8,1){};
        \vertex (E) at (.75,1){};
        \vertex (F) at (.75,0){};
        \vertex (G) at (1,0.5){\( i \epsilon \)};
        \vertex (H) at (1.7,2.35){\(\text{exact int.+ cont.} \)};
	\graph{(A)--[fermion, with arrow=1.](C),(B)--[fermion, with arrow=1.](D),  (E)--[plain, with arrow=.97](F)--[plain, with arrow=.97](E),  (R)--[plain, with arrow=0.99, in =90, out=0](S)};
	\end{feynhand}
	\end{tikzpicture}
 \\[-12pt]
 &
 \begin{tikzpicture}[baseline=-\the\dimexpr\fontdimen22\textfont2\relax]
\filldraw[color=Blue!50, fill= Blue!30] (-3.05,.-2.5) rectangle (-0.5 , 2.5);
\filldraw[color=Red!50, fill= Red!30] (.45,-2.5) rectangle (3,2.5);
\draw (-3,1) node[cross,Blue] {};
\draw (-3,1.5) node[cross,Blue] {};
\draw (-3,2) node[cross,Blue] {};
\draw (-3,2.5) node[cross,Blue] {};
\draw (-3,0.5) node[cross,Blue] {};
\draw (-3,0) node[cross,Blue] {};
\draw (-2.5,1) node[cross,Blue] {};
\draw (-2.5,1.5) node[cross,Blue] {};
\draw (-2.5,2) node[cross,Blue] {};
\draw (-2.5,2.5) node[cross,Blue] {};
\draw (-2.5,0.5) node[cross,Blue] {};
\draw (-2.5,0) node[cross,Blue] {};
\draw (-2,1) node[cross,Blue] {};
\draw (-2,1.5) node[cross,Blue] {};
\draw (-2,2) node[cross,Blue] {};
\draw (-2,2.5) node[cross,Blue] {};
\draw (-2,0.5) node[cross,Blue] {};
\draw (-2,0) node[cross,Blue] {};
\draw (-1,1) node[cross,Blue] {};
\draw (-1,1.5) node[cross,Blue] {};
\draw (-1,2) node[cross,Blue] {};
\draw (-1,2.5) node[cross,Blue] {};
\draw (-1,0.5) node[cross,Blue] {};
\draw (-1,0) node[cross,Blue] {};
\draw (-1.5,1) node[cross,Blue] {};
\draw (-1.5,1.5) node[cross,Blue] {};
\draw (-1.5,2) node[cross,Blue] {};
\draw (-1.5,2.5) node[cross,Blue] {};
\draw (-1.5,0.5) node[cross,Blue] {};
\draw (-1.5,0) node[cross,Blue] {};
\draw (-0.5,1) node[cross,Blue] {};
\draw (-0.5,1.5) node[cross,Blue] {};
\draw (-0.5,2) node[cross,Blue] {};
\draw (-0.5,2.5) node[cross,Blue] {};
\draw (-0.5,0.5) node[cross,Blue] {};
\draw (-0.5,0) node[cross,Blue] {};
\draw[Red,fill=Red] (.5,0) circle (2pt);
\draw[Red,fill=Red] (1,0) circle (2pt);
\draw[Red,fill=Red] (1.5,0) circle (2pt);
\draw[Red,fill=Red] (2,0) circle (2pt);
\draw[Red,fill=Red] (2.5,0) circle (2pt);
\draw[Red,fill=Red] (3,0) circle (2pt);
\draw[Red,fill=Red] (.5,0.5) circle (2pt);
\draw[Red,fill=Red] (1,0.5) circle (2pt);
\draw[Red,fill=Red] (1.5,0.5) circle (2pt);
\draw[Red,fill=Red] (2,0.5) circle (2pt);
\draw[Red,fill=Red] (2.5,0.5) circle (2pt);
\draw[Red,fill=Red] (3,0.5) circle (2pt);
\draw[Red,fill=Red] (.5,1) circle (2pt);
\draw[Red,fill=Red] (1,1) circle (2pt);
\draw[Red,fill=Red] (1.5,1) circle (2pt);
\draw[Red,fill=Red] (2,1) circle (2pt);
\draw[Red,fill=Red] (2.5,1) circle (2pt);
\draw[Red,fill=Red] (3,1) circle (2pt);
\draw[Red,fill=Red] (.5,1.5) circle (2pt);
\draw[Red,fill=Red] (1,1.5) circle (2pt);
\draw[Red,fill=Red] (1.5,1.5) circle (2pt);
\draw[Red,fill=Red] (2,1.5) circle (2pt);
\draw[Red,fill=Red] (2.5,1.5) circle (2pt);
\draw[Red,fill=Red] (3,1.5) circle (2pt);
\draw[Red,fill=Red] (.5,2) circle (2pt);
\draw[Red,fill=Red] (1,2) circle (2pt);
\draw[Red,fill=Red] (1.5,2) circle (2pt);
\draw[Red,fill=Red] (2,2) circle (2pt);
\draw[Red,fill=Red] (2.5,2) circle (2pt);
\draw[Red,fill=Red] (3,2) circle (2pt);
\draw[Red,fill=Red] (.5,2.5) circle (2pt);
\draw[Red,fill=Red] (1,2.5) circle (2pt);
\draw[Red,fill=Red] (1.5,2.5) circle (2pt);
\draw[Red,fill=Red] (2,2.5) circle (2pt);
\draw[Red,fill=Red] (2.5,2.5) circle (2pt);
\draw[Red,fill=Red] (3,2.5) circle (2pt);
\draw[Red,fill=Red] (.5,-.5) circle (2pt);
\draw[Red,fill=Red] (1,-.5) circle (2pt);
\draw[Red,fill=Red] (1.5,-.5) circle (2pt);
\draw[Red,fill=Red] (2,-.5) circle (2pt);
\draw[Red,fill=Red] (2.5,-.5) circle (2pt);
\draw[Red,fill=Red] (3,-.5) circle (2pt);
\draw[Red,fill=Red] (.5,-1) circle (2pt);
\draw[Red,fill=Red] (1,-1) circle (2pt);
\draw[Red,fill=Red] (1.5,-1) circle (2pt);
\draw[Red,fill=Red] (2,-1) circle (2pt);
\draw[Red,fill=Red] (2.5,-1) circle (2pt);
\draw[Red,fill=Red] (3,-1) circle (2pt);
\draw[Red,fill=Red] (.5,-1.5) circle (2pt);
\draw[Red,fill=Red] (1,-1.5) circle (2pt);
\draw[Red,fill=Red] (1.5,-1.5) circle (2pt);
\draw[Red,fill=Red] (2,-1.5) circle (2pt);
\draw[Red,fill=Red] (2.5,-1.5) circle (2pt);
\draw[Red,fill=Red] (3,-1.5) circle (2pt);
\draw[Red,fill=Red] (.5,-2) circle (2pt);
\draw[Red,fill=Red] (1,-2) circle (2pt);
\draw[Red,fill=Red] (1.5,-2) circle (2pt);
\draw[Red,fill=Red] (2,-2) circle (2pt);
\draw[Red,fill=Red] (2.5,-2) circle (2pt);
\draw[Red,fill=Red] (3,-2) circle (2pt);
\draw[Red,fill=Red] (.5,-2.5) circle (2pt);
\draw[Red,fill=Red] (1,-2.5) circle (2pt);
\draw[Red,fill=Red] (1.5,-2.5) circle (2pt);
\draw[Red,fill=Red] (2,-2.5) circle (2pt);
\draw[Red,fill=Red] (2.5,-2.5) circle (2pt);
\draw[Red,fill=Red] (3,-2.5) circle (2pt);
	\begin{feynhand}
        \vertex (A) at (-3,-3){};
        \vertex (B) at (-3.1,-0.0){};
        \vertex (C) at (-3,3){\(\phantom{i\omega}\)};
        \vertex (C11) at (-3,3.1){\( i\omega \)};
        \vertex (C1) at (-3,3.5){\( \textbf{(b)} \)};
        \vertex (D) at (0,0){\( |\pv| \)};
	\graph{(A)--[fermion, with arrow=1.](C),(B)--[fermion, with arrow=1.](D)};
        \vertex (E) at (-1.5,3);
        \vertex (E1) at (2,3);      
        \vertex (A1) at (.5,-3){};
        \vertex (B1) at (.4,-0.0){};
        \vertex (C1) at (.5,3){\( \phantom{\Omega} \)};
        \vertex (C111) at (.5,3.1){\( \Omega \)};
        \vertex (C2) at (3.2,2.8){\( (\Omega_{\text{max}}, p_{\text{max}}) \)};
        \vertex (C21) at (3.2,-2.8){\( (\Omega_{\text{min}}, p_{\text{max}}) \)};
        \vertex (D1) at (3.5,0){\( |\pv| \)};
        \vertex (H) at (0.25,3.5){\(\text{exact int.+ cont.} \)};
	\graph{(A1)--[fermion, with arrow=1.](C1),(B1)--[fermion, with arrow=1.](D1),(E)--[plain, with arrow=.99, in=90, out =90](E1)};
	\end{feynhand}
	\end{tikzpicture}
 \end{align*}
 \caption{Schematic diagram of the complex plane and the interpolation space. The complex frequency plane is shown in \textbf{(a)}, with the Matsubara frequencies along the vertical axis (blue shaded) and the real frequencies along the horizontal. The theory originally operates on the Matsubara frequencies and the inverse Green's functions $G_{\sigma,k}^{-1}(\omega,\pv)$ are parametrized by laying out a grid $(i \omega_i,p_j)$ in frequency and momentum space (blue crosses in \textbf{(b)}) and interpolating between the grid points (blue shaded region in \textbf{(b)}), using also the symmetry $G_{\sigma,k}^{-1}(-\omega,\pv)=G_{\sigma,k}^{-1}(\omega,\pv)^*$. After exact Matsubara integration and exact continuation, the RG is defined on a horizontal line $\mathbb{R}+ i \epsilon$ (red shaded region in \textbf{(a)}) and the retarded inverse Green's functions $G_{\sigma,k}^{R,-1}(\Omega + i \epsilon,\pv)$ are parametrized by laying out a grid $(\Omega_i+i \epsilon,p_j)$ (red dots in \textbf{(b)}) and interpolating between the grid points (red shaded region in \textbf{(b)}). For $\Omega<\Omega_{\text{min}}$ or $p>p_{\text{max}}$ the retarded Green's functions are approximated by asymptotic functions $G_{>,\sigma,k}^{R,-1}$  in the flow equations (see \cref{gphiasymp,gtasymp}), while $\Omega>\Omega_{\text{max}}$ is never accessed due to the structure of the flow equations \eqref{Gphiflowfinal} and \eqref{Gtflowfinal}. By comparison in Ref. \cite{Schmidt2011}, the RG equations are solved on a grid of Matsubara frequencies (blue in \textbf{(b)}) and only afterwards are the results  continued to real frequencies using numerical analytic continuation.}
 \label{fig:schematic_cont}
\end{figure}
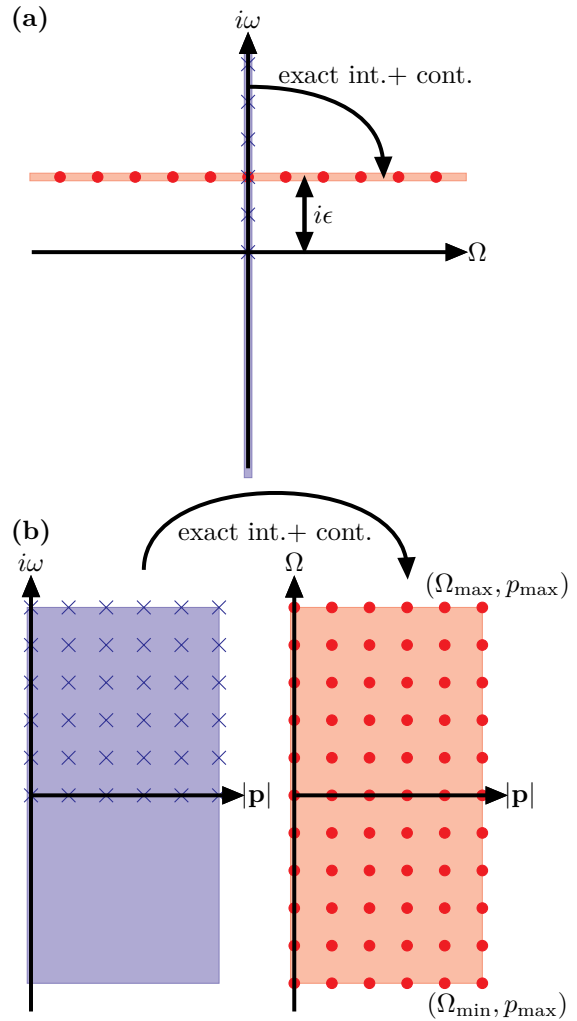

While this choice of regulator functions allows for simple comparison to different approximations, it holds a further advantage that is not immediately obvious. In the following, we will see how its trivial dependence on frequency and its simple structure allow for an \emph{exact} evaluation of the Matsubara integration in \cref{Gphiflow,Gpsiflow,Gtflow} and an \emph{exact} analytical continuation of the obtained Green's function to a horizontal line in the complex frequency plane (see \cref{fig:schematic_cont}).

\section{Exact Matsubara integration}
\label{sect:Exact_integration}
So far, our treatment of the Fermi polaron problem in 3D is exactly equivalent to the treatment developed in Ref.~\cite{Schmidt2011}. A treatment in 2D can be achieved as a natural extension of that work using  Ref.~\cite{vonMilczewski2022}. In Ref.~\cite{Schmidt2011} the flowing inverse Green's functions $G_{\sigma,k}^{-1}$ are parametrized by laying out a grid in Matsubara frequencies  and momenta $(\omega_i, p_j)$ (see \cref{fig:schematic_cont}). The full frequency- and momentum-dependence of  $G_{\sigma,k}^{-1}$ is then obtained interpolating over the function values at these points $C_{\sigma,k}^{i,j}\equiv  G_{\sigma,k}^{-1} (\omega_i, p_j)$. In Ref.~\cite{Schmidt2011} the flow of these coefficients is computed as a coupled differential equation and at the end of the flow the full Green's function in terms of Matsubara frequencies is obtained as an interpolation over these coefficients. To obtain the retarded Green's function just above the real axis, in Ref.~\cite{Schmidt2011} this function is then continued analytically using a Padé approximation (see \cref{fig:schematic_cont}). 

During the course of the evaluation of the flow equations, however, in Ref.~\cite{Schmidt2011} a costly integration over the Matsubara frequencies is performed numerically. Due to the slow convergence rate of this integration, its evaluation yields only moderate precision for reasonable computation times. For ground-state properties, this yields reasonable results. However, for the study of excited state properties the points of interest in the complex frequency plane lie further away from the points at which the fRG was performed. Thus, the numerical error incurred from the Matsubara integration is propagated during the numerical analytical continuation, rendering the obtained results for excited states highly unstable. This  may lead to  
misleading results such as a $p^2$ dispersion with positive effective mass of the attractive polaron in a region where it is known to have a negative effective mass \cite{Trefzger_2012,Private_Schmidt_Enss} (see also \cref{fig:Disp+life}).

Leveraging the analytical structure of the flowing propagators in the single-impurity system we will now demonstrate  how these two problems can be circumvented in a simply maneuver by performing the Matsubara integration \emph{exactly} which yields an \emph{exact} analytical continuation of the propagator functions to the whole complex frequency plane. 

\subsection{Analytical structure of zero-density propagators and the residue theorem}\label{analyticalstructure}

To begin, we recall general analytical properties of the Green's functions at hand \cite{FetterBook,Altland2010}. In the complex frequency plane, the Matsubara frequencies $\omega\in \mathbb{R}$ lie along the imaginary axis $z=i \omega$ and the flowing Matsubara Green's functions $G_{\sigma,k}$ are evaluated along this axis. Along this axis in the upper half of the complex frequency plane (UCP), they correspond to the retarded Green's functions $G^{\phantom{R}}_{\sigma,k}(\omega>0, \pv)= G^{R}_{\sigma,k}(i \omega, \pv)$ and since the retarded Green's function  $G^{R}_{\sigma,k}(z, \pv)$ is analytic for $\Im(z)>0$, the Matsubara Green's function can be continued to the retarded Green's function here. The analogous statement holds for the advanced Green's function $G^{A}_{\sigma,k}(z,\pv)$ for $\Im(z)<0$. Along the real axis, the retarded and the advanced Green's functions fulfill the relations $\Re G^{R}_{\sigma,k}(\Omega + i0^+,\pv)=\Re G^{A}_{\sigma,k}(\Omega - i0^+,\pv)$ and $\Im G^{R}_{\sigma,k}(\Omega + i0^+,\pv)=-\Im G^{A}_{\sigma,k}(\Omega - i0^+,\pv)$ for $\Omega\in \mathbb{R}$. Furthermore, from the retarded Green's function one can obtain the flowing spectral function 
\begin{align}
    \mathcal{A}_{\sigma,k}(\Omega, \pv )&=G^{R}_{\sigma,k}(\Omega + i0^+,\pv)-G^{A}_{\sigma,k}(\Omega - i0^+,\pv)  \nnl
    &=2 \Im\left[G^{R}_{\sigma,k}(\Omega + i0^+,\pv)\right]
\end{align}from which the occupation of states with momentum $\pv$ can be obtained as 
\begin{align}
    n_{\sigma,k}(\pv)=\int_{\Omega} n_{B/F}(\Omega)\mathcal{A}_{\sigma,k}(\Omega, \pv ), \label{nsigmacondition}
\end{align} where depending on the  statistics of the $\sigma$-field $n_{B/F}(\Omega)= 1/(\exp(\Omega/T)\mp 1)$ denotes a Bose- or Fermi-distribution function and $n_{B/F}\mathcal{A}_{\sigma,k} \geq 0 $.

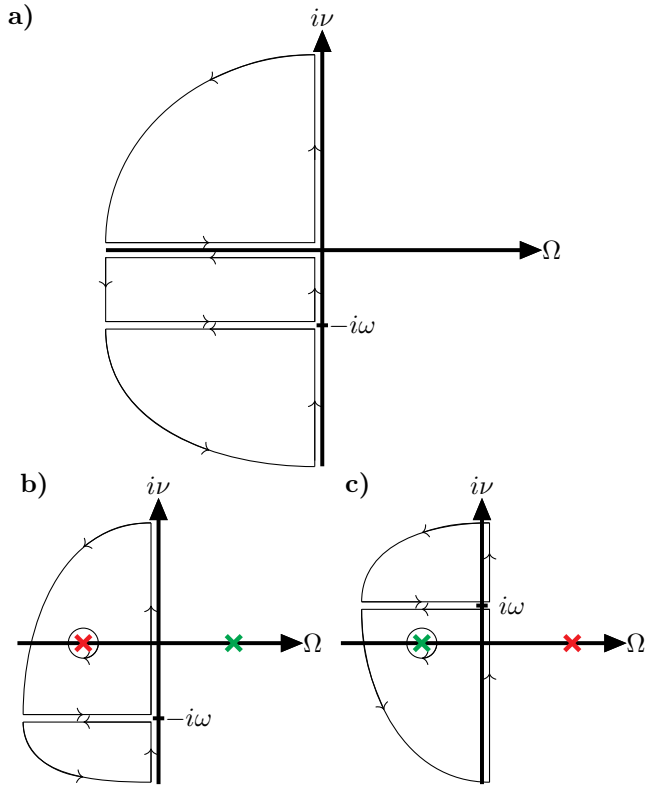
\begin{figure}[ht]
\normalsize
\begin{align*}& 
\begin{tikzpicture}[baseline=-\the\dimexpr\fontdimen22\textfont2\relax]
 \draw (-.1,-2.8) -- (-.1,-1.05) -- (-2.88,-1.05);
  \draw (-2.88,-.95) -- (-.1,-.95) -- (-.1,-.1) -- (-2.88,-.1)--(-2.88,-.95) ;
  \draw (-2.88,.1) -- (-.1,.1) -- (-.1,2.6) ;
  \draw[->] (-2.88,.1) -- (-1.5,.1);
  \draw[->] (-.1,1.2) -- (-.1,1.4) ;
\draw[->] (-.1,-2.88) -- (-.1,-2) ;
\draw[->] (-.1,-1.05) -- (-1.5,-1.05) ;
\draw[->] (-2.88,-.95) -- (-1.5,-.95);
\draw[->] (-.1,-.95) -- (-.1,-.5) ;
\draw[->] (-.1,-.1) -- (-1.5,-.1);
\draw[->] (-2.88,-.1)--(-2.88,-.5);
\draw (-2.88,-1.05) to[out=-90,in=180] (-.1,-2.88);
\draw[->]  (-2.88,-1.05) to[out=-90,in=160] (-1.5,-2.663);
\draw (-.1,2.6) to[out=180,in=90] (-2.88,.1);
\draw[->] (-.1,2.6) to[out=180,in=25] (-1.5,2.26);
	\begin{feynhand}
        \vertex (A) at (-0,-3){};
        \vertex (B) at (-3,-0){};
        \vertex (C1) at (-3+.35-1.35,3.1){\( \textbf{a)} \)};
        \vertex (C) at (-0,3){\( \phantom{i\nu} \)};
        \vertex (C111) at (-0,3.1){\( i\nu \)};
        \vertex (D) at (3,0){\( \phantom{\Omega} \)};
        \vertex (D1) at (3.05,0){\( \Omega \)};
        \vertex (R) at (-.1,2.2){};
        \vertex (S) at (1.8,1){};
        \vertex (E) at (.75,1){};
        \vertex (E1) at (-.2,-1){};
        \vertex (E2) at (.2,-1){};
        \vertex (E3) at (.4,-1){\( -i\omega \)};
        \vertex (E4) at (-.1,-3){};
        \vertex (E5) at (-.1,-1.05){};
        \vertex (E6) at (-3,-1.05){};
        \vertex (F) at (.75,0){};
	\graph{(A)--[fermion, with arrow=1.](C),(B)--[fermion, with arrow=1.](D), (E1)--[plain](E2)};
	\end{feynhand}
	\end{tikzpicture}
 \\[-12pt] &
\begin{tikzpicture}[baseline=-\the\dimexpr\fontdimen22\textfont2\relax]
\draw (-2.1,-1.85)--(-2.1,-1.05)--(-3.8,-1.05);
\draw[->] (-2.1,-1.85)--(-2.1,-1.4);
\draw[->] (-2.1,-1.05)--(-3,-1.05);
\draw (-2.1,-1.85) to[out=180, in =-90] (-3.8,-1.05);
\draw (-3.8,-.95)--(-2.1,-.95)--(-2.1,1.6);
\draw[->] (-3.8,-.95)--(-3,-.95);
\draw[->] (-2.1,-.95)--(-2.1,.5);
\draw (-3.8,-.95)  to [out=90, in = 180](-2.1,1.6);
\draw[->] (-2.1,1.6) to [in=37, out = 180]  (-3,1.28) ;
\draw[->]  (-3.8,-1.05) to[in=173, out =-90] (-3,-1.78);
\draw (-2.1+4.5,-1.85)--(-2.1+4.5,-1.05+1.5)--(-3.8+4.5,-1.05+1.5);
\draw[->] (-2.1+4.5,-1.85)--(-2.1+4.5,-1.4+1.);
\draw[->] (-2.1+4.5,-1.05+1.5)--(-3+4.5,-1.05+1.5);
\draw (-2.1+4.5,-1.85) to[out=180, in =-90] (-3.8+4.5,-1.05+1.5);
\draw (-3.8+4.5,-.95+1.5)--(-2.1+4.5,-.95+1.5)--(-2.1+4.5,1.6);
\draw[->] (-3.8+4.5,-.95+1.5)--(-3+4.5,-.95+1.5);
\draw[->] (-2.1+4.5,-.95+1.5)--(-2.1+4.5,1.2);
\draw (-3.8+4.5,-.95+1.5)  to [out=90, in = 180](-2.1+4.5,1.6);
\draw[->] (-2.1+4.5,1.6) to [in=18, out = 180]  (-3+4.5,1.47) ;
\draw[->]  (-3.8+4.5,-1.05+1.5) to[in=114, out =-90] (-3.5+4.5,-.888);
\draw[->]  (-2.8,.0) to[in=0, out =-90] (-3,-.2);
\draw[->]  (-2.8+4.5,.0) to[in=0, out =-90] (-3+4.5,-.2);
\draw (-3+4.5,0) circle (.2);
\draw (-3,0) circle (.2);
	\begin{feynhand}
        \vertex (A) at (-2,-2){};
        \vertex (B) at (-4,-0){};
        \vertex (C11) at (-3.65,2.1){\( \textbf{b)} \)};
        \vertex (C) at (-2,2){\( \phantom{i\nu} \)};
        \vertex (Cs1) at (-2,2.1){\( i\nu \)};
        \vertex (D) at (0,0){\( \phantom{\Omega} \)};
        \vertex (Ds) at (0.05,0){\( \Omega \)};
        \vertex (E1) at (-2.2,-1){};
        \vertex (E2) at (-1.8,-1){};
        \vertex (E3) at (-1.6,-1){\( -i\omega \)};
	\graph{(A)--[fermion, with arrow=1.](C),(B)--[fermion, with arrow=1.](D), (E1)--[plain](E2)};
        \vertex (A1) at (-2+4.3,-2){};
        \vertex (B1) at (-4+4.3,-0){};
        \vertex (C111) at (-3.65+4.3,2.1){\( \textbf{c)} \)};
        \vertex (C1) at (-2+4.3,2){\( \phantom{i\nu} \)};
        \vertex (Cs1) at (-2+4.3,2.1){\( i\nu \)};
        \vertex (D1) at (0+4.3,0){\( \phantom{\Omega} \)};
        \vertex (D1s) at (0+4.3+.05,0){\( \Omega \)};
        \vertex (E11) at (-2.2+4.3,.5){};
        \vertex (E21) at (-1.8+4.3,.5){};
        \vertex (E31) at (-1.6+4.3,.5){\( i\omega \)};
	\graph{(A1)--[fermion, with arrow=1.](C1),(B1)--[fermion, with arrow=1.](D1), (E11)--[plain](E21)}; 
	\end{feynhand}
 \draw[line width=0.6mm,Red] (-3+.1, .1)--(-3.1,-.1); 
\draw[line width=0.6mm,Red] (-3+.1, -.1)--(-3.1,.1); 
\draw[line width=0.6mm,Green] (-3+.1+2, .1)--(-3.1+2,-.1); 
\draw[line width=0.6mm,Green] (-3+.1+2, -.1)--(-3.1+2,.1); 
\draw[line width=0.6mm,Green] (-3+.1+4.5, .1)--(-3.1+4.5,-.1); 
\draw[line width=0.6mm,Green] (-3+.1+4.5, -.1)--(-3.1+4.5,.1); 
\draw[line width=0.6mm,Red] (-3+.1+2+4.5, .1)--(-3.1+2+4.5,-.1); 
\draw[line width=0.6mm,Red] (-3+.1+2+4.5, -.1)--(-3.1+2+4.5,.1); 
	\end{tikzpicture}\end{align*}
 \caption{Schematic drawing of the contours in the complex plane used to obtain \cref{eq:majorityflowzero,Gtflowfinal,Gphiflowfinal}. \textbf{(a)} The Matsubara summation  on the lhs. of \cref{eq_majorityflowzerobefore} is broken up into the three pieces shown along the y-axis, where the Matsubara Green's functions can be replaced with the corresponding retarded/advanced Green's functions. Using the analyticity of the integrands, the integral along the whole contour vanishes and the integral along the arcs to infinity vanishes due to the decay of the Green's functions. As result, the vertical components of this contour integration (\cref{eq_majorityflowzerobefore}) can be inferred from the horizontal components shown in \cref{eq:majorityflowzero}, which vanish due to the single-impurity limit (see \cref{Asigmacondition}). In \textbf{(b)} and \textbf{(c)} the contours used to obtain \cref{Gphiflowfinal,Gtflowfinal}, respectively, are shown. For $\qv-\epsilon_F<0$ (red crosses) and $\qv-\epsilon_F>0$ (green crosses) the position of the pole in the bath propagator is shown and it contributes to the integral if it lies within the contour.  }
 \label{fig:contoura}
 \end{figure}

Since we work in the single-impurity limit, the occupation of impurity and molecule states must vanish at all times: $n_{\sigma,k}(\pv)=0$ for $\sigma=\phi, t$ and for all $k, \pv$. Thus, from \cref{nsigmacondition} it is easy to see that at $T=0$ for $\Omega<0$, irrespective of the  statistic of the impurity, it holds that 
\begin{align}
\mathcal{A}_{\phi/ t,k}(\Omega<0, \pv )=0  . \label{Asigmacondition}
\end{align}
This has striking consequences: while the functional form of the impurity and molecule Green's function is generally unknown (it is exactly these functions that we are solving for), the bath Green's function is known exactly as it does not flow. Suppressing the momentum-dependencies for now and using the analytic properties for $\omega>0$,  \cref{Gpsiflow} can be rewritten as
\begin{align}
    \int_{\nu} G^c_{t,k}( \omega+\nu) G^c_{\phi,k}(\nu) &= \int_{0}^{\infty} \frac{d \nu}{2\pi} G^{c,R}_{t,k}( i\omega+i\nu) G^{c,R}_{\phi,k}(i\nu) \nonumber\\
   &+ \int_{-\omega}^{0} \frac{d \nu}{2\pi} G^{c,R}_{t,k}( i\omega+i\nu) G^{c,A}_{\phi,k}(i\nu) \nonumber\\
   &+ \int_{-\infty}^{-\omega} \frac{d \nu}{2\pi} G^{c,A}_{t,k}( i\omega+i\nu) G^{c,A}_{\phi,k}(i\nu) .\label{eq_majorityflowzerobefore}
\end{align}
After (\romannumeral 1) performing contour integration along the paths shown in \cref{fig:contoura}a), (\romannumeral 2) using that the integrands are analytic in the interior of these paths, and (\romannumeral 3) respecting that the integrand vanishes along the arcs to infinity, this is equivalent to 
\begin{align}
    =- \int_0^{-\infty}\frac{d\Omega}{2\pi}\Big[&G^{c,R}_{t,k}( i\omega+\Omega) \mathcal{A}^{c}_{\phi,k}(\Omega)\nonumber\\ 
    &+ \mathcal{A}^{c}_{t,k}(\Omega) G^{c,A}_{\phi,k}( -i\omega+\Omega) \Big] =0. \label{eq:majorityflowzero}
\end{align}
Here $G^{c,R}_{\sigma,k}$ and $\mathcal{A}^{c}_{\sigma,k}$ are defined analogous to the regulated flowing propagators in \cref{regulatorPsi,regulatorPhi,regulatorT}. As a result $\partial_k G^{-1}_{\psi,k}=0$ and $G^{-1}_{\psi,k}(\omega,\pv)= - i \omega + \pv^2 -\epsilon_F$ such that $G^{R,-1}_{\psi,k}(z,\pv)=G^{A,-1}_{\psi,k}(z,\pv)= - z + \pv^2 - \epsilon_F $, which can be used to significantly simplify the remaining flow equations. 

In \cref{Gphiflow,Gtflow}, the appearing bath propagators have poles at $\nu=-iz= -i(\qv^2 - \epsilon_F)$ and $\nu=-iz= i(\qv^2 - \epsilon_F)$, respectively, which each lie in the left half of the complex plane for $\qv^2- \epsilon_F<0$ and $\qv^2- \epsilon_F>0$, respectively. Replacing the integrand in \cref{Gphiflow,Gtflow} with the corresponding advanced and retarded propagators and carrying out a contour integration along the contours shown in \cref{fig:contoura}b) and \cref{fig:contoura}c), while taking into account the pole of the bath propagator and the vanishing of the spectral functions described above for $\omega>0$, one thus obtains  \begin{align}
\partial_k G^{-1}_{\phi,k}&(\omega,\pv) \nonumber \\
&= -h^2 \tilde{\partial}_k \int_{\qv} \frac{\Theta(|\pv+\qv|-k) \Theta(\epsilon_F-\qv^2-k^2)}{G_{t,k}^{-1,R}(i \omega+ \qv^2 - \epsilon_F,\pv+\qv)} ,\\
\partial_k G^{-1}_{t,k}&(\omega,\pv)\nonumber \\
&= -h^2 \tilde{\partial}_k \int_{\qv} \frac{\Theta(|\pv+\qv|-k)\Theta(\qv^2-\epsilon_F-k^2)}{G_{\phi,k}^{R,-1}(i\omega -\qv^2 + \epsilon_F,\pv+\qv)} .
\end{align}
Finally, the flow of the imaginary-time Green's function can be continued to an arbitrary horizontal line in the upper complex plane $i\omega \to\Omega + i \epsilon$ to arrive at
\begin{align}
\partial_k G^{R,-1}_{\phi,k}&(\Omega+ i \epsilon,\pv) \nonumber \\
&= -h^2 \tilde{\partial}_k \int_{\qv} \frac{\Theta(|\pv+\qv|-k) \Theta(\epsilon_F-\qv^2-k^2)}{G_{t,k}^{-1,R}(\Omega + i\epsilon+ \qv^2 - \epsilon_F,\pv+\qv)}, \label{Gphiflowfinal}\\
\partial_k G^{R,-1}_{t,k}&(\Omega+ i \epsilon,\pv) \nonumber \\
&= -h^2 \tilde{\partial}_k \int_{\qv} \frac{\Theta(|\pv+\qv|-k)\Theta(\qv^2-\epsilon_F-k^2)}{G_{\phi,k}^{R,-1}(\Omega+ i\epsilon -\qv^2 + \epsilon_F,\pv+\qv)}, \label{Gtflowfinal}
\end{align}
where $\epsilon>0$ is a positive number that does not necessarily have to be close to 0. The $\tilde{\partial}_k$ acts only on the Heaviside functions and under suitable parametrization, the rhs. of \cref{Gphiflowfinal,Gtflowfinal} contains only an integral over the angle between $\pv$ and $\qv$. The Matsubara integration has been eliminated completely and the coupled differential equation in Matsubara frequencies has been mapped to a coupled differential equation within a horizontal line in the complex frequency plane. 

\section{Solution of the coupled flow equations}
\label{sect:Solution_fRG}

After the elimination of the Matsubara integration along with the analytical continuation, we can now solve the coupled differential equation system in \cref{Gphiflowfinal,Gtflowfinal}. Importantly, upon choosing a horizontal line in the complex plane (see \cref{fig:schematic_cont}) these differential equations only couple the retarded impurity and molecule Green's functions within the given horizontal line, without coupling to other horizontal lines. 

\subsection{Parametrization of inverse retarded Green's functions}

To parametrize the flowing inverse retarded Green's functions we lay out a grid consisting of momenta $\pv_i$ and frequencies $\Omega_j + i \epsilon$ on which we store the function values of the Green's functions in form of the coefficients $D_{\sigma,k}^{i,j}\equiv  G_{\sigma,k}^{R,-1} ( \Omega_j + i \epsilon, p_i)$ for $\Omega \in \mathbb{R}$ and $ \epsilon>0$. The momenta and frequencies in this grid need to be chosen such that they 
\begin{enumerate}
    \item resolve well the regions of interest in the retarded Green's function, and
    \item enable a good resolution in the regions that are integrated over in the evaluation of the flow equations \eqref{Gphiflowfinal} and \eqref{Gtflowfinal}, such the interpolating function approximates the actual Green's function well.
\end{enumerate}

From \cref{Gphiflowfinal,Gtflowfinal} it can easily be seen that for a point of interest $\Omega+ i \epsilon$ only retarded Green's functions at points $\Omega'+ i \epsilon$ with $\Omega'<\Omega$ are evaluated. Furthermore, all Green's functions have spherical symmetry in their momentum component such that $G^{R,-1}_{\sigma,k}(\Omega+ i\epsilon,\pv)= G^{R,-1}_{\sigma,k}(\Omega+ i\epsilon,|\pv|)$, enabling a parametrization by the modulus of the momentum component. Thus the grid is contained within $(p_i, \Omega_j + i \epsilon)\in [0,p_{\text{max}} ] \times [\Omega_{\text{min}}+ i \epsilon,\Omega_{\text{max}}+i\epsilon]$ where $\Omega_{\text{max}}$ is chosen according to interest in physical properties and $p_{\text{max}}, \Omega_{\text{min}}$ are chosen to enable integration during the evaluation of flow equations. The choice of the value of $\epsilon$ follows from a compromise: It needs to be chosen such that $\mathbb{R}+ i \epsilon $ is close  enough to the real axis to  yield a good approximation for the spectral function \cref{spectralfunction}. However, if the chosen value of $\epsilon$ is too small, the integration of the flow equations will be over strongly peaked functions which requires small step sizes as the differential equation is solved along the flow parameter $k$.

Within the grid, the $G^{R,-1}_{\sigma,k}$ are obtained from the coefficients $D_{\sigma,k}^{i,j}$ using a bivariate cubic spline interpolation, while for values outside the grid we use that asympotically for $p\to \infty, \Omega\to -\infty$ the $G^{R,-1}_{\sigma,k}$ take on their bare form. Thus, ensuring continuity at the boundaries of the grid, for $|\pv|>p_{\text{max}}$ or $\Omega< \Omega_{\text{min}}$ they are approximated by functions of the functional form of their vacuum solutions \cite{Schmidt2011}
\begin{align}
    G^{R,-1}_{>,\phi,k}(z,\pv)&= -z+ \pv^2- \mu_\phi + f^1_{\text{cont}}, \label{gphiasymp}\\
   G^{R,-1}_{,>t,k}(z,\pv)&= \frac{h^2}{8 \pi} \left(-\frac{1}{a}+\sqrt{-\frac{ z}{2} + \frac{\pv^2}{4} + f^2_{\text{cont}}}\right)\label{gtasymp}, 
\end{align}
where $f^{1,2}_{\text{cont}}$  ensure continuity at the boundary.

\subsection{Initial conditions of the flow}

The initial conditions for the flow of the impurity at the  cutoff scale $k=\Lambda$  are given by the bare impurity propagator 
$G_{\phi,k=\Lambda}^{R,-1}( \Omega + i \epsilon,\pv)= -(\Omega + i \epsilon)+ \pv^2- \mu_\phi$. The initial condition of the molecule 
propagator is chosen such that for a flow in the vacuum two-body limit (obtained by tuning the chemical potentials accordingly 
\cite{Schmidt2011,vonMilczewski2022}) it reproduces   the vacuum molecule propagators at $k=0$. The initial condition of the bath
fermions is given by $G_{\psi,k=\Lambda}^{R,-1}(\Omega + i \epsilon,\pv)= -(\Omega + i \epsilon)+ \pv^2- \epsilon_F$.

From the flow equations in \cref{Gtflowfinal,Gphiflowfinal} one can see that the impurity propagator does not flow for $k^2>\epsilon_F$.  Thus for $\epsilon_F<k^2<\Lambda^2$, the impurity propagator remains in its bare form and \cref{Gtflowfinal,Gphiflowfinal} can be integrated analytically from $k=\Lambda$ down to $k=\sqrt{\epsilon_F}$. Hence the actual numerical solution of the flow equations begins at $k=\sqrt{\epsilon_F}$ with the initial condition
\begin{align}
G_{\phi,k=\sqrt{\epsilon_F}}^{R,-1}(\Omega + i \epsilon,\pv)= G_{\phi,k=\Lambda}^{R,-1}(\Omega + i \epsilon,\pv)
\end{align}
for the impurity and 
\begin{align}
G_{t,k=\sqrt{\epsilon_F}}^{R,-1}(\Omega + i \epsilon,\pv)&= G_{t,k=\Lambda}^{R,-1}(\Omega + i \epsilon,\pv)\nonumber \\ &-\int_{\sqrt{\epsilon_F}}^\Lambda dk' \left[\partial_k G_{t,k=k'}^{R,-1}\right](\Omega + i \epsilon,\pv) \label{initialconditionepsilonF}
\end{align}
for the molecule. A detailed expression for the molecule initial condition is given in \cref{app:molinitial}. Due to the start of the flow not at $k=\Lambda$, but  rather at $k=\sqrt{\epsilon_F}$ we can safely take the limit $\Lambda\to\infty$ during the computation of the molecular initial condition such that the solution of the flow equations is entirely independent of the upper cutoff scale.

\section{Results}
\label{sect:Results}
From the numerical evaluation of the flow equations down to $k=0$, we obtain the renormalized retarded Green's functions of the molecule and the impurity along a horizontal line in the complex frequency plane $G^{R,-1}_{\phi/t} (\Omega+ i \epsilon,\pv)=G^{R,-1}_{\phi/t, k=0} (\Omega+ i \epsilon,\pv)$. Performing the same calculation several times for different horizontal lines (characterized by the value of $\epsilon\in \mathbb{R}, \epsilon>0$), one then obtains a discretized parametrization of these Green's functions in the whole upper half of the complex frequency plane \footnote{Similarly, performing the calculation for $\epsilon<0$ one obtains the advanced Green's function in the lower half of the complex plane.}. 

Several quantities can be deduced from this data via analytical continuation of the retarded Green's function. The single particle spectral function of the molecule and the impurity can be obtained by analytical continuation to the real axis 
\begin{align}
\mathcal{A}_{\phi/t}(\Omega, \pv)=\lim_{\epsilon\to 0} \Im G_{\phi/t}^R(\Omega + i \epsilon, \pv) \ . \label{spectralfunction}
\end{align}
Here, in practice, a small but finite value of $\epsilon$ is sufficient, such that the results of our flow solution can be used without further analytical continuation. 

To obtain the exact energies and lifetimes of the quasiparticles visible as sharp peaks in the spectral function, one needs to find the poles of the retarded Green's function in the lower half of the complex plane (LCP) via analytic continuation of the retarded Green's function across the real axis. At such a pole the inverse retarded Green's functions vanish 
\begin{align}
G^{R,-1}_{\phi/t}(\Omega_{\phi/t}'(\pv)- i\Gamma_{\phi/t}(\pv), \pv)=0 , \Gamma_{\phi/t}>0 \label{PoleCondition}
\end{align}
and the momentum-dependent quasiparticle energy and decay width of the respective quasiparticle  are given by $E_{\phi/t}(\pv)=\Omega_{\phi/t}'(\pv)+ \mu_\phi$ and $\Gamma_{\phi/t}$, respectively. Finally, the inverse quasiparticle weight can be obtained as 
\begin{align}
Z_{\phi/t}^{-1}= - \frac{\partial}{\partial \Omega} G^{R,-1}_{\phi/t}\left(\Omega_{\phi/t}'(\pv)- i\Gamma_{\phi/t}(\pv), \pv\right). 
\end{align}

The analytic continuation to the LCP can be achieved using a Padé approximation in which data from the UCP is used as input. Alternatively, one can also employ an approximation to linear order making use of the Cauchy-Riemann equations to find the location of the quasiparticle poles, yielding very similar results. 

\subsection{Energies and lifetimes at zero momentum in 3D}
\begin{figure*}[ht!]
	\begin{center}
	\includegraphics[width=\textwidth]{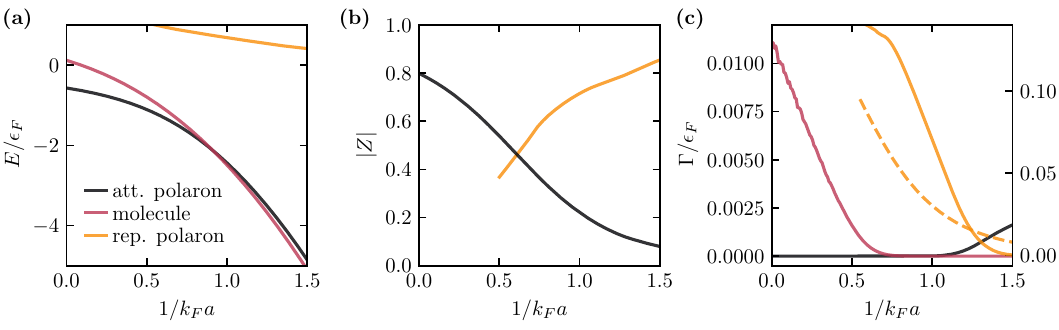}
	\end{center}
	\caption{Energy, quasiparticle weight and decay width  of the molecule as well as the attractive and repulsive polaron as a function of $1/k_F a$. \textbf{(a)} The zero-momentum energies are shown for the attractive polaron $E^{\text{att.}}_{\phi}(\pv=\zerov)$ (black), the repulsive polaron $E^{\text{rep.}}_{\phi}(\pv=\zerov)$ (yellow) and the molecule $E_{t}(\pv=\zerov)$ (red) in units of the Fermi energy $\epsilon_F$. A ground state transition at $1/(k_F a)_c\approx 0.9$ between the molecule and the attractive polaron can be seen, while the repulsive polaron is an excited state above the scattering threshold. For increasing $1/k_Fa$ the modulus of the quasiparticle weight $|Z|$ \textbf{(b)}  of the attractive polaron decreases and the spectral weight is thus shifted to the repulsive polaron, for which  $|Z|$ increases. \textbf{(c)} Approaching the transition, the decay width $\Gamma_{t}$ of the molecule, shown in units of the Fermi energy $\epsilon_F$, decreases and eventually turns to zero (within numerical accuracy) as the transition is reached. The decay width \textbf{(c)} of the attractive polaron $\Gamma^{\text{att.}}_{\phi}$, in turn is zero before the transition and begins to increase beyond it. The repulsive polaron has a decreasing decay width $\Gamma^{\text{rep.}}_{\phi}$ as its quasiparticle weight increases. While the decay widths of the attractive polaron $\Gamma^{\text{att.}}_{\phi}$ and the molecule $\Gamma_t$ are shown for the scale on the left side of \textbf{(c)}, the repulsive polaron decay width $\Gamma^{\text{rep.}}_{\phi}$  is shown with respect to the right scale. In addition to the results obtained from the fRG (solid), the decay width of the repulsive polaron is shown as obtained from a conventional non-selfconsistent $T$-matrix approach (dashed) \cite{Combescot2007,Combescot2009,Massignan2011}.}
	\label{fig:Fig1}	
\end{figure*}

To begin, we study the energies, quasiparticle weights and lifetimes of the attractive and the repulsive polaron as well as the molecule. In \cref{fig:Fig1} we show the  zero-momentum energies $E_{\phi/t}(\pv=0)$  as obtained in Ref.~\cite{Schmidt2011}. Below a critical interaction strength of $1/(k_F a)_c\approx 0.9$ \cite{Schmidt2011} the ground state is given by the attractive polaron while at the critical interaction strength the polaron-to-molecule transition \cite{Chevy2006,Lobo2006,Prokofev2008,Prokofev2008a,Punk2009,Schmidt2011} takes place, beyond which the ground state is given by a molecular state. The repulsive polaron exists as an excited state in the spectrum above the scattering threshold and its energy vanishes asymptotically  for $1/k_F a \to \infty$. The quasiparticle weight $Z$ of the attractive and the repulsive polaron is shown as well, and as expected \cite{Schmidt2011,Massignan2011} with increasing $1/k_F a$, the quasiparticle weight of the attractive polaron decreases while the quasiparticle weight of the repulsive polaron increases.

Additionally, in \cref{fig:Fig1} we show the decay widths of the zero-momentum attractive and repulsive polaron, $\Gamma^{\text{att.}}_{\phi}(\pv=0)$ and $\Gamma^{\text{rep.}}_{\phi}(\pv=0)$, as well as the molecule,  $\Gamma_{t}(\pv=0)$. Furthermore, the decay widths of the repulsive polaron as obtained from a non-selfconsistent $T$-matrix approach are shown \cite{Combescot2007,Combescot2009,Massignan2011}. As expected, the respective ground state particles have a decay width consistent with zero. In the regime where the attractive polaron or the molecule are excited state particles, their decay widths increase as one moves away from the polaron-to-molecule transition. With increasing quasiparticle weight, the decay width of the repulsive polaron $\Gamma^{\text{rep.}}_{\phi}(\pv=0)$ decreases.

Compared to previous work using a similar model (but a different method of solving the flow equations), we obtain decay widths about an order of magnitude larger than those obtained in \cite{Schmidt2011}, highlighting the delicacy of obtaining these roots and the need for a numerically stable method with many grid points and a small step size. For higher-excited states the decay widths are larger and the poles are further inside the LCP. As a result the numerical fluctuations of our method are clearly visible, but remain on the order of a few percent in contrast to previous work. 

We note that for most interaction strengths, the decay widths of the  attractive polaron and the molecule are not accessible in simple non-selfconsistent approaches, but rather approaches with some degree of self-consistency (such as a treatment in Fermi liquid theory, in self-consistent $T$-matrix theory \cite{Hu2023} or as in our work with fRG) are necessary to obtain access to these quantities. Compared to the decay widths of the repulsive polaron obtained from non-selfconsistent approaches, the fRG yields larger decay widths in the regime where the attractive polaron is the ground state, however the decay width of the fRG yields a more stable polaron as $1/k_Fa$ is increased.

\begin{figure}[ht!]
	\begin{center}
	\includegraphics[width=\linewidth]{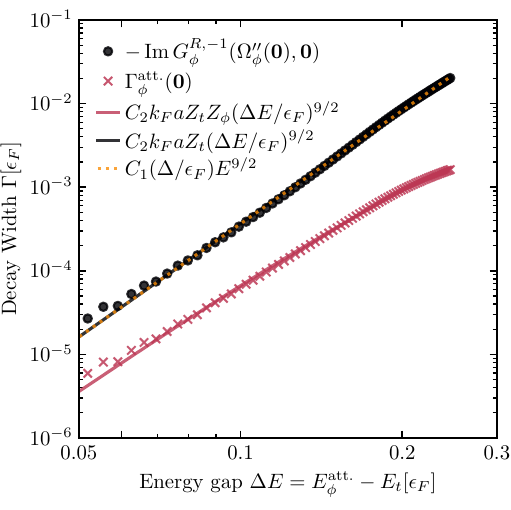}
	\end{center}
	\caption{Decay width of the attractive polaron as a function of the energy gap $\Delta E= E^{\text{att.}}_{\phi}- E_{t}$, both in units of the Fermi energy $\epsilon_F$. The decay width $\Gamma^{\text{att.}}_{\phi} (\pv=\zerov)$  of the attractive polaron as obtained from \cref{PoleCondition} is shown (red crosses) along with the imaginary part of the inverse polaron propagator $G_\phi^{R,-1}(\Omega''_\phi(\zerov),\zerov)$ at the pole position as obtained from \cref{PoleCondition1} (black dots). Note, that along the real axis, the imaginary parts of the self-energy and  the inverse propagator coincide. A curve proportional to $\Delta E^{9/2}$ is shown in yellow dots and fits the imaginary part of the self-energy. Furthermore, a fit according to \cref{PredictionBM1} is shown (solid black line). Multiplying the power law shown in \cref{PredictionBM1} with the quasiparticle weight of the attractive polaron $Z_\phi$, in analogy to \cref{decayapprox}, closely matches the decay width as obtained form \cref{PoleCondition} (solid red line).}
	\label{fig:Fig2}	
\end{figure}
In Ref.~\cite{Bruun2010} the decay width of the attractive polaron in the excited state was predicted to follow a $\Delta E^{9/2}$ scaling where $\Delta E= E_\phi (\pv=0)- E_t (\pv=0)>0$ denotes the energy gap between the attractive polaron and the molecule. To be precise, it was predicted that the imaginary part of the retarded self-energy follows a scaling
\begin{align}
\Im \Sigma^R_\phi \left(\Omega''_\phi (\pv=0),\pv=0\right) \propto Z''_t k_F a \left(\frac{\Delta E}{\epsilon_F}\right)^{\frac{9}{2}} \epsilon_F, \label{PredictionBM1}
\end{align}
where in contrast to \cref{PoleCondition}, $\Omega''$ is defined as 
\begin{align}
\Re G^{R,-1}_{\phi/t}(\Omega_{\phi/t}''(\pv), \pv)=0 \label{PoleCondition1}
\end{align}
and $Z''_{\phi/t}$ is evaluated at $\Omega_{\phi/t}''(\pv=\zerov)$. In this scheme one can then approximate the decay width as 
\begin{align}
\Gamma''_{\phi/t}\approx \Re (Z''_{\phi/t}) \Im \left[\Sigma^R_{\phi/t}  \left(\Omega_{\phi/t}''(\pv=\zerov), \pv=\zerov\right) \right]\label{decayapprox}.
\end{align}

Using \cref{PoleCondition1}, in \cref{fig:Fig2} we show the imaginary part of the inverse polaron propagator $G_\phi^{R,-1}$ at $\Omega''_\phi(\pv=\zerov)$ and $\pv=\zerov$  as a function of the energy gap for $1/k_Fa > 1/k_Fa_c$. Note that the self-energy and the inverse propagator are related by $G_{\phi/t}^{R,-1}= G^{R,-1}_{\phi/t,k=\Lambda}- \Sigma^R_{\phi/t}$. Furthermore we show the polaron  quasiparticle decay width as obtained from \cref{PoleCondition}. As it can be seen the imaginary parts as obtained using \cref{PoleCondition1} fit well with a power law scaling of $\Delta E^{9/2}$, obtained by fitting a function of the form $C_1 (\Delta E/\epsilon_F)^{9/2}$, where $C_1 \in \mathbb{R}$. Furthermore, they fit well with the scaling proposed in Ref.~\cite{Bruun2010}, obtained by fitting the function $C_2 Z_t k_F a \left(\Delta E/\epsilon_F\right)^{9/2}$, $C_2 \in \mathbb{R}$. Multiplying that same curve with the polaron quasiparticle weight $Z_\phi$ results in a curve that fits well with the quasiparticle decay widths computed according to \cref{PoleCondition}. This relation between the imaginary part of the self-energy and the decay width remains accurate for all the results shown in this work. At small energy gaps the value of $\epsilon=10^{-4}$ we used becomes larger than the decay widths and thus the decay widths become inaccurate and begin to fluctuate.

\begin{figure}[ht!]
	\begin{center}
	\includegraphics[width=\linewidth]{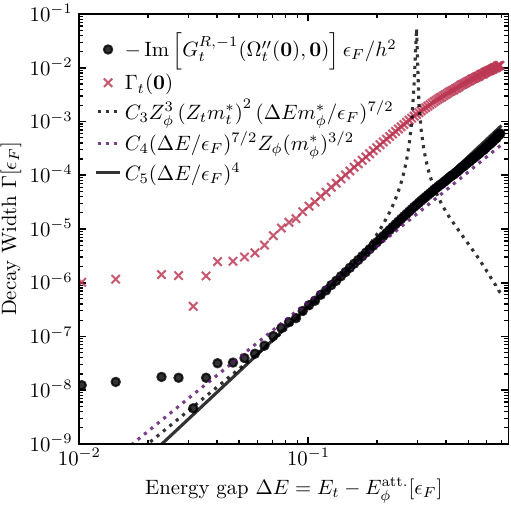}
	\end{center}
	\caption{Decay width of the molecule as a function of the energy gap $\Delta E= E_{t}- E^{\textrm{att.}}_{\phi}$, both in units of the Fermi energy $\epsilon_F$. Like in \cref{fig:Fig2}, the decay width $\Gamma_{t} (\pv=\zerov)$  of the molecule obtained from \cref{PoleCondition} is shown (red crosses) along with the imaginary part of $G_t^{R,-1}(\Omega''_t(\zerov),\zerov)\epsilon_F/h^2$ at the pole position as obtained from \cref{PoleCondition1} (black dots). A curve fit proportional to $\Delta E^{4}$ is shown as a solid line and fits the imaginary part of the self-energy. Furthermore, a fit following a ${\sim}Z_\phi^3 Z_t^2 (m^*_{\phi})^{7/2} (m^*_{t})^{2} \Delta E^{7/2}$ power law is shown (dotted black line) along with a simplified scaling ${\sim}Z_\phi (m^*_{\phi})^{3/2} \Delta E^{7/2} $ (dotted purple line), for detail see \cref{app_decay_mol}.}
	\label{fig:Fig3}	
\end{figure}

\begin{figure*}[t]
	\begin{center}
	\includegraphics[width=\linewidth]{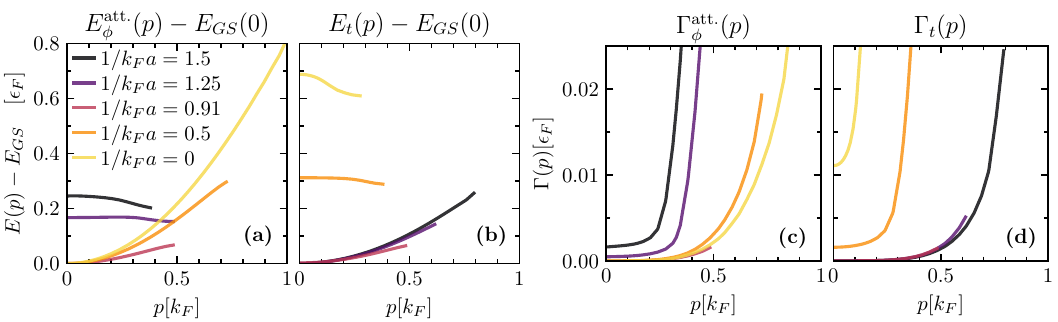}
	\end{center}
	\caption{Momentum-dependent dispersion relations and decay widths of the attractive polaron and the molecule at different interaction parameters. The momentum-dependent energies of the attractive polaron $E_\phi^{\text{att.}}(p)$ \textbf{(a)} and the molecule $E_t(p)$ \textbf{(b)} are shown in units of $\epsilon_F$ as a function of momentum $p=|\pv|$ for interaction strengths, $1/k_F a= 1.5$ (black), $1.25$ (purple), $0.91$ (red), $0.5$ (orange) and $0$ (yellow). The zero-momentum ground-state energy, $E_{\phi}^{\text{att.}}(0)$ for $1/k_Fa<0.91$ and $E_t(0)$ for $1/k_Fa>0.91$ is subtracted for reference. The corresponding decay widths, $\Gamma_\phi^{\text{att.}}(p)$ and $\Gamma_t(p)$ are shown in \textbf{(c)} and \textbf{(d)}, respectively. As can be seen, away from the transition, the ground state develops a quadratic dispersion relation, while the excited state acquires a negative effective mass. In both cases, increasing the momentum leads to increasing decay widths.}
	\label{fig:Disp+life}	
\end{figure*}

Conducting the same analysis for the molecule for $1/k_F a < 1/k_Fa_c$, in \cref{fig:Fig3} we show the imaginary part of the molecule self-energy  along with the molecule decay widths. As before, at small energy gaps the decay widths and imaginary parts fluctuate, but for $\Delta E>0.06 \epsilon_F$ they are stable. As can be seen, the imaginary parts fit well a $C_3 (\Delta E/ \epsilon_F)^4$ scaling with $C_3 \in \mathbb{R}$ which is notably different from the $\Delta E^{9/2}$ scaling proposed in \cite{Bruun2010}. While the diagrammatics in Ref.~\cite{Bruun2010} does not include decay processes to infinite order like our fRG, there is also a fundamental difference in the diagrammatics used. Due to the coupling of the impurity-majority interaction into a molecule channel, crossed diagrams are excluded in our approach at all orders. In Ref.~\cite{Bruun2010}, however, a low-order diagrammatic expansion is employed that includes crossed diagrams. Within that diagram, two $T$-matrices appear which contain no crossed diagrams within them (see \cref{app_decay_mol}). As a result, as one approaches the transition, in the diagrammatics in Ref.~\cite{Bruun2010} the available phase space for decay processes vanishes as $\Delta E^{7/2}$, while the corresponding matrix element vanishes as $\Delta E$. The vanishing of the matrix element in that approach, however, is entirely due to the use of a non-crossed $T$-matrix within a crossed diagrammatics. Performing a similar analysis as in Ref.~\cite{Bruun2010}, but excluding crossed diagrams we analytically obtain a scaling proportional to 
${\sim}Z_\phi^3 Z_t^2 (m^*_{\phi})^{7/2} (m^*_{t})^{2} \Delta E^{7/2} $
(see \cref{app_decay_mol} for detail). This scaling is shown in \cref{fig:Fig3} as well, but it fits the data points only for $0.05 \epsilon_F< \Delta E <0.2\epsilon_F$, as the effective mass of the molecule eventually diverges and turns negative (see \cref{fig:Disp+life}), and thus the pure $\Delta E^4$ scaling fits more accurately.

\subsection{Momentum-dependent energies and decay widths}

Using the precision available within our numerical approach it is possible to obtain not only zero-momentum properties but also momentum-resolved energies (i.e. the full dispersion relation, including effective mass) as well as lifetimes and weight. 
 In \cref{fig:Disp+life}, we show the momentum-dependent attractive polaron and molecule dispersion relations with respect to the energy of the ground state. As can be seen, for $1/k_Fa< 1/k_Fa_c$, the polaron energies at $\pv=0$ coincide with the ground state energies. The dispersion relations follow a close to quadratic behavior with $|\pv|$. Approaching and crossing the transition at $1/k_Fa_c$ this quadratic dependence becomes weaker as the effective polaron mass increases and eventually diverges, as can be seen from the polaron dispersions at $1/k_Fa=1.25$ and $1/k_Fa=1.5$ \cite{Trefzger_2012}. Accordingly, the decay width of the attractive polaron has $\Gamma^{\text{att.}}_{\phi}(\pv=0)\approx 0$  for   $1/k_Fa< 1/k_Fa_c$ and for $1/k_Fa> 1/k_Fa_c$ it has $\Gamma^{\text{att.}}_{\phi}(\pv=0)> 0$. In both regimes the decay width of the attractive polaron increases monotonously as $|\pv|$ increases, see \cref{fig:Disp+life}(c,d).

Similarly, the dispersion of the molecule is gapped for $1/k_Fa< 1/k_Fa_c$ and exhibits a negative effective mass at sufficient detuning from $1/k_Fa_c$. Approaching the transition the effective mass diverges and turns towards a quadratic dispersion with positive effective mass before the transition is crossed. Beyond the transition, the dispersion is ungapped and the effective mass is always positive. As expected, the decay width of the zero-momentum molecule vanishes for $1/k_Fa> 1/k_Fa_c$, while it is finite for $1/k_Fa< 1/k_Fa_c$. As for the polaron, the decay width of the molecule increases as the momentum $|\pv|$ increases. 

The momentum-dependent decay widths observed in \cref{fig:Disp+life} are qualitatively different from the decay described in \cref{fig:Fig1,fig:Fig2,fig:Fig3}: There, the decay is from a zero-momentum excited state such as the attractive polaron to a lower-lying ground state manifold such as the molecule. In \cref{fig:Disp+life} on the other hand, the decay may take place within the ground-state manifold from higher to lower momenta \cite{Sadeghzadeh2011}. For example, as can be seen from \cref{fig:Disp+life}, at $1/k_F a=0$ the attractive polaron with $|\pv|=0.5 k_F$ lies lower in energy than the molecule state and the respective particle-particle continuum. As a result, the attractive polaron with $|\pv|=0.5 k_F$ decays to attractive polaron states with $|\pv'|<0.5 k_F$, necessitating at least a minimal degree of self-consistency to capture this process.

As can be seen in \cref{fig:Disp+life}, for $1/k_Fa\ll 1/k_Fa_c$ the polaron exhibits a near quadratic dispersion relation, while the molecule exhibits a near quadratic dispersion relation for $1/k_Fa\gg 1/k_Fa_c$.  This suggests that the decay width within the ground state manifold may follow a simple behavior with respect to its dependence on momentum. In the following, we investigate the momentum-dependent decay widths of the attractive polaron and the molecule, in regions where they are the ground state and where their dispersion relations suggest that a treatment of the particle within Fermi liquid theory may be appropriate.

\begin{figure}[ht!]
	\centering 
 \begin{tikzpicture}
    \node[anchor=south west,inner sep=0] (image) at (0,0) {\includegraphics[width=\linewidth]{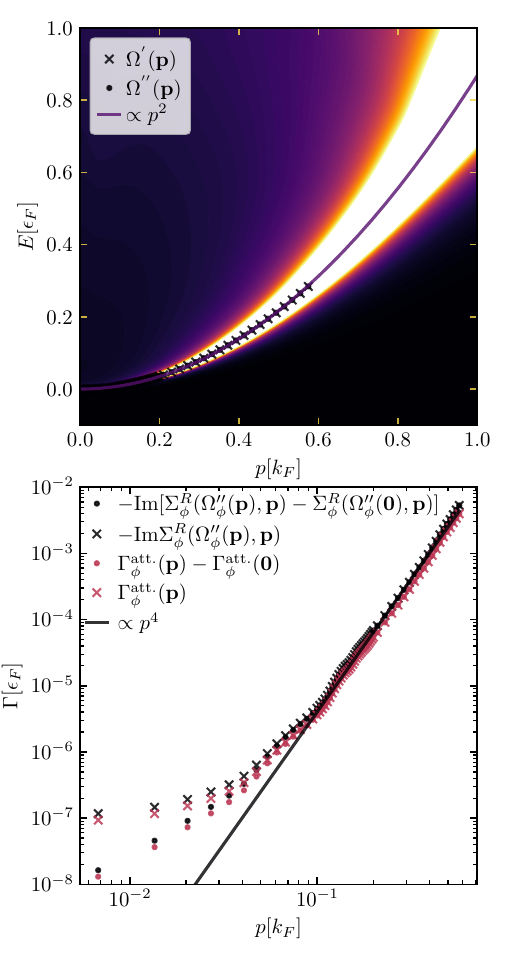}};
    \begin{scope}[x={(image.south east)},y={(image.north west)}]
            \node[align=right] at (0.03,0.49) {$\textbf{(b)}$};
        \node[align=right] at (0.03,0.965) {$\textbf{(a)}$};
\end{scope}
\end{tikzpicture}
	\caption{Impurity spectral function and momentum-dependent decay width of the attractive polaron at unitarity. The impurity  spectral function $\mathcal{A}_\phi(\Omega, \pv)$ is shown in \textbf{(a)}, along with a quadratic fit to the dispersion relation, which coincides with both energies  $\Omega'_\phi (\pv)$ and $\Omega''_\phi (\pv)$ obtained from different criteria, \eqref{PoleCondition}  and \eqref{PoleCondition1}. In \textbf{(b)} the momentum-dependent decay width  $\Gamma^{\text{att.}}_{\phi}$ (red crosses) as well as the self-energy contribution $\Im \Sigma^R_\phi (\Omega''_\phi (\pv),\pv)$ (black crosses), each also offset by their zero-momentum contribution (dots) are shown. For $p\gtrsim 0.1 k_F$ they all follow a ${\sim} p^4$ scaling (solid black line). A value of $\epsilon=10^{-4} \epsilon_F$ was used.}
	\label{fig:p4plot}	
\end{figure}

In \cref{fig:p4plot}(a) we show the momentum-resolved impurity spectral function $\mathcal{A_\phi}$ (see \cref{spectralfunction})  at unitarity. The attractive polaron is the dominant feature of the plot and its energy as obtained from \cref{PoleCondition,PoleCondition1} shows a quadratic dependence $\sim p^2 $ with respect to momentum with effective mass $m^*/m \approx 1.15$. Moreover, the attractive polaron shows a continuously increasing broadening for increasing momentum. This is directly reflected in the behavior of the momentum-dependent decay width of the attractive polaron shown in \cref{fig:p4plot}(b). In this figure we show both $\Gamma^{\text{att.}}_{\phi}(\pv)$ as evaluated from \cref{PoleCondition} as well as the imaginary part of the self-energy $\Im \Sigma_\phi (\Omega''_\phi (\pv),\pv)$. Both evaluations yield consistent results indicating a $\propto p^4 $ scaling for $p\gtrsim 0.1 k_F$. 

The $p^4$ scaling can be obtained from an analysis within Fermi liquid theory (see \cref{app_decay_pol}) \cite{Bruun2008}. In this analysis  the attractive polaron at small momenta is treated as a free particle with quasiparticle properties such as energy, effective mass, quasiparticle weight and decay width that are modified compared to the original bare particle. 
In this picture one thus makes full use of the quasiparticle picture of the attractive polaron that despite strong renormalization by strong-coupling at unitarity still behaves as essentially a free particle (building the basis of Fermi liquid theory).

In \cref{fig:p4plot}, it can be seen that for $p\lesssim 0.1 k_F$ the decay width and the self-energy depart from the $\propto p^4$ scaling. At this point, the decay width has become so small that it is comparable to the distance from the real axis ($ \epsilon = 10^{-4} \epsilon_F$) and thus the numerical continuation of the obtained grid data from $z= \Omega+ i \epsilon$ to $z= \Omega- i \Gamma$ incurs errors that are comparable to $i \epsilon$. At the same time, lowering the value of $i \epsilon$ further slows down the integration over the renormalization group scale $k$ and the momentum $\qv$ within \cref{Gtflowfinal,Gphiflowfinal} as effectively a narrowly shaped Lorentzian curve needs to be integrated over numerically, which requires an increasing amount of computational effort as the Lorentzian becomes sharper. Thus, it can be seen that the decay width of the zero-momentum attractive polaron $\Gamma^{\text{att.}}_{\phi}(\pv=0)$ does not tend to zero (the expected behaviour for a ground state) but rather approaches a small, but finite value. Subtracting the contribution of the decay width and the self-energy at zero momentum, we see that both are closer to the $\propto p^4 $ scaling, but there is still residual error left. 

\begin{figure}[ht!]
\centering 
 \begin{tikzpicture}
    \node[anchor=south west,inner sep=0] (image) at (0,0) {
	\includegraphics[width=\linewidth]{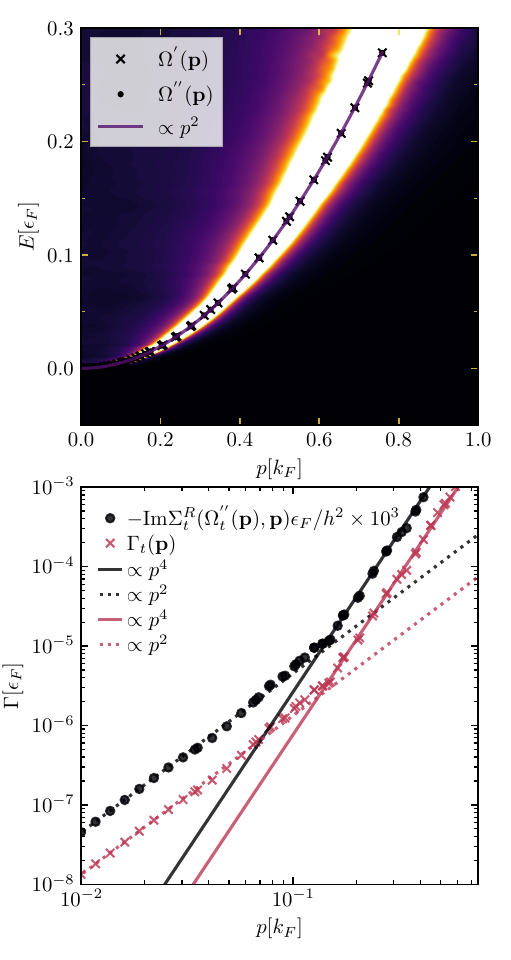}};
    \begin{scope}[x={(image.south east)},y={(image.north west)}]
            \node[align=right] at (0.03,0.49) {$\textbf{(b)}$};
        \node[align=right] at (0.03,0.965) {$\textbf{(a)}$};
\end{scope}
\end{tikzpicture}
	\caption{Molecule spectral function and momentum-dependent decay width of the molecule state at $1/k_Fa=3$. The molecule  spectral function $\mathcal{A}_t(\Omega, \pv)$ is shown in \textbf{(a)}, along with a quadratic fit to the dispersion relation, which reproduces both energies  $\Omega'_t (\pv)$ and $\Omega''_t (\pv)$ obtained from \cref{PoleCondition,PoleCondition1}. The momentum-dependent decay width $\Gamma_t(\pv)$  (red crosses) as well as the self-energy contribution $\Im \Sigma^R_t (\Omega''_\phi (\pv),\pv)$ (black dots, rescaled by a factor of $10^3 \epsilon_F/h^2$) are shown in \textbf{(b)} and follow a ${\sim} p^4$ scaling (black and red solid lines)  for $p\gtrsim 0.1 k_F$. Interestingly, for $p \lesssim 0.12$ a ${\sim} p^2 $ scaling is observed (black and red dotted lines). A value of $\epsilon=10^{-5} \epsilon_F$ was used.}
	\label{fig:p4plotmol}	
\end{figure}

In \cref{fig:p4plotmol}(a) we show the spectral function of the molecule $\mathcal{A}_t$ (see \cref{spectralfunction}) for $1/k_Fa=3$ as well as its dispersion relations. Again, both methods to determine the energy coincide and the dispersion is well characterized by a $\propto p^2$ scaling. In \cref{fig:p4plotmol}(b) in turn   the momentum-dependent decay widths and self-energy evaluations of the molecule are shown. As for the polaron, the $\propto p^2$ dispersion  suggests a $\propto p^4$ scaling in decay width and its self-energy contribution. In  \cref{fig:p4plotmol}(b) such a scaling can be seen to develop for $p \gtrsim 0.12$.
At smaller momenta the value of $\epsilon$ dominates the results. In this calculation $ \epsilon= 10^{-5}\epsilon_F$ was used. Interestingly, the values obtained for $\Gamma_t (\zerov)$ and also those obtained for the corresponding imaginary self-energy contribution are so small that subtracting them does not alter the shown results significantly. Instead, for $p \lesssim 0.12$ a $\propto p^2$ scaling is observed. A similar observation was noted in Ref.~\cite{Bruun2008}, where for a strongly population-imbalanced mixture of two Fermi gases, the decay width scaled quadratically with impurity momentum, when the impurity momentum was below the impurity Fermi wavevector, representing the well-known scaling of fermionic quasiparticles in Fermi liquid theory. Of course, the impurity Fermi level vanishes in our work (and therefore so does the impurity Fermi wavevector), however it is possible that the error incurred from a small, but non-vanishing $\epsilon$ and the ensuing analytical continuation from the horizontal line $\mathbb{R}+ i \epsilon$ effectively results in a small, effective pseudo impurity Fermi wavevector, leading to an analogous quadratic scaling at very small momenta.

\section{Conclusion}
\label{sect:concl}

In this paper we have presented a modified fRG treatment of the Fermi polaron problem which not only avoids the necessity to carry out a numerical integration over imaginary Matsubara frequencies but also the need of continuing analytically  to real frequencies.  This is achieved by leveraging the analytical structure of the Fermi polaron problem to carry out the integration and continuation exactly. As a result, the fRG in imaginary frequencies is mapped onto an equivalent fRG on a horizontal line above the real axis, which can be shifted arbitrarily close to the real axis. The resulting fRG is significantly simpler to solve and allows to consider quasiparticle properties that  either may  not be accessible to previous treatments due to a lack in stability and precision or that  fundamentally cannot be accessed in these treatments. 

Using this modified fRG, the Fermi polaron problem was solved and the quasiparticle properties of the attractive polaron, the repulsive polaron and the molecule were revisited. We showed that energy and quasiparticle width are in accordance with previous findings and the decay width of the attractive polaron does follow a scaling of $\Delta E^{9/2}$ with respect to the energy gap to the molecule. For the decay width of the molecule near the polaron-to-molecule transition, however, the applicability of the  $\Delta E^{9/2}$ scaling is less clear and further research in this direction is necessary. One of the significant improvements of the method presented in this paper is that it allows to investigate momentum-dependent decay widths which are small for states near the ground state. We find that both the attractive polaron and the molecule seem to be captured rather accurately within Fermi liquid theory.

The measurement of these  quasiparticle properties is within experimental reach, using for instance Raman transfers of impurities to finite momentum states \cite{Ness2020}. The decay of such states is then observable using Ramsey interferometry \cite{Cetina2015,Cetina2016}. Similarly, such properties may be accessed using implementations relying on a constantly driven many-body system  \cite{Vivanco2023}. This may be of particular  relevance as momentum relaxation seems to play an important role in the decay of Rabi oscillations \cite{Scazza2017,Mulkerin2023}. 

As polarons may now be controlled so reliably that even induced interactions between polarons can be measured \cite{Baroni2023}, extensions of our fRG method may be of interest where for small impurity concentrations the interaction between polarons may be derived from an additional polaron-polaron scattering vertex. At larger impurity concentrations further modifications may be in order where some of the exact frequency integrations are replaced by contour integrals along horizontal lines above the real axis, which may prove as a promising method of self-consistently investigating strongly-coupled Bose-Fermi and Fermi-Fermi mixtures. 

\textbf{Note added.} During completion of this manuscript, several other works appeared addressing similar analytical approaches within standard resummation techniques \cite{Johansen2023,Hu2023,Enss2023,Dizer2023}.

\begin{acknowledgments}
We thank Aileen Durst, F\'{e}lix Rose, Haydn Adlong, Eugen Dizer and Rudi Grimm for inspiring discussions. This work was supported by the Deutsche Forschungsgemeinschaft under Germany's Excellence Strategy EXC 2181/1 - 390900948 (the Heidelberg STRUCTURES Excellence Cluster). J.v.M. is also supported by a fellowship of the International Max Planck Research School for Quantum Science and Technology (IMPRS-QST).
\end{acknowledgments}


\begin{thebibliography}{91}%
\makeatletter
\providecommand \@ifxundefined [1]{%
 \@ifx{#1\undefined}
}%
\providecommand \@ifnum [1]{%
 \ifnum #1\expandafter \@firstoftwo
 \else \expandafter \@secondoftwo
 \fi
}%
\providecommand \@ifx [1]{%
 \ifx #1\expandafter \@firstoftwo
 \else \expandafter \@secondoftwo
 \fi
}%
\providecommand \natexlab [1]{#1}%
\providecommand \enquote  [1]{``#1''}%
\providecommand \bibnamefont  [1]{#1}%
\providecommand \bibfnamefont [1]{#1}%
\providecommand \citenamefont [1]{#1}%
\providecommand \href@noop [0]{\@secondoftwo}%
\providecommand \href [0]{\begingroup \@sanitize@url \@href}%
\providecommand \@href[1]{\@@startlink{#1}\@@href}%
\providecommand \@@href[1]{\endgroup#1\@@endlink}%
\providecommand \@sanitize@url [0]{\catcode `\\12\catcode `\$12\catcode
  `\&12\catcode `\#12\catcode `\^12\catcode `\_12\catcode `\%12\relax}%
\providecommand \@@startlink[1]{}%
\providecommand \@@endlink[0]{}%
\providecommand \url  [0]{\begingroup\@sanitize@url \@url }%
\providecommand \@url [1]{\endgroup\@href {#1}{\urlprefix }}%
\providecommand \urlprefix  [0]{URL }%
\providecommand \Eprint [0]{\href }%
\providecommand \doibase [0]{https://doi.org/}%
\providecommand \selectlanguage [0]{\@gobble}%
\providecommand \bibinfo  [0]{\@secondoftwo}%
\providecommand \bibfield  [0]{\@secondoftwo}%
\providecommand \translation [1]{[#1]}%
\providecommand \BibitemOpen [0]{}%
\providecommand \bibitemStop [0]{}%
\providecommand \bibitemNoStop [0]{.\EOS\space}%
\providecommand \EOS [0]{\spacefactor3000\relax}%
\providecommand \BibitemShut  [1]{\csname bibitem#1\endcsname}%
\let\auto@bib@innerbib\@empty
\bibitem [{\citenamefont {Gezerlis}\ and\ \citenamefont
  {Carlson}(2008)}]{Gezerlis2008}%
  \BibitemOpen
  \bibfield  {author} {\bibinfo {author} {\bibfnamefont {A.}~\bibnamefont
  {Gezerlis}}\ and\ \bibinfo {author} {\bibfnamefont {J.}~\bibnamefont
  {Carlson}},\ }\bibfield  {title} {\bibinfo {title} {Strongly paired fermions:
  Cold atoms and neutron matter},\ }\href
  {https://doi.org/10.1103/PhysRevC.77.032801} {\bibfield  {journal} {\bibinfo
  {journal} {Phys. Rev. C}\ }\textbf {\bibinfo {volume} {77}},\ \bibinfo
  {pages} {032801} (\bibinfo {year} {2008})}\BibitemShut {NoStop}%
\bibitem [{\citenamefont {Forbes}\ \emph {et~al.}(2014)\citenamefont {Forbes},
  \citenamefont {Gezerlis}, \citenamefont {Hebeler}, \citenamefont {Lesinski},\
  and\ \citenamefont {Schwenk}}]{Forbes2014}%
  \BibitemOpen
  \bibfield  {author} {\bibinfo {author} {\bibfnamefont {M.~M.}\ \bibnamefont
  {Forbes}}, \bibinfo {author} {\bibfnamefont {A.}~\bibnamefont {Gezerlis}},
  \bibinfo {author} {\bibfnamefont {K.}~\bibnamefont {Hebeler}}, \bibinfo
  {author} {\bibfnamefont {T.}~\bibnamefont {Lesinski}},\ and\ \bibinfo
  {author} {\bibfnamefont {A.}~\bibnamefont {Schwenk}},\ }\bibfield  {title}
  {\bibinfo {title} {Neutron polaron as a constraint on nuclear density
  functionals},\ }\href {https://doi.org/10.1103/PhysRevC.89.041301} {\bibfield
   {journal} {\bibinfo  {journal} {Phys. Rev. C}\ }\textbf {\bibinfo {volume}
  {89}},\ \bibinfo {pages} {041301} (\bibinfo {year} {2014})}\BibitemShut
  {NoStop}%
\bibitem [{\citenamefont {Landau}(1933)}]{landau1933electron}%
  \BibitemOpen
  \bibfield  {author} {\bibinfo {author} {\bibfnamefont {L.~D.}\ \bibnamefont
  {Landau}},\ }\bibfield  {title} {\bibinfo {title} {Electron motion in crystal
  lattices},\ }\href@noop {} {\bibfield  {journal} {\bibinfo  {journal} {Phys.
  Z. Sowjet.}\ }\textbf {\bibinfo {volume} {3}},\ \bibinfo {pages} {664}
  (\bibinfo {year} {1933})}\BibitemShut {NoStop}%
\bibitem [{\citenamefont {Fr\"{o}hlich}(1954)}]{Froehlich1954}%
  \BibitemOpen
  \bibfield  {author} {\bibinfo {author} {\bibfnamefont {H.}~\bibnamefont
  {Fr\"{o}hlich}},\ }\bibfield  {title} {\bibinfo {title} {Electrons in lattice
  fields},\ }\href {https://doi.org/10.1080/00018735400101213} {\bibfield
  {journal} {\bibinfo  {journal} {Advances in Physics}\ }\textbf {\bibinfo
  {volume} {3}},\ \bibinfo {pages} {325} (\bibinfo {year} {1954})}\BibitemShut
  {NoStop}%
\bibitem [{\citenamefont {Sidler}\ \emph {et~al.}(2016)\citenamefont {Sidler},
  \citenamefont {Back}, \citenamefont {Cotlet}, \citenamefont {Srivastava},
  \citenamefont {Fink}, \citenamefont {Kroner}, \citenamefont {Demler},\ and\
  \citenamefont {Imamoglu}}]{Sidler2016}%
  \BibitemOpen
  \bibfield  {author} {\bibinfo {author} {\bibfnamefont {M.}~\bibnamefont
  {Sidler}}, \bibinfo {author} {\bibfnamefont {P.}~\bibnamefont {Back}},
  \bibinfo {author} {\bibfnamefont {O.}~\bibnamefont {Cotlet}}, \bibinfo
  {author} {\bibfnamefont {A.}~\bibnamefont {Srivastava}}, \bibinfo {author}
  {\bibfnamefont {T.}~\bibnamefont {Fink}}, \bibinfo {author} {\bibfnamefont
  {M.}~\bibnamefont {Kroner}}, \bibinfo {author} {\bibfnamefont
  {E.}~\bibnamefont {Demler}},\ and\ \bibinfo {author} {\bibfnamefont
  {A.}~\bibnamefont {Imamoglu}},\ }\bibfield  {title} {\bibinfo {title} {Fermi
  polaron-polaritons in charge-tunable atomically thin semiconductors},\ }\href
  {https://doi.org/10.1038/nphys3949} {\bibfield  {journal} {\bibinfo
  {journal} {Nat. Phys.}\ }\textbf {\bibinfo {volume} {13}},\ \bibinfo {pages}
  {255} (\bibinfo {year} {2016})}\BibitemShut {NoStop}%
\bibitem [{\citenamefont {Goldstein}\ \emph {et~al.}(2020)\citenamefont
  {Goldstein}, \citenamefont {Wu}, \citenamefont {Chen}, \citenamefont
  {Taniguchi}, \citenamefont {Watanabe}, \citenamefont {Varga},\ and\
  \citenamefont {Yan}}]{Goldstein2020}%
  \BibitemOpen
  \bibfield  {author} {\bibinfo {author} {\bibfnamefont {T.}~\bibnamefont
  {Goldstein}}, \bibinfo {author} {\bibfnamefont {Y.-C.}\ \bibnamefont {Wu}},
  \bibinfo {author} {\bibfnamefont {S.-Y.}\ \bibnamefont {Chen}}, \bibinfo
  {author} {\bibfnamefont {T.}~\bibnamefont {Taniguchi}}, \bibinfo {author}
  {\bibfnamefont {K.}~\bibnamefont {Watanabe}}, \bibinfo {author}
  {\bibfnamefont {K.}~\bibnamefont {Varga}},\ and\ \bibinfo {author}
  {\bibfnamefont {J.}~\bibnamefont {Yan}},\ }\bibfield  {title} {\bibinfo
  {title} {Ground and excited state exciton polarons in monolayer {MoSe}2},\
  }\href {https://doi.org/10.1063/5.0013092} {\bibfield  {journal} {\bibinfo
  {journal} {J Chem. Phys.}\ }\textbf {\bibinfo {volume} {153}},\ \bibinfo
  {pages} {070401} (\bibinfo {year} {2020})}\BibitemShut {NoStop}%
\bibitem [{\citenamefont {Xiao}\ \emph {et~al.}(2021)\citenamefont {Xiao},
  \citenamefont {Yan}, \citenamefont {Liu}, \citenamefont {Yang}, \citenamefont
  {Kan}, \citenamefont {Duan}, \citenamefont {Liu},\ and\ \citenamefont
  {Cui}}]{Xiao2021}%
  \BibitemOpen
  \bibfield  {author} {\bibinfo {author} {\bibfnamefont {K.}~\bibnamefont
  {Xiao}}, \bibinfo {author} {\bibfnamefont {T.}~\bibnamefont {Yan}}, \bibinfo
  {author} {\bibfnamefont {Q.}~\bibnamefont {Liu}}, \bibinfo {author}
  {\bibfnamefont {S.}~\bibnamefont {Yang}}, \bibinfo {author} {\bibfnamefont
  {C.}~\bibnamefont {Kan}}, \bibinfo {author} {\bibfnamefont {R.}~\bibnamefont
  {Duan}}, \bibinfo {author} {\bibfnamefont {Z.}~\bibnamefont {Liu}},\ and\
  \bibinfo {author} {\bibfnamefont {X.}~\bibnamefont {Cui}},\ }\bibfield
  {title} {\bibinfo {title} {Many-body effect on optical properties of
  monolayer molybdenum diselenide},\ }\href
  {https://doi.org/10.1021/acs.jpclett.1c00320} {\bibfield  {journal} {\bibinfo
   {journal} {J Phys. Chem. Lett.}\ }\textbf {\bibinfo {volume} {12}},\
  \bibinfo {pages} {2555} (\bibinfo {year} {2021})}\BibitemShut {NoStop}%
\bibitem [{\citenamefont {Liu}\ \emph {et~al.}(2021)\citenamefont {Liu},
  \citenamefont {van Baren}, \citenamefont {Lu}, \citenamefont {Taniguchi},
  \citenamefont {Watanabe}, \citenamefont {Smirnov}, \citenamefont {Chang},\
  and\ \citenamefont {Lui}}]{Liu2021}%
  \BibitemOpen
  \bibfield  {author} {\bibinfo {author} {\bibfnamefont {E.}~\bibnamefont
  {Liu}}, \bibinfo {author} {\bibfnamefont {J.}~\bibnamefont {van Baren}},
  \bibinfo {author} {\bibfnamefont {Z.}~\bibnamefont {Lu}}, \bibinfo {author}
  {\bibfnamefont {T.}~\bibnamefont {Taniguchi}}, \bibinfo {author}
  {\bibfnamefont {K.}~\bibnamefont {Watanabe}}, \bibinfo {author}
  {\bibfnamefont {D.}~\bibnamefont {Smirnov}}, \bibinfo {author} {\bibfnamefont
  {Y.-C.}\ \bibnamefont {Chang}},\ and\ \bibinfo {author} {\bibfnamefont
  {C.~H.}\ \bibnamefont {Lui}},\ }\bibfield  {title} {\bibinfo {title}
  {Exciton-polaron rydberg states in monolayer {MoSe}2 and {WSe}2},\ }\href
  {https://doi.org/10.1038/s41467-021-26304-w} {\bibfield  {journal} {\bibinfo
  {journal} {Nat. Comm.}\ }\textbf {\bibinfo {volume} {12}},\ \bibinfo {pages}
  {6131} (\bibinfo {year} {2021})}\BibitemShut {NoStop}%
\bibitem [{\citenamefont {Tan}\ \emph {et~al.}(2023)\citenamefont {Tan},
  \citenamefont {Diessel}, \citenamefont {Popert}, \citenamefont {Schmidt},
  \citenamefont {Imamoglu},\ and\ \citenamefont {Kroner}}]{tan2022bose}%
  \BibitemOpen
  \bibfield  {author} {\bibinfo {author} {\bibfnamefont {L.~B.}\ \bibnamefont
  {Tan}}, \bibinfo {author} {\bibfnamefont {O.~K.}\ \bibnamefont {Diessel}},
  \bibinfo {author} {\bibfnamefont {A.}~\bibnamefont {Popert}}, \bibinfo
  {author} {\bibfnamefont {R.}~\bibnamefont {Schmidt}}, \bibinfo {author}
  {\bibfnamefont {A.}~\bibnamefont {Imamoglu}},\ and\ \bibinfo {author}
  {\bibfnamefont {M.}~\bibnamefont {Kroner}},\ }\bibfield  {title} {\bibinfo
  {title} {Bose polaron interactions in a cavity-coupled monolayer
  semiconductor},\ }\href {https://doi.org/10.1103/PhysRevX.13.031036}
  {\bibfield  {journal} {\bibinfo  {journal} {Phys. Rev. X}\ }\textbf {\bibinfo
  {volume} {13}},\ \bibinfo {pages} {031036} (\bibinfo {year}
  {2023})}\BibitemShut {NoStop}%
\bibitem [{\citenamefont {Zipfel}\ \emph {et~al.}(2022)\citenamefont {Zipfel},
  \citenamefont {Wagner}, \citenamefont {Semina}, \citenamefont {Ziegler},
  \citenamefont {Taniguchi}, \citenamefont {Watanabe}, \citenamefont {Glazov},\
  and\ \citenamefont {Chernikov}}]{Zipfel2022}%
  \BibitemOpen
  \bibfield  {author} {\bibinfo {author} {\bibfnamefont {J.}~\bibnamefont
  {Zipfel}}, \bibinfo {author} {\bibfnamefont {K.}~\bibnamefont {Wagner}},
  \bibinfo {author} {\bibfnamefont {M.~A.}\ \bibnamefont {Semina}}, \bibinfo
  {author} {\bibfnamefont {J.~D.}\ \bibnamefont {Ziegler}}, \bibinfo {author}
  {\bibfnamefont {T.}~\bibnamefont {Taniguchi}}, \bibinfo {author}
  {\bibfnamefont {K.}~\bibnamefont {Watanabe}}, \bibinfo {author}
  {\bibfnamefont {M.~M.}\ \bibnamefont {Glazov}},\ and\ \bibinfo {author}
  {\bibfnamefont {A.}~\bibnamefont {Chernikov}},\ }\bibfield  {title} {\bibinfo
  {title} {Electron recoil effect in electrically tunable
  $\mathrm{Mo}{\mathrm{se}}_{2}$ monolayers},\ }\href
  {https://doi.org/10.1103/PhysRevB.105.075311} {\bibfield  {journal} {\bibinfo
   {journal} {Phys. Rev. B}\ }\textbf {\bibinfo {volume} {105}},\ \bibinfo
  {pages} {075311} (\bibinfo {year} {2022})}\BibitemShut {NoStop}%
\bibitem [{\citenamefont {Laussy}\ \emph {et~al.}(2010)\citenamefont {Laussy},
  \citenamefont {Kavokin},\ and\ \citenamefont {Shelykh}}]{Laussy2010}%
  \BibitemOpen
  \bibfield  {author} {\bibinfo {author} {\bibfnamefont {F.~P.}\ \bibnamefont
  {Laussy}}, \bibinfo {author} {\bibfnamefont {A.~V.}\ \bibnamefont
  {Kavokin}},\ and\ \bibinfo {author} {\bibfnamefont {I.~A.}\ \bibnamefont
  {Shelykh}},\ }\bibfield  {title} {\bibinfo {title} {Exciton-polariton
  mediated superconductivity},\ }\href
  {https://doi.org/10.1103/physrevlett.104.106402} {\bibfield  {journal}
  {\bibinfo  {journal} {Phys. Rev. Lett.}\ }\textbf {\bibinfo {volume} {104}},\
  \bibinfo {pages} {106402} (\bibinfo {year} {2010})}\BibitemShut {NoStop}%
\bibitem [{\citenamefont {Cotle\ifmmode~\mbox{\c{t}}\else \c{t}\fi{}}\ \emph
  {et~al.}(2016)\citenamefont {Cotle\ifmmode~\mbox{\c{t}}\else \c{t}\fi{}},
  \citenamefont {Zeytino\ifmmode~\check{g}\else \v{g}\fi{}lu}, \citenamefont
  {Sigrist}, \citenamefont {Demler},\ and\ \citenamefont
  {Imamo\ifmmode~\check{g}\else \v{g}\fi{}lu}}]{Cotlet2016}%
  \BibitemOpen
  \bibfield  {author} {\bibinfo {author} {\bibfnamefont {O.}~\bibnamefont
  {Cotle\ifmmode~\mbox{\c{t}}\else \c{t}\fi{}}}, \bibinfo {author}
  {\bibfnamefont {S.}~\bibnamefont {Zeytino\ifmmode~\check{g}\else
  \v{g}\fi{}lu}}, \bibinfo {author} {\bibfnamefont {M.}~\bibnamefont
  {Sigrist}}, \bibinfo {author} {\bibfnamefont {E.}~\bibnamefont {Demler}},\
  and\ \bibinfo {author} {\bibfnamefont {A.}~\bibnamefont
  {Imamo\ifmmode~\check{g}\else \v{g}\fi{}lu}},\ }\bibfield  {title} {\bibinfo
  {title} {Superconductivity and other collective phenomena in a hybrid
  bose-fermi mixture formed by a polariton condensate and an electron system in
  two dimensions},\ }\href {https://doi.org/10.1103/PhysRevB.93.054510}
  {\bibfield  {journal} {\bibinfo  {journal} {Phys. Rev. B}\ }\textbf {\bibinfo
  {volume} {93}},\ \bibinfo {pages} {054510} (\bibinfo {year}
  {2016})}\BibitemShut {NoStop}%
\bibitem [{\citenamefont {Kinnunen}\ \emph {et~al.}(2018)\citenamefont
  {Kinnunen}, \citenamefont {Wu},\ and\ \citenamefont {Bruun}}]{Kinnunen2018}%
  \BibitemOpen
  \bibfield  {author} {\bibinfo {author} {\bibfnamefont {J.~J.}\ \bibnamefont
  {Kinnunen}}, \bibinfo {author} {\bibfnamefont {Z.}~\bibnamefont {Wu}},\ and\
  \bibinfo {author} {\bibfnamefont {G.~M.}\ \bibnamefont {Bruun}},\ }\bibfield
  {title} {\bibinfo {title} {Induced p -wave pairing in bose-fermi mixtures},\
  }\href {https://doi.org/10.1103/physrevlett.121.253402} {\bibfield  {journal}
  {\bibinfo  {journal} {Phys. Rev. Lett.}\ }\textbf {\bibinfo {volume} {121}},\
  \bibinfo {pages} {253402} (\bibinfo {year} {2018})}\BibitemShut {NoStop}%
\bibitem [{\citenamefont {Julku}\ \emph {et~al.}(2022)\citenamefont {Julku},
  \citenamefont {Kinnunen}, \citenamefont {Camacho-Guardian},\ and\
  \citenamefont {Bruun}}]{Julku_2022}%
  \BibitemOpen
  \bibfield  {author} {\bibinfo {author} {\bibfnamefont {A.}~\bibnamefont
  {Julku}}, \bibinfo {author} {\bibfnamefont {J.~J.}\ \bibnamefont {Kinnunen}},
  \bibinfo {author} {\bibfnamefont {A.}~\bibnamefont {Camacho-Guardian}},\ and\
  \bibinfo {author} {\bibfnamefont {G.~M.}\ \bibnamefont {Bruun}},\ }\bibfield
  {title} {\bibinfo {title} {Light-induced topological superconductivity in
  transition metal dichalcogenide monolayers},\ }\href
  {https://doi.org/10.1103/PhysRevB.106.134510} {\bibfield  {journal} {\bibinfo
   {journal} {Phys. Rev. B}\ }\textbf {\bibinfo {volume} {106}},\ \bibinfo
  {pages} {134510} (\bibinfo {year} {2022})}\BibitemShut {NoStop}%
\bibitem [{\citenamefont {von Milczewski}\ \emph {et~al.}(2023)\citenamefont
  {von Milczewski}, \citenamefont {Chen}, \citenamefont {Imamoglu},\ and\
  \citenamefont {Schmidt}}]{vonMilczewski2023}%
  \BibitemOpen
  \bibfield  {author} {\bibinfo {author} {\bibfnamefont {J.}~\bibnamefont {von
  Milczewski}}, \bibinfo {author} {\bibfnamefont {X.}~\bibnamefont {Chen}},
  \bibinfo {author} {\bibfnamefont {A.}~\bibnamefont {Imamoglu}},\ and\
  \bibinfo {author} {\bibfnamefont {R.}~\bibnamefont {Schmidt}},\ }\bibfield
  {title} {\bibinfo {title} {Superconductivity induced by strong
  electron-exciton coupling in doped atomically thin semiconductor
  heterostructures},\ }\href {https://doi.org/10.48550/ARXIV.2310.10726}
  {\bibfield  {journal} {\bibinfo  {journal} {arXiv:2310.10726}\ } (\bibinfo
  {year} {2023})}\BibitemShut {NoStop}%
\bibitem [{\citenamefont {Zerba}\ \emph {et~al.}(2023)\citenamefont {Zerba},
  \citenamefont {Kuhlenkamp}, \citenamefont {Imamoğlu},\ and\ \citenamefont
  {Knap}}]{Zerba2023}%
  \BibitemOpen
  \bibfield  {author} {\bibinfo {author} {\bibfnamefont {C.}~\bibnamefont
  {Zerba}}, \bibinfo {author} {\bibfnamefont {C.}~\bibnamefont {Kuhlenkamp}},
  \bibinfo {author} {\bibfnamefont {A.}~\bibnamefont {Imamoğlu}},\ and\
  \bibinfo {author} {\bibfnamefont {M.}~\bibnamefont {Knap}},\ }\bibfield
  {title} {\bibinfo {title} {Realizing topological superconductivity in tunable
  bose-fermi mixtures with transition metal dichalcogenide heterostructures},\
  }\href {https://doi.org/10.48550/ARXIV.2310.10720} {\bibfield  {journal}
  {\bibinfo  {journal} {arXiv:2310.10720}\ } (\bibinfo {year}
  {2023})}\BibitemShut {NoStop}%
\bibitem [{\citenamefont {Ludwig}\ \emph {et~al.}(2011)\citenamefont {Ludwig},
  \citenamefont {Floerchinger}, \citenamefont {Moroz},\ and\ \citenamefont
  {Wetterich}}]{Ludwig2011}%
  \BibitemOpen
  \bibfield  {author} {\bibinfo {author} {\bibfnamefont {D.}~\bibnamefont
  {Ludwig}}, \bibinfo {author} {\bibfnamefont {S.}~\bibnamefont
  {Floerchinger}}, \bibinfo {author} {\bibfnamefont {S.}~\bibnamefont
  {Moroz}},\ and\ \bibinfo {author} {\bibfnamefont {C.}~\bibnamefont
  {Wetterich}},\ }\bibfield  {title} {\bibinfo {title} {Quantum phase
  transition in bose-fermi mixtures},\ }\href
  {https://doi.org/10.1103/PhysRevA.84.033629} {\bibfield  {journal} {\bibinfo
  {journal} {Phys. Rev. A}\ }\textbf {\bibinfo {volume} {84}},\ \bibinfo
  {pages} {033629} (\bibinfo {year} {2011})}\BibitemShut {NoStop}%
\bibitem [{\citenamefont {Bertaina}\ \emph {et~al.}(2013)\citenamefont
  {Bertaina}, \citenamefont {Fratini}, \citenamefont {Giorgini},\ and\
  \citenamefont {Pieri}}]{Bertaina2013}%
  \BibitemOpen
  \bibfield  {author} {\bibinfo {author} {\bibfnamefont {G.}~\bibnamefont
  {Bertaina}}, \bibinfo {author} {\bibfnamefont {E.}~\bibnamefont {Fratini}},
  \bibinfo {author} {\bibfnamefont {S.}~\bibnamefont {Giorgini}},\ and\
  \bibinfo {author} {\bibfnamefont {P.}~\bibnamefont {Pieri}},\ }\bibfield
  {title} {\bibinfo {title} {Quantum monte carlo study of a resonant bose-fermi
  mixture},\ }\href {https://doi.org/10.1103/PhysRevLett.110.115303} {\bibfield
   {journal} {\bibinfo  {journal} {Phys. Rev. Lett.}\ }\textbf {\bibinfo
  {volume} {110}},\ \bibinfo {pages} {115303} (\bibinfo {year}
  {2013})}\BibitemShut {NoStop}%
\bibitem [{\citenamefont {Guidini}\ \emph {et~al.}(2015)\citenamefont
  {Guidini}, \citenamefont {Bertaina}, \citenamefont {Galli},\ and\
  \citenamefont {Pieri}}]{Guidini2015}%
  \BibitemOpen
  \bibfield  {author} {\bibinfo {author} {\bibfnamefont {A.}~\bibnamefont
  {Guidini}}, \bibinfo {author} {\bibfnamefont {G.}~\bibnamefont {Bertaina}},
  \bibinfo {author} {\bibfnamefont {D.~E.}\ \bibnamefont {Galli}},\ and\
  \bibinfo {author} {\bibfnamefont {P.}~\bibnamefont {Pieri}},\ }\bibfield
  {title} {\bibinfo {title} {Condensed phase of bose-fermi mixtures with a
  pairing interaction},\ }\href {https://doi.org/10.1103/PhysRevA.91.023603}
  {\bibfield  {journal} {\bibinfo  {journal} {Phys. Rev. A}\ }\textbf {\bibinfo
  {volume} {91}},\ \bibinfo {pages} {023603} (\bibinfo {year}
  {2015})}\BibitemShut {NoStop}%
\bibitem [{\citenamefont {von Milczewski}\ \emph {et~al.}(2022)\citenamefont
  {von Milczewski}, \citenamefont {Rose},\ and\ \citenamefont
  {Schmidt}}]{vonMilczewski2022}%
  \BibitemOpen
  \bibfield  {author} {\bibinfo {author} {\bibfnamefont {J.}~\bibnamefont {von
  Milczewski}}, \bibinfo {author} {\bibfnamefont {F.}~\bibnamefont {Rose}},\
  and\ \bibinfo {author} {\bibfnamefont {R.}~\bibnamefont {Schmidt}},\
  }\bibfield  {title} {\bibinfo {title} {Functional-renormalization-group
  approach to strongly coupled bose-fermi mixtures in two dimensions},\ }\href
  {https://doi.org/10.1103/PhysRevA.105.013317} {\bibfield  {journal} {\bibinfo
   {journal} {Phys. Rev. A}\ }\textbf {\bibinfo {volume} {105}},\ \bibinfo
  {pages} {013317} (\bibinfo {year} {2022})}\BibitemShut {NoStop}%
\bibitem [{\citenamefont {Diessel}\ \emph {et~al.}(2022)\citenamefont
  {Diessel}, \citenamefont {von Milczewski}, \citenamefont {Christianen},\ and\
  \citenamefont {Schmidt}}]{Diessel2022}%
  \BibitemOpen
  \bibfield  {author} {\bibinfo {author} {\bibfnamefont {O.~K.}\ \bibnamefont
  {Diessel}}, \bibinfo {author} {\bibfnamefont {J.}~\bibnamefont {von
  Milczewski}}, \bibinfo {author} {\bibfnamefont {A.}~\bibnamefont
  {Christianen}},\ and\ \bibinfo {author} {\bibfnamefont {R.}~\bibnamefont
  {Schmidt}},\ }\bibfield  {title} {\bibinfo {title} {Probing molecular
  spectral functions and unconventional pairing using raman spectroscopy},\
  }\href {https://doi.org/10.48550/ARXIV.2209.11758} {\bibfield  {journal}
  {\bibinfo  {journal} {arXiv:2209.11758}\ } (\bibinfo {year}
  {2022})}\BibitemShut {NoStop}%
\bibitem [{\citenamefont {Duda}\ \emph {et~al.}(2023)\citenamefont {Duda},
  \citenamefont {Chen}, \citenamefont {Schindewolf}, \citenamefont {Bause},
  \citenamefont {von Milczewski}, \citenamefont {Schmidt}, \citenamefont
  {Bloch},\ and\ \citenamefont {Luo}}]{Duda2023}%
  \BibitemOpen
  \bibfield  {author} {\bibinfo {author} {\bibfnamefont {M.}~\bibnamefont
  {Duda}}, \bibinfo {author} {\bibfnamefont {X.-Y.}\ \bibnamefont {Chen}},
  \bibinfo {author} {\bibfnamefont {A.}~\bibnamefont {Schindewolf}}, \bibinfo
  {author} {\bibfnamefont {R.}~\bibnamefont {Bause}}, \bibinfo {author}
  {\bibfnamefont {J.}~\bibnamefont {von Milczewski}}, \bibinfo {author}
  {\bibfnamefont {R.}~\bibnamefont {Schmidt}}, \bibinfo {author} {\bibfnamefont
  {I.}~\bibnamefont {Bloch}},\ and\ \bibinfo {author} {\bibfnamefont {X.-Y.}\
  \bibnamefont {Luo}},\ }\bibfield  {title} {\bibinfo {title} {Transition from
  a polaronic condensate to a degenerate fermi gas of heteronuclear
  molecules},\ }\href {https://doi.org/10.1038/s41567-023-01948-1} {\bibfield
  {journal} {\bibinfo  {journal} {Nat. Phys.}\ }\textbf {\bibinfo {volume}
  {19}},\ \bibinfo {pages} {720} (\bibinfo {year} {2023})}\BibitemShut
  {NoStop}%
\bibitem [{\citenamefont {Ness}\ \emph {et~al.}(2020)\citenamefont {Ness},
  \citenamefont {Shkedrov}, \citenamefont {Florshaim}, \citenamefont {Diessel},
  \citenamefont {von Milczewski}, \citenamefont {Schmidt},\ and\ \citenamefont
  {Sagi}}]{Ness2020}%
  \BibitemOpen
  \bibfield  {author} {\bibinfo {author} {\bibfnamefont {G.}~\bibnamefont
  {Ness}}, \bibinfo {author} {\bibfnamefont {C.}~\bibnamefont {Shkedrov}},
  \bibinfo {author} {\bibfnamefont {Y.}~\bibnamefont {Florshaim}}, \bibinfo
  {author} {\bibfnamefont {O.~K.}\ \bibnamefont {Diessel}}, \bibinfo {author}
  {\bibfnamefont {J.}~\bibnamefont {von Milczewski}}, \bibinfo {author}
  {\bibfnamefont {R.}~\bibnamefont {Schmidt}},\ and\ \bibinfo {author}
  {\bibfnamefont {Y.}~\bibnamefont {Sagi}},\ }\bibfield  {title} {\bibinfo
  {title} {Observation of a smooth polaron-molecule transition in a degenerate
  fermi gas},\ }\href {https://doi.org/10.1103/PhysRevX.10.041019} {\bibfield
  {journal} {\bibinfo  {journal} {Phys. Rev. X}\ }\textbf {\bibinfo {volume}
  {10}},\ \bibinfo {pages} {041019} (\bibinfo {year} {2020})}\BibitemShut
  {NoStop}%
\bibitem [{\citenamefont {Fritsche}\ \emph {et~al.}(2021)\citenamefont
  {Fritsche}, \citenamefont {Baroni}, \citenamefont {Dobler}, \citenamefont
  {Kirilov}, \citenamefont {Huang}, \citenamefont {Grimm}, \citenamefont
  {Bruun},\ and\ \citenamefont {Massignan}}]{Fritsche2021}%
  \BibitemOpen
  \bibfield  {author} {\bibinfo {author} {\bibfnamefont {I.}~\bibnamefont
  {Fritsche}}, \bibinfo {author} {\bibfnamefont {C.}~\bibnamefont {Baroni}},
  \bibinfo {author} {\bibfnamefont {E.}~\bibnamefont {Dobler}}, \bibinfo
  {author} {\bibfnamefont {E.}~\bibnamefont {Kirilov}}, \bibinfo {author}
  {\bibfnamefont {B.}~\bibnamefont {Huang}}, \bibinfo {author} {\bibfnamefont
  {R.}~\bibnamefont {Grimm}}, \bibinfo {author} {\bibfnamefont {G.~M.}\
  \bibnamefont {Bruun}},\ and\ \bibinfo {author} {\bibfnamefont
  {P.}~\bibnamefont {Massignan}},\ }\bibfield  {title} {\bibinfo {title}
  {Stability and breakdown of fermi polarons in a strongly interacting
  fermi-bose mixture},\ }\href {https://doi.org/10.1103/PhysRevA.103.053314}
  {\bibfield  {journal} {\bibinfo  {journal} {Phys. Rev. A}\ }\textbf {\bibinfo
  {volume} {103}},\ \bibinfo {pages} {053314} (\bibinfo {year}
  {2021})}\BibitemShut {NoStop}%
\bibitem [{\citenamefont {Schirotzek}\ \emph {et~al.}(2009)\citenamefont
  {Schirotzek}, \citenamefont {Wu}, \citenamefont {Sommer},\ and\ \citenamefont
  {Zwierlein}}]{Schirotzek2009}%
  \BibitemOpen
  \bibfield  {author} {\bibinfo {author} {\bibfnamefont {A.}~\bibnamefont
  {Schirotzek}}, \bibinfo {author} {\bibfnamefont {C.-H.}\ \bibnamefont {Wu}},
  \bibinfo {author} {\bibfnamefont {A.}~\bibnamefont {Sommer}},\ and\ \bibinfo
  {author} {\bibfnamefont {M.~W.}\ \bibnamefont {Zwierlein}},\ }\bibfield
  {title} {\bibinfo {title} {Observation of fermi polarons in a tunable fermi
  liquid of ultracold atoms},\ }\href
  {https://doi.org/10.1103/PhysRevLett.102.230402} {\bibfield  {journal}
  {\bibinfo  {journal} {Phys. Rev. Lett.}\ }\textbf {\bibinfo {volume} {102}},\
  \bibinfo {pages} {230402} (\bibinfo {year} {2009})}\BibitemShut {NoStop}%
\bibitem [{\citenamefont {Koschorreck}\ \emph {et~al.}(2012)\citenamefont
  {Koschorreck}, \citenamefont {Pertot}, \citenamefont {Vogt}, \citenamefont
  {Fr\"{o}hlich}, \citenamefont {Feld},\ and\ \citenamefont
  {K\"{o}hl}}]{Koschorreck2012}%
  \BibitemOpen
  \bibfield  {author} {\bibinfo {author} {\bibfnamefont {M.}~\bibnamefont
  {Koschorreck}}, \bibinfo {author} {\bibfnamefont {D.}~\bibnamefont {Pertot}},
  \bibinfo {author} {\bibfnamefont {E.}~\bibnamefont {Vogt}}, \bibinfo {author}
  {\bibfnamefont {B.}~\bibnamefont {Fr\"{o}hlich}}, \bibinfo {author}
  {\bibfnamefont {M.}~\bibnamefont {Feld}},\ and\ \bibinfo {author}
  {\bibfnamefont {M.}~\bibnamefont {K\"{o}hl}},\ }\bibfield  {title} {\bibinfo
  {title} {Attractive and repulsive fermi polarons in two dimensions},\ }\href
  {https://doi.org/10.1038/nature11151} {\bibfield  {journal} {\bibinfo
  {journal} {Nature}\ }\textbf {\bibinfo {volume} {485}},\ \bibinfo {pages}
  {619} (\bibinfo {year} {2012})}\BibitemShut {NoStop}%
\bibitem [{\citenamefont {Hu}\ \emph {et~al.}(2016)\citenamefont {Hu},
  \citenamefont {Van~de Graaff}, \citenamefont {Kedar}, \citenamefont {Corson},
  \citenamefont {Cornell},\ and\ \citenamefont {Jin}}]{Hu2016}%
  \BibitemOpen
  \bibfield  {author} {\bibinfo {author} {\bibfnamefont {M.-G.}\ \bibnamefont
  {Hu}}, \bibinfo {author} {\bibfnamefont {M.~J.}\ \bibnamefont {Van~de
  Graaff}}, \bibinfo {author} {\bibfnamefont {D.}~\bibnamefont {Kedar}},
  \bibinfo {author} {\bibfnamefont {J.~P.}\ \bibnamefont {Corson}}, \bibinfo
  {author} {\bibfnamefont {E.~A.}\ \bibnamefont {Cornell}},\ and\ \bibinfo
  {author} {\bibfnamefont {D.~S.}\ \bibnamefont {Jin}},\ }\bibfield  {title}
  {\bibinfo {title} {Bose polarons in the strongly interacting regime},\ }\href
  {https://doi.org/10.1103/PhysRevLett.117.055301} {\bibfield  {journal}
  {\bibinfo  {journal} {Phys. Rev. Lett.}\ }\textbf {\bibinfo {volume} {117}},\
  \bibinfo {pages} {055301} (\bibinfo {year} {2016})}\BibitemShut {NoStop}%
\bibitem [{\citenamefont {J{\o}rgensen}\ \emph {et~al.}(2016)\citenamefont
  {J{\o}rgensen}, \citenamefont {Wacker}, \citenamefont {Skalmstang},
  \citenamefont {Parish}, \citenamefont {Levinsen}, \citenamefont
  {Christensen}, \citenamefont {Bruun},\ and\ \citenamefont
  {Arlt}}]{jorgensen2016}%
  \BibitemOpen
  \bibfield  {author} {\bibinfo {author} {\bibfnamefont {N.~B.}\ \bibnamefont
  {J{\o}rgensen}}, \bibinfo {author} {\bibfnamefont {L.}~\bibnamefont
  {Wacker}}, \bibinfo {author} {\bibfnamefont {K.~T.}\ \bibnamefont
  {Skalmstang}}, \bibinfo {author} {\bibfnamefont {M.~M.}\ \bibnamefont
  {Parish}}, \bibinfo {author} {\bibfnamefont {J.}~\bibnamefont {Levinsen}},
  \bibinfo {author} {\bibfnamefont {R.~S.}\ \bibnamefont {Christensen}},
  \bibinfo {author} {\bibfnamefont {G.~M.}\ \bibnamefont {Bruun}},\ and\
  \bibinfo {author} {\bibfnamefont {J.~J.}\ \bibnamefont {Arlt}},\ }\bibfield
  {title} {\bibinfo {title} {{Observation of attractive and repulsive polarons
  in a Bose-Einstein condensate}},\ }\href
  {https://doi.org/10.1103/PhysRevLett.117.055302} {\bibfield  {journal}
  {\bibinfo  {journal} {Phys. Rev. Lett.}\ }\textbf {\bibinfo {volume} {117}},\
  \bibinfo {pages} {055302} (\bibinfo {year} {2016})}\BibitemShut {NoStop}%
\bibitem [{\citenamefont {Yan}\ \emph {et~al.}(2020)\citenamefont {Yan},
  \citenamefont {Ni}, \citenamefont {Robens},\ and\ \citenamefont
  {Zwierlein}}]{yan2020bose}%
  \BibitemOpen
  \bibfield  {author} {\bibinfo {author} {\bibfnamefont {Z.~Z.}\ \bibnamefont
  {Yan}}, \bibinfo {author} {\bibfnamefont {Y.}~\bibnamefont {Ni}}, \bibinfo
  {author} {\bibfnamefont {C.}~\bibnamefont {Robens}},\ and\ \bibinfo {author}
  {\bibfnamefont {M.~W.}\ \bibnamefont {Zwierlein}},\ }\bibfield  {title}
  {\bibinfo {title} {Bose polarons near quantum criticality},\ }\href
  {https://doi.org/10.1126/science.aax5850} {\bibfield  {journal} {\bibinfo
  {journal} {Science}\ }\textbf {\bibinfo {volume} {368}},\ \bibinfo {pages}
  {190} (\bibinfo {year} {2020})}\BibitemShut {NoStop}%
\bibitem [{\citenamefont {Combescot}\ and\ \citenamefont
  {Giraud}(2008)}]{Combescot2008}%
  \BibitemOpen
  \bibfield  {author} {\bibinfo {author} {\bibfnamefont {R.}~\bibnamefont
  {Combescot}}\ and\ \bibinfo {author} {\bibfnamefont {S.}~\bibnamefont
  {Giraud}},\ }\bibfield  {title} {\bibinfo {title} {Normal state of highly
  polarized fermi gases: Full many-body treatment},\ }\href
  {https://doi.org/10.1103/physrevlett.101.050404} {\bibfield  {journal}
  {\bibinfo  {journal} {Phys. Rev. Lett.}\ }\textbf {\bibinfo {volume} {101}},\
  \bibinfo {pages} {050404} (\bibinfo {year} {2008})}\BibitemShut {NoStop}%
\bibitem [{\citenamefont {Mora}\ and\ \citenamefont {Chevy}(2009)}]{Mora2009}%
  \BibitemOpen
  \bibfield  {author} {\bibinfo {author} {\bibfnamefont {C.}~\bibnamefont
  {Mora}}\ and\ \bibinfo {author} {\bibfnamefont {F.}~\bibnamefont {Chevy}},\
  }\bibfield  {title} {\bibinfo {title} {Ground state of a tightly bound
  composite dimer immersed in a fermi sea},\ }\href
  {https://doi.org/10.1103/physreva.80.033607} {\bibfield  {journal} {\bibinfo
  {journal} {Phys. Rev. A}\ }\textbf {\bibinfo {volume} {80}},\ \bibinfo
  {pages} {033607} (\bibinfo {year} {2009})}\BibitemShut {NoStop}%
\bibitem [{\citenamefont {Parish}(2011)}]{Parish2011}%
  \BibitemOpen
  \bibfield  {author} {\bibinfo {author} {\bibfnamefont {M.~M.}\ \bibnamefont
  {Parish}},\ }\bibfield  {title} {\bibinfo {title} {Polaron-molecule
  transitions in a two-dimensional fermi gas},\ }\href
  {https://doi.org/10.1103/PhysRevA.83.051603} {\bibfield  {journal} {\bibinfo
  {journal} {Phys. Rev. A}\ }\textbf {\bibinfo {volume} {83}},\ \bibinfo
  {pages} {051603} (\bibinfo {year} {2011})}\BibitemShut {NoStop}%
\bibitem [{\citenamefont {Parish}\ and\ \citenamefont
  {Levinsen}(2013)}]{Parish2013}%
  \BibitemOpen
  \bibfield  {author} {\bibinfo {author} {\bibfnamefont {M.~M.}\ \bibnamefont
  {Parish}}\ and\ \bibinfo {author} {\bibfnamefont {J.}~\bibnamefont
  {Levinsen}},\ }\bibfield  {title} {\bibinfo {title} {Highly polarized fermi
  gases in two dimensions},\ }\href
  {https://doi.org/10.1103/physreva.87.033616} {\bibfield  {journal} {\bibinfo
  {journal} {Phys. Rev. A}\ }\textbf {\bibinfo {volume} {87}},\ \bibinfo
  {pages} {033616} (\bibinfo {year} {2013})}\BibitemShut {NoStop}%
\bibitem [{\citenamefont {Cui}(2020)}]{cui2020fermi}%
  \BibitemOpen
  \bibfield  {author} {\bibinfo {author} {\bibfnamefont {X.}~\bibnamefont
  {Cui}},\ }\bibfield  {title} {\bibinfo {title} {Fermi polaron revisited:
  Polaron-molecule transition and coexistence},\ }\href
  {https://doi.org/10.1103/PhysRevA.102.061301} {\bibfield  {journal} {\bibinfo
   {journal} {Phys. Rev. A}\ }\textbf {\bibinfo {volume} {102}},\ \bibinfo
  {pages} {061301} (\bibinfo {year} {2020})}\BibitemShut {NoStop}%
\bibitem [{\citenamefont {Parish}\ \emph {et~al.}(2021)\citenamefont {Parish},
  \citenamefont {Adlong}, \citenamefont {Liu},\ and\ \citenamefont
  {Levinsen}}]{Parish2020}%
  \BibitemOpen
  \bibfield  {author} {\bibinfo {author} {\bibfnamefont {M.~M.}\ \bibnamefont
  {Parish}}, \bibinfo {author} {\bibfnamefont {H.~S.}\ \bibnamefont {Adlong}},
  \bibinfo {author} {\bibfnamefont {W.~E.}\ \bibnamefont {Liu}},\ and\ \bibinfo
  {author} {\bibfnamefont {J.}~\bibnamefont {Levinsen}},\ }\bibfield  {title}
  {\bibinfo {title} {Thermodynamic signatures of the polaron-molecule
  transition in a fermi gas},\ }\href
  {https://doi.org/10.1103/PhysRevA.103.023312} {\bibfield  {journal} {\bibinfo
   {journal} {Phys. Rev. A}\ }\textbf {\bibinfo {volume} {103}},\ \bibinfo
  {pages} {023312} (\bibinfo {year} {2021})}\BibitemShut {NoStop}%
\bibitem [{\citenamefont {Chevy}(2006)}]{Chevy2006}%
  \BibitemOpen
  \bibfield  {author} {\bibinfo {author} {\bibfnamefont {F.}~\bibnamefont
  {Chevy}},\ }\bibfield  {title} {\bibinfo {title} {Universal phase diagram of
  a strongly interacting fermi gas with unbalanced spin populations},\ }\href
  {https://doi.org/10.1103/physreva.74.063628} {\bibfield  {journal} {\bibinfo
  {journal} {Phys. Rev. A}\ }\textbf {\bibinfo {volume} {74}},\ \bibinfo
  {pages} {063628} (\bibinfo {year} {2006})}\BibitemShut {NoStop}%
\bibitem [{\citenamefont {Punk}\ \emph {et~al.}(2009)\citenamefont {Punk},
  \citenamefont {Dumitrescu},\ and\ \citenamefont {Zwerger}}]{Punk2009}%
  \BibitemOpen
  \bibfield  {author} {\bibinfo {author} {\bibfnamefont {M.}~\bibnamefont
  {Punk}}, \bibinfo {author} {\bibfnamefont {P.~T.}\ \bibnamefont
  {Dumitrescu}},\ and\ \bibinfo {author} {\bibfnamefont {W.}~\bibnamefont
  {Zwerger}},\ }\bibfield  {title} {\bibinfo {title} {Polaron-to-molecule
  transition in a strongly imbalanced fermi gas},\ }\href
  {https://doi.org/10.1103/physreva.80.053605} {\bibfield  {journal} {\bibinfo
  {journal} {Phys. Rev. A}\ }\textbf {\bibinfo {volume} {80}},\ \bibinfo
  {pages} {053605} (\bibinfo {year} {2009})}\BibitemShut {NoStop}%
\bibitem [{\citenamefont {Z\"ollner}\ \emph {et~al.}(2011)\citenamefont
  {Z\"ollner}, \citenamefont {Bruun},\ and\ \citenamefont
  {Pethick}}]{Zoellner2011}%
  \BibitemOpen
  \bibfield  {author} {\bibinfo {author} {\bibfnamefont {S.}~\bibnamefont
  {Z\"ollner}}, \bibinfo {author} {\bibfnamefont {G.~M.}\ \bibnamefont
  {Bruun}},\ and\ \bibinfo {author} {\bibfnamefont {C.~J.}\ \bibnamefont
  {Pethick}},\ }\bibfield  {title} {\bibinfo {title} {Polarons and molecules in
  a two-dimensional fermi gas},\ }\href
  {https://doi.org/10.1103/PhysRevA.83.021603} {\bibfield  {journal} {\bibinfo
  {journal} {Phys. Rev. A}\ }\textbf {\bibinfo {volume} {83}},\ \bibinfo
  {pages} {021603} (\bibinfo {year} {2011})}\BibitemShut {NoStop}%
\bibitem [{\citenamefont {Trefzger}\ and\ \citenamefont
  {Castin}(2012)}]{Trefzger_2012}%
  \BibitemOpen
  \bibfield  {author} {\bibinfo {author} {\bibfnamefont {C.}~\bibnamefont
  {Trefzger}}\ and\ \bibinfo {author} {\bibfnamefont {Y.}~\bibnamefont
  {Castin}},\ }\bibfield  {title} {\bibinfo {title} {Impurity in a fermi sea on
  a narrow feshbach resonance: A variational study of the polaronic and
  dimeronic branches},\ }\href {https://doi.org/10.1103/PhysRevA.85.053612}
  {\bibfield  {journal} {\bibinfo  {journal} {Phys. Rev. A}\ }\textbf {\bibinfo
  {volume} {85}},\ \bibinfo {pages} {053612} (\bibinfo {year}
  {2012})}\BibitemShut {NoStop}%
\bibitem [{\citenamefont {Schmidt}\ \emph {et~al.}(2012)\citenamefont
  {Schmidt}, \citenamefont {Enss}, \citenamefont {Pietil\"a},\ and\
  \citenamefont {Demler}}]{Schmidt2012}%
  \BibitemOpen
  \bibfield  {author} {\bibinfo {author} {\bibfnamefont {R.}~\bibnamefont
  {Schmidt}}, \bibinfo {author} {\bibfnamefont {T.}~\bibnamefont {Enss}},
  \bibinfo {author} {\bibfnamefont {V.}~\bibnamefont {Pietil\"a}},\ and\
  \bibinfo {author} {\bibfnamefont {E.}~\bibnamefont {Demler}},\ }\bibfield
  {title} {\bibinfo {title} {Fermi polarons in two dimensions},\ }\href
  {https://doi.org/10.1103/physreva.85.021602} {\bibfield  {journal} {\bibinfo
  {journal} {Phys. Rev. A}\ }\textbf {\bibinfo {volume} {85}},\ \bibinfo
  {pages} {021602} (\bibinfo {year} {2012})}\BibitemShut {NoStop}%
\bibitem [{\citenamefont {Combescot}\ \emph {et~al.}(2007)\citenamefont
  {Combescot}, \citenamefont {Recati}, \citenamefont {Lobo},\ and\
  \citenamefont {Chevy}}]{Combescot2007}%
  \BibitemOpen
  \bibfield  {author} {\bibinfo {author} {\bibfnamefont {R.}~\bibnamefont
  {Combescot}}, \bibinfo {author} {\bibfnamefont {A.}~\bibnamefont {Recati}},
  \bibinfo {author} {\bibfnamefont {C.}~\bibnamefont {Lobo}},\ and\ \bibinfo
  {author} {\bibfnamefont {F.}~\bibnamefont {Chevy}},\ }\bibfield  {title}
  {\bibinfo {title} {Normal state of highly polarized fermi gases: Simple
  many-body approaches},\ }\href
  {https://doi.org/10.1103/PhysRevLett.98.180402} {\bibfield  {journal}
  {\bibinfo  {journal} {Phys. Rev. Lett.}\ }\textbf {\bibinfo {volume} {98}},\
  \bibinfo {pages} {180402} (\bibinfo {year} {2007})}\BibitemShut {NoStop}%
\bibitem [{\citenamefont {Combescot}\ \emph {et~al.}(2009)\citenamefont
  {Combescot}, \citenamefont {Giraud},\ and\ \citenamefont
  {Leyronas}}]{Combescot2009}%
  \BibitemOpen
  \bibfield  {author} {\bibinfo {author} {\bibfnamefont {R.}~\bibnamefont
  {Combescot}}, \bibinfo {author} {\bibfnamefont {S.}~\bibnamefont {Giraud}},\
  and\ \bibinfo {author} {\bibfnamefont {X.}~\bibnamefont {Leyronas}},\
  }\bibfield  {title} {\bibinfo {title} {Analytical theory of the dressed bound
  state in highly polarized fermi gases},\ }\href
  {https://doi.org/10.1209/0295-5075/88/60007} {\bibfield  {journal} {\bibinfo
  {journal} {{EPL}}\ }\textbf {\bibinfo {volume} {88}},\ \bibinfo {pages}
  {60007} (\bibinfo {year} {2009})}\BibitemShut {NoStop}%
\bibitem [{\citenamefont {Bruun}\ and\ \citenamefont
  {Massignan}(2010)}]{Bruun2010}%
  \BibitemOpen
  \bibfield  {author} {\bibinfo {author} {\bibfnamefont {G.~M.}\ \bibnamefont
  {Bruun}}\ and\ \bibinfo {author} {\bibfnamefont {P.}~\bibnamefont
  {Massignan}},\ }\bibfield  {title} {\bibinfo {title} {Decay of polarons and
  molecules in a strongly polarized fermi gas},\ }\href
  {https://doi.org/10.1103/physrevlett.105.020403} {\bibfield  {journal}
  {\bibinfo  {journal} {Phys. Rev. Lett.}\ }\textbf {\bibinfo {volume} {105}},\
  \bibinfo {pages} {020403} (\bibinfo {year} {2010})}\BibitemShut {NoStop}%
\bibitem [{\citenamefont {Massignan}\ and\ \citenamefont
  {Bruun}(2011)}]{Massignan2011}%
  \BibitemOpen
  \bibfield  {author} {\bibinfo {author} {\bibfnamefont {P.}~\bibnamefont
  {Massignan}}\ and\ \bibinfo {author} {\bibfnamefont {G.~M.}\ \bibnamefont
  {Bruun}},\ }\bibfield  {title} {\bibinfo {title} {Repulsive polarons and
  itinerant ferromagnetism in strongly polarized fermi gases},\ }\href
  {https://doi.org/10.1140/epjd/e2011-20084-5} {\bibfield  {journal} {\bibinfo
  {journal} {EPJ D}\ }\textbf {\bibinfo {volume} {65}},\ \bibinfo {pages} {83}
  (\bibinfo {year} {2011})}\BibitemShut {NoStop}%
\bibitem [{\citenamefont {Lobo}\ \emph {et~al.}(2006)\citenamefont {Lobo},
  \citenamefont {Recati}, \citenamefont {Giorgini},\ and\ \citenamefont
  {Stringari}}]{Lobo2006}%
  \BibitemOpen
  \bibfield  {author} {\bibinfo {author} {\bibfnamefont {C.}~\bibnamefont
  {Lobo}}, \bibinfo {author} {\bibfnamefont {A.}~\bibnamefont {Recati}},
  \bibinfo {author} {\bibfnamefont {S.}~\bibnamefont {Giorgini}},\ and\
  \bibinfo {author} {\bibfnamefont {S.}~\bibnamefont {Stringari}},\ }\bibfield
  {title} {\bibinfo {title} {Normal state of a polarized fermi gas at
  unitarity},\ }\href {https://doi.org/10.1103/physrevlett.97.200403}
  {\bibfield  {journal} {\bibinfo  {journal} {Phys. Rev. Lett.}\ }\textbf
  {\bibinfo {volume} {97}},\ \bibinfo {pages} {200403} (\bibinfo {year}
  {2006})}\BibitemShut {NoStop}%
\bibitem [{\citenamefont {Prokof'ev}\ and\ \citenamefont
  {Svistunov}(2008{\natexlab{a}})}]{Prokofev2008}%
  \BibitemOpen
  \bibfield  {author} {\bibinfo {author} {\bibfnamefont {N.}~\bibnamefont
  {Prokof'ev}}\ and\ \bibinfo {author} {\bibfnamefont {B.}~\bibnamefont
  {Svistunov}},\ }\bibfield  {title} {\bibinfo {title} {Fermi-polaron problem:
  Diagrammatic monte carlo method for divergent sign-alternating series},\
  }\href {https://doi.org/10.1103/PhysRevB.77.020408} {\bibfield  {journal}
  {\bibinfo  {journal} {Phys. Rev. B}\ }\textbf {\bibinfo {volume} {77}},\
  \bibinfo {pages} {020408} (\bibinfo {year} {2008}{\natexlab{a}})}\BibitemShut
  {NoStop}%
\bibitem [{\citenamefont {Prokof'ev}\ and\ \citenamefont
  {Svistunov}(2008{\natexlab{b}})}]{Prokofev2008a}%
  \BibitemOpen
  \bibfield  {author} {\bibinfo {author} {\bibfnamefont {N.~V.}\ \bibnamefont
  {Prokof'ev}}\ and\ \bibinfo {author} {\bibfnamefont {B.~V.}\ \bibnamefont
  {Svistunov}},\ }\bibfield  {title} {\bibinfo {title} {Bold diagrammatic
  {Monte Carlo}: A generic sign-problem tolerant technique for polaron models
  and possibly interacting many-body problems},\ }\href
  {https://doi.org/10.1103/physrevb.77.125101} {\bibfield  {journal} {\bibinfo
  {journal} {Phys. Rev. B}\ }\textbf {\bibinfo {volume} {77}},\ \bibinfo
  {pages} {125101} (\bibinfo {year} {2008}{\natexlab{b}})}\BibitemShut
  {NoStop}%
\bibitem [{\citenamefont {Bertaina}(2012)}]{Bertaina2012}%
  \BibitemOpen
  \bibfield  {author} {\bibinfo {author} {\bibfnamefont {G.}~\bibnamefont
  {Bertaina}},\ }\bibfield  {title} {\bibinfo {title} {{BCS}-{BEC} crossover in
  two dimensions: A quantum monte carlo study},\ }\href
  {https://doi.org/10.1063/1.4755827} {\bibfield  {journal} {\bibinfo
  {journal} {AIP Conference Proceedings}\ }\textbf {\bibinfo {volume} {1485}},\
  \bibinfo {pages} {286} (\bibinfo {year} {2012})}\BibitemShut {NoStop}%
\bibitem [{\citenamefont {Kroiss}\ and\ \citenamefont
  {Pollet}(2014)}]{Kroiss2014}%
  \BibitemOpen
  \bibfield  {author} {\bibinfo {author} {\bibfnamefont {P.}~\bibnamefont
  {Kroiss}}\ and\ \bibinfo {author} {\bibfnamefont {L.}~\bibnamefont
  {Pollet}},\ }\bibfield  {title} {\bibinfo {title} {Diagrammatic monte carlo
  study of quasi-two-dimensional fermi polarons},\ }\href
  {https://doi.org/10.1103/physrevb.90.104510} {\bibfield  {journal} {\bibinfo
  {journal} {Phys. Rev. B}\ }\textbf {\bibinfo {volume} {90}},\ \bibinfo
  {pages} {104510} (\bibinfo {year} {2014})}\BibitemShut {NoStop}%
\bibitem [{\citenamefont {Vlietinck}\ \emph {et~al.}(2014)\citenamefont
  {Vlietinck}, \citenamefont {Ryckebusch},\ and\ \citenamefont
  {Van~Houcke}}]{Vlietinck2014}%
  \BibitemOpen
  \bibfield  {author} {\bibinfo {author} {\bibfnamefont {J.}~\bibnamefont
  {Vlietinck}}, \bibinfo {author} {\bibfnamefont {J.}~\bibnamefont
  {Ryckebusch}},\ and\ \bibinfo {author} {\bibfnamefont {K.}~\bibnamefont
  {Van~Houcke}},\ }\bibfield  {title} {\bibinfo {title} {Diagrammatic monte
  carlo study of the fermi polaron in two dimensions},\ }\href
  {https://doi.org/10.1103/physrevb.89.085119} {\bibfield  {journal} {\bibinfo
  {journal} {Phys. Rev. B}\ }\textbf {\bibinfo {volume} {89}},\ \bibinfo
  {pages} {085119} (\bibinfo {year} {2014})}\BibitemShut {NoStop}%
\bibitem [{\citenamefont {Nikoli\ifmmode~\acute{c}\else \'{c}\fi{}}\ and\
  \citenamefont {Sachdev}(2007)}]{Nikolic2007}%
  \BibitemOpen
  \bibfield  {author} {\bibinfo {author} {\bibfnamefont {P.}~\bibnamefont
  {Nikoli\ifmmode~\acute{c}\else \'{c}\fi{}}}\ and\ \bibinfo {author}
  {\bibfnamefont {S.}~\bibnamefont {Sachdev}},\ }\bibfield  {title} {\bibinfo
  {title} {Renormalization-group fixed points, universal phase diagram, and
  $1/n$ expansion for quantum liquids with interactions near the unitarity
  limit},\ }\href {https://doi.org/10.1103/PhysRevA.75.033608} {\bibfield
  {journal} {\bibinfo  {journal} {Phys. Rev. A}\ }\textbf {\bibinfo {volume}
  {75}},\ \bibinfo {pages} {033608} (\bibinfo {year} {2007})}\BibitemShut
  {NoStop}%
\bibitem [{\citenamefont {Gubbels}\ and\ \citenamefont
  {Stoof}(2008)}]{Gubbels2008}%
  \BibitemOpen
  \bibfield  {author} {\bibinfo {author} {\bibfnamefont {K.~B.}\ \bibnamefont
  {Gubbels}}\ and\ \bibinfo {author} {\bibfnamefont {H.~T.~C.}\ \bibnamefont
  {Stoof}},\ }\bibfield  {title} {\bibinfo {title} {Renormalization group
  theory for the imbalanced fermi gas},\ }\href
  {https://doi.org/10.1103/physrevlett.100.140407} {\bibfield  {journal}
  {\bibinfo  {journal} {Phys. Rev. Lett.}\ }\textbf {\bibinfo {volume} {100}},\
  \bibinfo {pages} {140407} (\bibinfo {year} {2008})}\BibitemShut {NoStop}%
\bibitem [{\citenamefont {Schmidt}\ and\ \citenamefont
  {Enss}(2011)}]{Schmidt2011}%
  \BibitemOpen
  \bibfield  {author} {\bibinfo {author} {\bibfnamefont {R.}~\bibnamefont
  {Schmidt}}\ and\ \bibinfo {author} {\bibfnamefont {T.}~\bibnamefont {Enss}},\
  }\bibfield  {title} {\bibinfo {title} {Excitation spectra and rf response
  near the polaron-to-molecule transition from the functional renormalization
  group},\ }\href {https://doi.org/10.1103/physreva.83.063620} {\bibfield
  {journal} {\bibinfo  {journal} {Phys. Rev. A}\ }\textbf {\bibinfo {volume}
  {83}},\ \bibinfo {pages} {063620} (\bibinfo {year} {2011})}\BibitemShut
  {NoStop}%
\bibitem [{\citenamefont {Rath}\ and\ \citenamefont
  {Schmidt}(2013)}]{rath2013}%
  \BibitemOpen
  \bibfield  {author} {\bibinfo {author} {\bibfnamefont {S.~P.}\ \bibnamefont
  {Rath}}\ and\ \bibinfo {author} {\bibfnamefont {R.}~\bibnamefont {Schmidt}},\
  }\bibfield  {title} {\bibinfo {title} {Field-theoretical study of the bose
  polaron},\ }\href {https://doi.org/10.1103/PhysRevA.88.053632} {\bibfield
  {journal} {\bibinfo  {journal} {Phys. Rev. A}\ }\textbf {\bibinfo {volume}
  {88}},\ \bibinfo {pages} {053632} (\bibinfo {year} {2013})}\BibitemShut
  {NoStop}%
\bibitem [{\citenamefont {Kamikado}\ \emph {et~al.}(2017)\citenamefont
  {Kamikado}, \citenamefont {Kanazawa},\ and\ \citenamefont
  {Uchino}}]{Kamikado2017}%
  \BibitemOpen
  \bibfield  {author} {\bibinfo {author} {\bibfnamefont {K.}~\bibnamefont
  {Kamikado}}, \bibinfo {author} {\bibfnamefont {T.}~\bibnamefont {Kanazawa}},\
  and\ \bibinfo {author} {\bibfnamefont {S.}~\bibnamefont {Uchino}},\
  }\bibfield  {title} {\bibinfo {title} {Mobile impurity in a fermi sea from
  the functional renormalization group analytically continued to real time},\
  }\href {https://doi.org/10.1103/physreva.95.013612} {\bibfield  {journal}
  {\bibinfo  {journal} {Phys. Rev. A}\ }\textbf {\bibinfo {volume} {95}},\
  \bibinfo {pages} {013612} (\bibinfo {year} {2017})}\BibitemShut {NoStop}%
\bibitem [{\citenamefont {Isaule}\ \emph {et~al.}(2021)\citenamefont {Isaule},
  \citenamefont {Morera}, \citenamefont {Massignan},\ and\ \citenamefont
  {Juli\'a-D\'{\i}az}}]{Isaule2021}%
  \BibitemOpen
  \bibfield  {author} {\bibinfo {author} {\bibfnamefont {F.}~\bibnamefont
  {Isaule}}, \bibinfo {author} {\bibfnamefont {I.}~\bibnamefont {Morera}},
  \bibinfo {author} {\bibfnamefont {P.}~\bibnamefont {Massignan}},\ and\
  \bibinfo {author} {\bibfnamefont {B.}~\bibnamefont {Juli\'a-D\'{\i}az}},\
  }\bibfield  {title} {\bibinfo {title} {Renormalization-group study of bose
  polarons},\ }\href {https://doi.org/10.1103/PhysRevA.104.023317} {\bibfield
  {journal} {\bibinfo  {journal} {Phys. Rev. A}\ }\textbf {\bibinfo {volume}
  {104}},\ \bibinfo {pages} {023317} (\bibinfo {year} {2021})}\BibitemShut
  {NoStop}%
\bibitem [{\citenamefont {Tajima}\ and\ \citenamefont
  {Uchino}(2018)}]{Tajima2018}%
  \BibitemOpen
  \bibfield  {author} {\bibinfo {author} {\bibfnamefont {H.}~\bibnamefont
  {Tajima}}\ and\ \bibinfo {author} {\bibfnamefont {S.}~\bibnamefont
  {Uchino}},\ }\bibfield  {title} {\bibinfo {title} {Many fermi polarons at
  nonzero temperature},\ }\href {https://doi.org/10.1088/1367-2630/aad1e7}
  {\bibfield  {journal} {\bibinfo  {journal} {NJP}\ }\textbf {\bibinfo {volume}
  {20}},\ \bibinfo {pages} {073048} (\bibinfo {year} {2018})}\BibitemShut
  {NoStop}%
\bibitem [{\citenamefont {Adlong}\ \emph {et~al.}(2020)\citenamefont {Adlong},
  \citenamefont {Liu}, \citenamefont {Scazza}, \citenamefont {Zaccanti},
  \citenamefont {Oppong}, \citenamefont {F\"olling}, \citenamefont {Parish},\
  and\ \citenamefont {Levinsen}}]{Adlong2020}%
  \BibitemOpen
  \bibfield  {author} {\bibinfo {author} {\bibfnamefont {H.~S.}\ \bibnamefont
  {Adlong}}, \bibinfo {author} {\bibfnamefont {W.~E.}\ \bibnamefont {Liu}},
  \bibinfo {author} {\bibfnamefont {F.}~\bibnamefont {Scazza}}, \bibinfo
  {author} {\bibfnamefont {M.}~\bibnamefont {Zaccanti}}, \bibinfo {author}
  {\bibfnamefont {N.~D.}\ \bibnamefont {Oppong}}, \bibinfo {author}
  {\bibfnamefont {S.}~\bibnamefont {F\"olling}}, \bibinfo {author}
  {\bibfnamefont {M.~M.}\ \bibnamefont {Parish}},\ and\ \bibinfo {author}
  {\bibfnamefont {J.}~\bibnamefont {Levinsen}},\ }\bibfield  {title} {\bibinfo
  {title} {Quasiparticle lifetime of the repulsive fermi polaron},\ }\href
  {https://doi.org/10.1103/PhysRevLett.125.133401} {\bibfield  {journal}
  {\bibinfo  {journal} {Phys. Rev. Lett.}\ }\textbf {\bibinfo {volume} {125}},\
  \bibinfo {pages} {133401} (\bibinfo {year} {2020})}\BibitemShut {NoStop}%
\bibitem [{\citenamefont {Hu}\ \emph {et~al.}(2018)\citenamefont {Hu},
  \citenamefont {Mulkerin}, \citenamefont {Wang},\ and\ \citenamefont
  {Liu}}]{Hu2018}%
  \BibitemOpen
  \bibfield  {author} {\bibinfo {author} {\bibfnamefont {H.}~\bibnamefont
  {Hu}}, \bibinfo {author} {\bibfnamefont {B.~C.}\ \bibnamefont {Mulkerin}},
  \bibinfo {author} {\bibfnamefont {J.}~\bibnamefont {Wang}},\ and\ \bibinfo
  {author} {\bibfnamefont {X.-J.}\ \bibnamefont {Liu}},\ }\bibfield  {title}
  {\bibinfo {title} {Attractive fermi polarons at nonzero temperatures with a
  finite impurity concentration},\ }\href
  {https://doi.org/10.1103/PhysRevA.98.013626} {\bibfield  {journal} {\bibinfo
  {journal} {Phys. Rev. A}\ }\textbf {\bibinfo {volume} {98}},\ \bibinfo
  {pages} {013626} (\bibinfo {year} {2018})}\BibitemShut {NoStop}%
\bibitem [{\citenamefont {Hu}\ and\ \citenamefont {Liu}(2022)}]{Hu2022}%
  \BibitemOpen
  \bibfield  {author} {\bibinfo {author} {\bibfnamefont {H.}~\bibnamefont
  {Hu}}\ and\ \bibinfo {author} {\bibfnamefont {X.-J.}\ \bibnamefont {Liu}},\
  }\bibfield  {title} {\bibinfo {title} {Fermi polarons at finite temperature:
  Spectral function and rf spectroscopy},\ }\href
  {https://doi.org/10.1103/PhysRevA.105.043303} {\bibfield  {journal} {\bibinfo
   {journal} {Phys. Rev. A}\ }\textbf {\bibinfo {volume} {105}},\ \bibinfo
  {pages} {043303} (\bibinfo {year} {2022})}\BibitemShut {NoStop}%
\bibitem [{\citenamefont {Hu}\ and\ \citenamefont {Liu}(2023)}]{Hu2023}%
  \BibitemOpen
  \bibfield  {author} {\bibinfo {author} {\bibfnamefont {H.}~\bibnamefont
  {Hu}}\ and\ \bibinfo {author} {\bibfnamefont {X.-J.}\ \bibnamefont {Liu}},\
  }\bibfield  {title} {\bibinfo {title} {Spectral function of fermi polarons at
  finite temperature from a self-consistent many-body $t$-matrix approach in
  real frequency},\ }\href {https://doi.org/10.48550/ARXIV.2311.11554}
  {\bibfield  {journal} {\bibinfo  {journal} {arXiv:2311.11554}\ } (\bibinfo
  {year} {2023})}\BibitemShut {NoStop}%
\bibitem [{\citenamefont {Baym}\ and\ \citenamefont
  {Pethick}(1991)}]{Baym1991}%
  \BibitemOpen
  \bibfield  {author} {\bibinfo {author} {\bibfnamefont {G.}~\bibnamefont
  {Baym}}\ and\ \bibinfo {author} {\bibfnamefont {C.}~\bibnamefont {Pethick}},\
  }\href {https://doi.org/10.1002/9783527617159} {\emph {\bibinfo {title}
  {Landau Fermi-Liquid Theory}}}\ (\bibinfo  {publisher} {Wiley},\ \bibinfo
  {year} {1991})\BibitemShut {NoStop}%
\bibitem [{\citenamefont {Bruun}\ \emph {et~al.}(2008)\citenamefont {Bruun},
  \citenamefont {Recati}, \citenamefont {Pethick}, \citenamefont {Smith},\ and\
  \citenamefont {Stringari}}]{Bruun2008}%
  \BibitemOpen
  \bibfield  {author} {\bibinfo {author} {\bibfnamefont {G.~M.}\ \bibnamefont
  {Bruun}}, \bibinfo {author} {\bibfnamefont {A.}~\bibnamefont {Recati}},
  \bibinfo {author} {\bibfnamefont {C.~J.}\ \bibnamefont {Pethick}}, \bibinfo
  {author} {\bibfnamefont {H.}~\bibnamefont {Smith}},\ and\ \bibinfo {author}
  {\bibfnamefont {S.}~\bibnamefont {Stringari}},\ }\bibfield  {title} {\bibinfo
  {title} {Collisional properties of a polarized fermi gas with resonant
  interactions},\ }\href {https://doi.org/10.1103/PhysRevLett.100.240406}
  {\bibfield  {journal} {\bibinfo  {journal} {Phys. Rev. Lett.}\ }\textbf
  {\bibinfo {volume} {100}},\ \bibinfo {pages} {240406} (\bibinfo {year}
  {2008})}\BibitemShut {NoStop}%
\bibitem [{\citenamefont {Sadeghzadeh}\ \emph {et~al.}(2011)\citenamefont
  {Sadeghzadeh}, \citenamefont {Bruun}, \citenamefont {Lobo}, \citenamefont
  {Massignan},\ and\ \citenamefont {Recati}}]{Sadeghzadeh2011}%
  \BibitemOpen
  \bibfield  {author} {\bibinfo {author} {\bibfnamefont {K.}~\bibnamefont
  {Sadeghzadeh}}, \bibinfo {author} {\bibfnamefont {G.~M.}\ \bibnamefont
  {Bruun}}, \bibinfo {author} {\bibfnamefont {C.}~\bibnamefont {Lobo}},
  \bibinfo {author} {\bibfnamefont {P.}~\bibnamefont {Massignan}},\ and\
  \bibinfo {author} {\bibfnamefont {A.}~\bibnamefont {Recati}},\ }\bibfield
  {title} {\bibinfo {title} {Metastability in spin-polarized fermi gases and
  quasiparticle decays},\ }\href
  {https://doi.org/10.1088/1367-2630/13/5/055011} {\bibfield  {journal}
  {\bibinfo  {journal} {NJP}\ }\textbf {\bibinfo {volume} {13}},\ \bibinfo
  {pages} {055011} (\bibinfo {year} {2011})}\BibitemShut {NoStop}%
\bibitem [{\citenamefont {Ngampruetikorn}\ \emph {et~al.}(2012)\citenamefont
  {Ngampruetikorn}, \citenamefont {Levinsen},\ and\ \citenamefont
  {Parish}}]{Ngampruetikorn2012}%
  \BibitemOpen
  \bibfield  {author} {\bibinfo {author} {\bibfnamefont {V.}~\bibnamefont
  {Ngampruetikorn}}, \bibinfo {author} {\bibfnamefont {J.}~\bibnamefont
  {Levinsen}},\ and\ \bibinfo {author} {\bibfnamefont {M.~M.}\ \bibnamefont
  {Parish}},\ }\bibfield  {title} {\bibinfo {title} {Repulsive polarons in
  two-dimensional fermi gases},\ }\href
  {https://doi.org/10.1209/0295-5075/98/30005} {\bibfield  {journal} {\bibinfo
  {journal} {EPL}\ }\textbf {\bibinfo {volume} {98}},\ \bibinfo {pages} {30005}
  (\bibinfo {year} {2012})}\BibitemShut {NoStop}%
\bibitem [{\citenamefont {Trefzger}\ and\ \citenamefont
  {Castin}(2014)}]{Trefzger2014}%
  \BibitemOpen
  \bibfield  {author} {\bibinfo {author} {\bibfnamefont {C.}~\bibnamefont
  {Trefzger}}\ and\ \bibinfo {author} {\bibfnamefont {Y.}~\bibnamefont
  {Castin}},\ }\bibfield  {title} {\bibinfo {title} {Self-energy of an impurity
  in an ideal fermi gas to second order in the interaction strength},\ }\href
  {https://doi.org/10.1103/PhysRevA.90.033619} {\bibfield  {journal} {\bibinfo
  {journal} {Phys. Rev. A}\ }\textbf {\bibinfo {volume} {90}},\ \bibinfo
  {pages} {033619} (\bibinfo {year} {2014})}\BibitemShut {NoStop}%
\bibitem [{\citenamefont {Cetina}\ \emph {et~al.}(2015)\citenamefont {Cetina},
  \citenamefont {Jag}, \citenamefont {Lous}, \citenamefont {Walraven},
  \citenamefont {Grimm}, \citenamefont {Christensen},\ and\ \citenamefont
  {Bruun}}]{Cetina2015}%
  \BibitemOpen
  \bibfield  {author} {\bibinfo {author} {\bibfnamefont {M.}~\bibnamefont
  {Cetina}}, \bibinfo {author} {\bibfnamefont {M.}~\bibnamefont {Jag}},
  \bibinfo {author} {\bibfnamefont {R.~S.}\ \bibnamefont {Lous}}, \bibinfo
  {author} {\bibfnamefont {J.~T.~M.}\ \bibnamefont {Walraven}}, \bibinfo
  {author} {\bibfnamefont {R.}~\bibnamefont {Grimm}}, \bibinfo {author}
  {\bibfnamefont {R.~S.}\ \bibnamefont {Christensen}},\ and\ \bibinfo {author}
  {\bibfnamefont {G.~M.}\ \bibnamefont {Bruun}},\ }\bibfield  {title} {\bibinfo
  {title} {Decoherence of impurities in a fermi sea of ultracold atoms},\
  }\href {https://doi.org/10.1103/PhysRevLett.115.135302} {\bibfield  {journal}
  {\bibinfo  {journal} {Phys. Rev. Lett.}\ }\textbf {\bibinfo {volume} {115}},\
  \bibinfo {pages} {135302} (\bibinfo {year} {2015})}\BibitemShut {NoStop}%
\bibitem [{\citenamefont {Luri{\'{e}}}\ and\ \citenamefont
  {Macfarlane}(1964)}]{Lurie1964}%
  \BibitemOpen
  \bibfield  {author} {\bibinfo {author} {\bibfnamefont {D.}~\bibnamefont
  {Luri{\'{e}}}}\ and\ \bibinfo {author} {\bibfnamefont {A.~J.}\ \bibnamefont
  {Macfarlane}},\ }\bibfield  {title} {\bibinfo {title} {Equivalence between
  four-fermion and yukawa coupling, and {theZ}3=0condition for composite
  bosons},\ }\href {https://doi.org/10.1103/physrev.136.b816} {\bibfield
  {journal} {\bibinfo  {journal} {Phys. Rev.}\ }\textbf {\bibinfo {volume}
  {136}},\ \bibinfo {pages} {B816} (\bibinfo {year} {1964})}\BibitemShut
  {NoStop}%
\bibitem [{\citenamefont {Holland}\ \emph {et~al.}(2001)\citenamefont
  {Holland}, \citenamefont {Kokkelmans}, \citenamefont {Chiofalo},\ and\
  \citenamefont {Walser}}]{Holland2001}%
  \BibitemOpen
  \bibfield  {author} {\bibinfo {author} {\bibfnamefont {M.}~\bibnamefont
  {Holland}}, \bibinfo {author} {\bibfnamefont {S.~J. J. M.~F.}\ \bibnamefont
  {Kokkelmans}}, \bibinfo {author} {\bibfnamefont {M.~L.}\ \bibnamefont
  {Chiofalo}},\ and\ \bibinfo {author} {\bibfnamefont {R.}~\bibnamefont
  {Walser}},\ }\bibfield  {title} {\bibinfo {title} {{Resonance Superfluidity
  in a Quantum Degenerate Fermi Gas}},\ }\href
  {https://doi.org/10.1103/PhysRevLett.87.120406} {\bibfield  {journal}
  {\bibinfo  {journal} {Phys. Rev. Lett.}\ }\textbf {\bibinfo {volume} {87}},\
  \bibinfo {pages} {120406} (\bibinfo {year} {2001})}\BibitemShut {NoStop}%
\bibitem [{\citenamefont {{Timmermans}}\ \emph {et~al.}(2001)\citenamefont
  {{Timmermans}}, \citenamefont {{Furuya}}, \citenamefont {{Milonni}},\ and\
  \citenamefont {{Kerman}}}]{Timmermans2001}%
  \BibitemOpen
  \bibfield  {author} {\bibinfo {author} {\bibfnamefont {E.}~\bibnamefont
  {{Timmermans}}}, \bibinfo {author} {\bibfnamefont {K.}~\bibnamefont
  {{Furuya}}}, \bibinfo {author} {\bibfnamefont {P.~W.}\ \bibnamefont
  {{Milonni}}},\ and\ \bibinfo {author} {\bibfnamefont {A.~K.}\ \bibnamefont
  {{Kerman}}},\ }\bibfield  {title} {\bibinfo {title} {{Prospect of creating a
  composite Fermi-Bose superfluid}},\ }\href
  {https://doi.org/10.1016/S0375-9601(01)00346-2} {\bibfield  {journal}
  {\bibinfo  {journal} {Phys. Lett. A}\ }\textbf {\bibinfo {volume} {285}},\
  \bibinfo {pages} {228} (\bibinfo {year} {2001})}\BibitemShut {NoStop}%
\bibitem [{\citenamefont {Bruun}\ and\ \citenamefont
  {Pethick}(2004)}]{Bruun2004}%
  \BibitemOpen
  \bibfield  {author} {\bibinfo {author} {\bibfnamefont {G.~M.}\ \bibnamefont
  {Bruun}}\ and\ \bibinfo {author} {\bibfnamefont {C.~J.}\ \bibnamefont
  {Pethick}},\ }\bibfield  {title} {\bibinfo {title} {Effective theory of
  feshbach resonances and many-body properties of fermi gases},\ }\href
  {https://doi.org/10.1103/physrevlett.92.140404} {\bibfield  {journal}
  {\bibinfo  {journal} {Phys. Rev. Lett.}\ }\textbf {\bibinfo {volume} {92}},\
  \bibinfo {pages} {140404} (\bibinfo {year} {2004})}\BibitemShut {NoStop}%
\bibitem [{\citenamefont {Bloch}\ \emph {et~al.}(2008)\citenamefont {Bloch},
  \citenamefont {Dalibard},\ and\ \citenamefont {Zwerger}}]{Bloch2008}%
  \BibitemOpen
  \bibfield  {author} {\bibinfo {author} {\bibfnamefont {I.}~\bibnamefont
  {Bloch}}, \bibinfo {author} {\bibfnamefont {J.}~\bibnamefont {Dalibard}},\
  and\ \bibinfo {author} {\bibfnamefont {W.}~\bibnamefont {Zwerger}},\
  }\bibfield  {title} {\bibinfo {title} {Many-body physics with ultracold
  gases},\ }\href {https://doi.org/10.1103/revmodphys.80.885} {\bibfield
  {journal} {\bibinfo  {journal} {Rev. Mod. Phys.}\ }\textbf {\bibinfo {volume}
  {80}},\ \bibinfo {pages} {885} (\bibinfo {year} {2008})}\BibitemShut
  {NoStop}%
\bibitem [{\citenamefont {Berges}\ \emph {et~al.}(2002)\citenamefont {Berges},
  \citenamefont {Tetradis},\ and\ \citenamefont {Wetterich}}]{Berges2002}%
  \BibitemOpen
  \bibfield  {author} {\bibinfo {author} {\bibfnamefont {J.}~\bibnamefont
  {Berges}}, \bibinfo {author} {\bibfnamefont {N.}~\bibnamefont {Tetradis}},\
  and\ \bibinfo {author} {\bibfnamefont {C.}~\bibnamefont {Wetterich}},\
  }\bibfield  {title} {\bibinfo {title} {Non-perturbative renormalization flow
  in quantum field theory and statistical physics},\ }\href
  {https://doi.org/10.1016/s0370-1573(01)00098-9} {\bibfield  {journal}
  {\bibinfo  {journal} {Physics Reports}\ }\textbf {\bibinfo {volume} {363}},\
  \bibinfo {pages} {223} (\bibinfo {year} {2002})}\BibitemShut {NoStop}%
\bibitem [{\citenamefont {Delamotte}(2012)}]{Delamotte2012}%
  \BibitemOpen
  \bibfield  {author} {\bibinfo {author} {\bibfnamefont {B.}~\bibnamefont
  {Delamotte}},\ }\bibfield  {title} {\bibinfo {title} {An introduction to the
  nonperturbative renormalization group},\ }in\ \href
  {https://doi.org/10.1007/978-3-642-27320-9_2} {\emph {\bibinfo {booktitle}
  {Renormalization Group and Effective Field Theory Approaches to Many-Body
  Systems}}}\ (\bibinfo  {publisher} {Springer Berlin Heidelberg},\ \bibinfo
  {year} {2012})\ pp.\ \bibinfo {pages} {49--132}\BibitemShut {NoStop}%
\bibitem [{\citenamefont {Gies}(2012)}]{Gies2012}%
  \BibitemOpen
  \bibfield  {author} {\bibinfo {author} {\bibfnamefont {H.}~\bibnamefont
  {Gies}},\ }\bibfield  {title} {\bibinfo {title} {Introduction to the
  functional {RG} and applications to gauge theories},\ }in\ \href
  {https://doi.org/10.1007/978-3-642-27320-9_6} {\emph {\bibinfo {booktitle}
  {Renormalization Group and Effective Field Theory Approaches to Many-Body
  Systems}}}\ (\bibinfo  {publisher} {Springer Berlin Heidelberg},\ \bibinfo
  {year} {2012})\ pp.\ \bibinfo {pages} {287--348}\BibitemShut {NoStop}%
\bibitem [{\citenamefont {Dupuis}\ \emph {et~al.}(2021)\citenamefont {Dupuis},
  \citenamefont {Canet}, \citenamefont {Eichhorn}, \citenamefont {Metzner},
  \citenamefont {Pawlowski}, \citenamefont {Tissier},\ and\ \citenamefont
  {Wschebor}}]{Dupuis2020}%
  \BibitemOpen
  \bibfield  {author} {\bibinfo {author} {\bibfnamefont {N.}~\bibnamefont
  {Dupuis}}, \bibinfo {author} {\bibfnamefont {L.}~\bibnamefont {Canet}},
  \bibinfo {author} {\bibfnamefont {A.}~\bibnamefont {Eichhorn}}, \bibinfo
  {author} {\bibfnamefont {W.}~\bibnamefont {Metzner}}, \bibinfo {author}
  {\bibfnamefont {J.}~\bibnamefont {Pawlowski}}, \bibinfo {author}
  {\bibfnamefont {M.}~\bibnamefont {Tissier}},\ and\ \bibinfo {author}
  {\bibfnamefont {N.}~\bibnamefont {Wschebor}},\ }\bibfield  {title} {\bibinfo
  {title} {The nonperturbative functional renormalization group and its
  applications},\ }\href
  {https://doi.org/https://doi.org/10.1016/j.physrep.2021.01.001} {\bibfield
  {journal} {\bibinfo  {journal} {Phys. Rep.}\ }\textbf {\bibinfo {volume}
  {910}},\ \bibinfo {pages} {1} (\bibinfo {year} {2021})}\BibitemShut {NoStop}%
\bibitem [{\citenamefont {Metzner}\ \emph {et~al.}(2012)\citenamefont
  {Metzner}, \citenamefont {Salmhofer}, \citenamefont {Honerkamp},
  \citenamefont {Meden},\ and\ \citenamefont {Sch\"onhammer}}]{Metzner2012}%
  \BibitemOpen
  \bibfield  {author} {\bibinfo {author} {\bibfnamefont {W.}~\bibnamefont
  {Metzner}}, \bibinfo {author} {\bibfnamefont {M.}~\bibnamefont {Salmhofer}},
  \bibinfo {author} {\bibfnamefont {C.}~\bibnamefont {Honerkamp}}, \bibinfo
  {author} {\bibfnamefont {V.}~\bibnamefont {Meden}},\ and\ \bibinfo {author}
  {\bibfnamefont {K.}~\bibnamefont {Sch\"onhammer}},\ }\bibfield  {title}
  {\bibinfo {title} {Functional renormalization group approach to correlated
  fermion systems},\ }\href {https://doi.org/10.1103/revmodphys.84.299}
  {\bibfield  {journal} {\bibinfo  {journal} {Rev. Mod. Phys.}\ }\textbf
  {\bibinfo {volume} {84}},\ \bibinfo {pages} {299} (\bibinfo {year}
  {2012})}\BibitemShut {NoStop}%
\bibitem [{\citenamefont {Wetterich}(1993)}]{Wetterich1993}%
  \BibitemOpen
  \bibfield  {author} {\bibinfo {author} {\bibfnamefont {C.}~\bibnamefont
  {Wetterich}},\ }\bibfield  {title} {\bibinfo {title} {Exact evolution
  equation for the effective potential},\ }\href
  {https://doi.org/10.1016/0370-2693(93)90726-x} {\bibfield  {journal}
  {\bibinfo  {journal} {Phys. Lett. B}\ }\textbf {\bibinfo {volume} {301}},\
  \bibinfo {pages} {90} (\bibinfo {year} {1993})}\BibitemShut {NoStop}%
\bibitem [{\citenamefont {Pawlowski}\ \emph {et~al.}(2017)\citenamefont
  {Pawlowski}, \citenamefont {Scherer}, \citenamefont {Schmidt},\ and\
  \citenamefont {Wetzel}}]{Pawlowski2017}%
  \BibitemOpen
  \bibfield  {author} {\bibinfo {author} {\bibfnamefont {J.~M.}\ \bibnamefont
  {Pawlowski}}, \bibinfo {author} {\bibfnamefont {M.~M.}\ \bibnamefont
  {Scherer}}, \bibinfo {author} {\bibfnamefont {R.}~\bibnamefont {Schmidt}},\
  and\ \bibinfo {author} {\bibfnamefont {S.~J.}\ \bibnamefont {Wetzel}},\
  }\bibfield  {title} {\bibinfo {title} {Physics and the choice of regulators
  in functional renormalisation group flows},\ }\href
  {https://doi.org/10.1016/j.aop.2017.06.017} {\bibfield  {journal} {\bibinfo
  {journal} {Annals of Physics}\ }\textbf {\bibinfo {volume} {384}},\ \bibinfo
  {pages} {165} (\bibinfo {year} {2017})}\BibitemShut {NoStop}%
\bibitem [{\citenamefont {Schmidt}\ and\ \citenamefont
  {Enss}(2022)}]{Private_Schmidt_Enss}%
  \BibitemOpen
  \bibfield  {author} {\bibinfo {author} {\bibfnamefont {R.}~\bibnamefont
  {Schmidt}}\ and\ \bibinfo {author} {\bibfnamefont {T.}~\bibnamefont {Enss}},\
  }\href@noop {} {}\bibinfo {howpublished} {private communication} (\bibinfo
  {year} {2022})\BibitemShut {NoStop}%
\bibitem [{\citenamefont {Fetter}\ and\ \citenamefont
  {Walecka}(2012)}]{FetterBook}%
  \BibitemOpen
  \bibfield  {author} {\bibinfo {author} {\bibfnamefont {A.~L.}\ \bibnamefont
  {Fetter}}\ and\ \bibinfo {author} {\bibfnamefont {J.~D.}\ \bibnamefont
  {Walecka}},\ }\href@noop {} {\emph {\bibinfo {title} {Quantum Theory of
  Many-Particle Systems -}}}\ (\bibinfo  {publisher} {Courier Corporation},\
  \bibinfo {address} {New York},\ \bibinfo {year} {2012})\BibitemShut {NoStop}%
\bibitem [{\citenamefont {Altland}\ and\ \citenamefont
  {Simons}(2010)}]{Altland2010}%
  \BibitemOpen
  \bibfield  {author} {\bibinfo {author} {\bibfnamefont {A.}~\bibnamefont
  {Altland}}\ and\ \bibinfo {author} {\bibfnamefont {B.~D.}\ \bibnamefont
  {Simons}},\ }\href {https://doi.org/10.1017/cbo9780511789984} {\emph
  {\bibinfo {title} {Condensed Matter Field Theory}}}\ (\bibinfo  {publisher}
  {Cambridge University Press},\ \bibinfo {year} {2010})\BibitemShut {NoStop}%
\bibitem [{Note1()}]{Note1}%
  \BibitemOpen
  \bibinfo {note} {Similarly, performing the calculation for $\epsilon <0$ one
  obtains the advanced Green's function in the lower half of the complex
  plane.}\BibitemShut {Stop}%
\bibitem [{\citenamefont {Cetina}\ \emph {et~al.}(2016)\citenamefont {Cetina},
  \citenamefont {Jag}, \citenamefont {Lous}, \citenamefont {Fritsche},
  \citenamefont {Walraven}, \citenamefont {Grimm}, \citenamefont {Levinsen},
  \citenamefont {Parish}, \citenamefont {Schmidt}, \citenamefont {Knap},\ and\
  \citenamefont {Demler}}]{Cetina2016}%
  \BibitemOpen
  \bibfield  {author} {\bibinfo {author} {\bibfnamefont {M.}~\bibnamefont
  {Cetina}}, \bibinfo {author} {\bibfnamefont {M.}~\bibnamefont {Jag}},
  \bibinfo {author} {\bibfnamefont {R.~S.}\ \bibnamefont {Lous}}, \bibinfo
  {author} {\bibfnamefont {I.}~\bibnamefont {Fritsche}}, \bibinfo {author}
  {\bibfnamefont {J.~T.~M.}\ \bibnamefont {Walraven}}, \bibinfo {author}
  {\bibfnamefont {R.}~\bibnamefont {Grimm}}, \bibinfo {author} {\bibfnamefont
  {J.}~\bibnamefont {Levinsen}}, \bibinfo {author} {\bibfnamefont {M.~M.}\
  \bibnamefont {Parish}}, \bibinfo {author} {\bibfnamefont {R.}~\bibnamefont
  {Schmidt}}, \bibinfo {author} {\bibfnamefont {M.}~\bibnamefont {Knap}},\ and\
  \bibinfo {author} {\bibfnamefont {E.}~\bibnamefont {Demler}},\ }\bibfield
  {title} {\bibinfo {title} {Ultrafast many-body interferometry of impurities
  coupled to a fermi sea},\ }\href {https://doi.org/10.1126/science.aaf5134}
  {\bibfield  {journal} {\bibinfo  {journal} {Science}\ }\textbf {\bibinfo
  {volume} {354}},\ \bibinfo {pages} {96–99} (\bibinfo {year}
  {2016})}\BibitemShut {NoStop}%
\bibitem [{\citenamefont {Vivanco}\ \emph {et~al.}(2023)\citenamefont
  {Vivanco}, \citenamefont {Schuckert}, \citenamefont {Huang}, \citenamefont
  {Schumacher}, \citenamefont {Assump{\c{c}}{\~a}o}, \citenamefont {Ji},
  \citenamefont {Chen}, \citenamefont {Knap},\ and\ \citenamefont
  {Navon}}]{Vivanco2023}%
  \BibitemOpen
  \bibfield  {author} {\bibinfo {author} {\bibfnamefont {F.~J.}\ \bibnamefont
  {Vivanco}}, \bibinfo {author} {\bibfnamefont {A.}~\bibnamefont {Schuckert}},
  \bibinfo {author} {\bibfnamefont {S.}~\bibnamefont {Huang}}, \bibinfo
  {author} {\bibfnamefont {G.~L.}\ \bibnamefont {Schumacher}}, \bibinfo
  {author} {\bibfnamefont {G.~G.}\ \bibnamefont {Assump{\c{c}}{\~a}o}},
  \bibinfo {author} {\bibfnamefont {Y.}~\bibnamefont {Ji}}, \bibinfo {author}
  {\bibfnamefont {J.}~\bibnamefont {Chen}}, \bibinfo {author} {\bibfnamefont
  {M.}~\bibnamefont {Knap}},\ and\ \bibinfo {author} {\bibfnamefont
  {N.}~\bibnamefont {Navon}},\ }\bibfield  {title} {\bibinfo {title} {The
  strongly driven fermi polaron},\ }\href@noop {} {\bibfield  {journal}
  {\bibinfo  {journal} {arXiv:2308.05746}\ } (\bibinfo {year}
  {2023})}\BibitemShut {NoStop}%
\bibitem [{\citenamefont {Scazza}\ \emph {et~al.}(2017)\citenamefont {Scazza},
  \citenamefont {Valtolina}, \citenamefont {Massignan}, \citenamefont {Recati},
  \citenamefont {Amico}, \citenamefont {Burchianti}, \citenamefont {Fort},
  \citenamefont {Inguscio}, \citenamefont {Zaccanti},\ and\ \citenamefont
  {Roati}}]{Scazza2017}%
  \BibitemOpen
  \bibfield  {author} {\bibinfo {author} {\bibfnamefont {F.}~\bibnamefont
  {Scazza}}, \bibinfo {author} {\bibfnamefont {G.}~\bibnamefont {Valtolina}},
  \bibinfo {author} {\bibfnamefont {P.}~\bibnamefont {Massignan}}, \bibinfo
  {author} {\bibfnamefont {A.}~\bibnamefont {Recati}}, \bibinfo {author}
  {\bibfnamefont {A.}~\bibnamefont {Amico}}, \bibinfo {author} {\bibfnamefont
  {A.}~\bibnamefont {Burchianti}}, \bibinfo {author} {\bibfnamefont
  {C.}~\bibnamefont {Fort}}, \bibinfo {author} {\bibfnamefont {M.}~\bibnamefont
  {Inguscio}}, \bibinfo {author} {\bibfnamefont {M.}~\bibnamefont {Zaccanti}},\
  and\ \bibinfo {author} {\bibfnamefont {G.}~\bibnamefont {Roati}},\ }\bibfield
   {title} {\bibinfo {title} {Repulsive fermi polarons in a resonant mixture of
  ultracold $^{6}\mathrm{Li}$ atoms},\ }\href
  {https://doi.org/10.1103/PhysRevLett.118.083602} {\bibfield  {journal}
  {\bibinfo  {journal} {Phys. Rev. Lett.}\ }\textbf {\bibinfo {volume} {118}},\
  \bibinfo {pages} {083602} (\bibinfo {year} {2017})}\BibitemShut {NoStop}%
\bibitem [{\citenamefont {Mulkerin}\ \emph {et~al.}(2023)\citenamefont
  {Mulkerin}, \citenamefont {Levinsen},\ and\ \citenamefont
  {Parish}}]{Mulkerin2023}%
  \BibitemOpen
  \bibfield  {author} {\bibinfo {author} {\bibfnamefont {B.~C.}\ \bibnamefont
  {Mulkerin}}, \bibinfo {author} {\bibfnamefont {J.}~\bibnamefont {Levinsen}},\
  and\ \bibinfo {author} {\bibfnamefont {M.~M.}\ \bibnamefont {Parish}},\
  }\bibfield  {title} {\bibinfo {title} {Rabi oscillations and magnetization of
  a mobile spin-1/2 impurity in a fermi sea},\ }\href
  {https://doi.org/10.48550/ARXIV.2308.06659} {\bibfield  {journal} {\bibinfo
  {journal} {arXiv:2308.06659}\ } (\bibinfo {year} {2023})}\BibitemShut
  {NoStop}%
\bibitem [{\citenamefont {Baroni}\ \emph {et~al.}(2023)\citenamefont {Baroni},
  \citenamefont {Huang}, \citenamefont {Fritsche}, \citenamefont {Dobler},
  \citenamefont {Anich}, \citenamefont {Kirilov}, \citenamefont {Grimm},
  \citenamefont {Bastarrachea-Magnani}, \citenamefont {Massignan},\ and\
  \citenamefont {Bruun}}]{Baroni2023}%
  \BibitemOpen
  \bibfield  {author} {\bibinfo {author} {\bibfnamefont {C.}~\bibnamefont
  {Baroni}}, \bibinfo {author} {\bibfnamefont {B.}~\bibnamefont {Huang}},
  \bibinfo {author} {\bibfnamefont {I.}~\bibnamefont {Fritsche}}, \bibinfo
  {author} {\bibfnamefont {E.}~\bibnamefont {Dobler}}, \bibinfo {author}
  {\bibfnamefont {G.}~\bibnamefont {Anich}}, \bibinfo {author} {\bibfnamefont
  {E.}~\bibnamefont {Kirilov}}, \bibinfo {author} {\bibfnamefont
  {R.}~\bibnamefont {Grimm}}, \bibinfo {author} {\bibfnamefont {M.~A.}\
  \bibnamefont {Bastarrachea-Magnani}}, \bibinfo {author} {\bibfnamefont
  {P.}~\bibnamefont {Massignan}},\ and\ \bibinfo {author} {\bibfnamefont
  {G.~M.}\ \bibnamefont {Bruun}},\ }\bibfield  {title} {\bibinfo {title}
  {Mediated interactions between fermi polarons and the role of impurity
  quantum statistics},\ }\href {https://doi.org/10.1038/s41567-023-02248-4}
  {\bibfield  {journal} {\bibinfo  {journal} {Nat Phys.}\ } (\bibinfo {year}
  {2023})}\BibitemShut {NoStop}%
\bibitem [{\citenamefont {Johansen}\ \emph {et~al.}(2023)\citenamefont
  {Johansen}, \citenamefont {Frank},\ and\ \citenamefont
  {Lang}}]{Johansen2023}%
  \BibitemOpen
  \bibfield  {author} {\bibinfo {author} {\bibfnamefont {C.~H.}\ \bibnamefont
  {Johansen}}, \bibinfo {author} {\bibfnamefont {B.}~\bibnamefont {Frank}},\
  and\ \bibinfo {author} {\bibfnamefont {J.}~\bibnamefont {Lang}},\ }\bibfield
  {title} {\bibinfo {title} {Spectral functions of the strongly interacting 3d
  fermi gas},\ }\href {https://doi.org/10.48550/ARXIV.2311.03953} {\bibfield
  {journal} {\bibinfo  {journal} {arXiv:2311.03953}\ } (\bibinfo {year}
  {2023})}\BibitemShut {NoStop}%
\bibitem [{\citenamefont {Enss}(2023)}]{Enss2023}%
  \BibitemOpen
  \bibfield  {author} {\bibinfo {author} {\bibfnamefont {T.}~\bibnamefont
  {Enss}},\ }\bibfield  {title} {\bibinfo {title} {Particle and pair spectra
  for strongly correlated fermi gases: a real-frequency solver},\ }\href
  {https://doi.org/10.48550/ARXIV.2311.05443} {\bibfield  {journal} {\bibinfo
  {journal} {arXiv:2311.05443}\ } (\bibinfo {year} {2023})}\BibitemShut
  {NoStop}%
\bibitem [{\citenamefont {Dizer}\ \emph {et~al.}(2023)\citenamefont {Dizer},
  \citenamefont {Horak},\ and\ \citenamefont {Pawlowski}}]{Dizer2023}%
  \BibitemOpen
  \bibfield  {author} {\bibinfo {author} {\bibfnamefont {E.}~\bibnamefont
  {Dizer}}, \bibinfo {author} {\bibfnamefont {J.}~\bibnamefont {Horak}},\ and\
  \bibinfo {author} {\bibfnamefont {J.~M.}\ \bibnamefont {Pawlowski}},\
  }\bibfield  {title} {\bibinfo {title} {Spectral properties and observables in
  ultracold fermi gases},\ }\href {https://doi.org/10.48550/ARXIV.2311.16788}
  {\bibfield  {journal} {\bibinfo  {journal} {arXiv:2311.16788}\ } (\bibinfo
  {year} {2023})}\BibitemShut {NoStop}%
\end{thebibliography}
%

\appendix
\begin{widetext}

\section{Molecule initial condition}\label{app:molinitial}
The initial condition of the flowing retarded molecular Green's function at the cutoff scale is given by \cite{Schmidt2011}
    \begin{align}
        G^{R,-1}_{t,k= \Lambda} (\Omega+ i \epsilon,\pv)&= - \frac{h^2}{8 \pi a }+ \frac{h^2 \Lambda}{4 \pi^2}- h^2 \int_\qv \left[\frac{\Theta(|\pv + \qv|- \Lambda)\Theta(|\qv|-\Lambda)}{-\Omega - i \epsilon+ \qv^2 + (\pv+ \qv)^2}- \frac{\Theta (|\qv|- \Lambda)}{2 \qv^2}\right] \label{gtinitial},
    \end{align} such that using \cref{initialconditionepsilonF}
    \begin{align}
        G_{t,k=\sqrt{\epsilon_F}}^{R,-1}(\Omega + i \epsilon,\pv) &= G_{t,k=\Lambda}^{R,-1}(\Omega + i \epsilon,\pv) +h^2 \!\!  \int_{\qv}\! \!\left[ \frac{\Theta(|\pv+\qv|-\Lambda)\Theta(\qv^2-\epsilon_F-\Lambda^2)}{G_{\phi,\Lambda}^{R,-1}(\Omega+ i\epsilon -\qv^2 + \epsilon_F,\pv+\qv)} - \frac{\Theta(|\pv+\qv|^2-\epsilon_F)\Theta(\qv^2-2\epsilon_F)}{G_{\phi,\Lambda}^{R,-1}(\Omega+ i\epsilon -\qv^2 + \epsilon_F,\pv+\qv)}\right]\label{gtintermediate} \\
        &= - \frac{h^2}{8 \pi a }+ h^2  \int_{\qv} \left[\frac{1}{2 \qv^2}- \frac{1}{G_{\phi,\Lambda}^{R,-1}(\Omega+ i\epsilon -\qv^2 + \epsilon_F,\pv+\qv)}- \frac{\Theta(|\pv+\qv|-\epsilon_F)\Theta(\qv^2-2\epsilon_F)-1}{G_{\phi,\Lambda}^{R,-1}(\Omega+ i\epsilon -\qv^2 + \epsilon_F,\pv+\qv)} \right], \label{gtfinalinitial}
    \end{align}
    where we have cancelled the third term in \cref{gtinitial} against the second term in \cref{gtintermediate}. The integrals in \cref{gtfinalinitial} can be solved analytically.

\section{Decay width scaling from Fermi liquid theory}

\begin{figure*}[ht]
    \normalsize
\begin{align*}
 \begin{tikzpicture}[baseline=-\the\dimexpr\fontdimen22\textfont2\relax]
\begin{feynhand}
\vertex[squaredot] (a) at (-1,0){};
\vertex[squaredot](b) at (-1,-1){};
\vertex[squaredot] (c) at (-1.707-.2,0.707){};
\vertex (d) at (0,.5){};
\vertex (e) at (0,-.5){};
\vertex (f) at (0,-1.5){};
\vertex (g) at (0,1.5){};
\vertex (h) at (-1.707-1.2,0.707){};
\vertex (i) at (-1.707-.9,0.707+.8){$\mathbf{(a)}$};
\graph{(h)--[fermion](c) -- [scalar,half right,relative=false, out=60, in=180,looseness=1.2,with arrow=1](g), (c)--[chabos](a)--[fermion](b)-- [charged scalar,half right,relative=false, out=45, in=180,looseness=1.2, with arrow=1](e), (b)--[boson, half left,relative=false, out=-45, in=180,looseness=1.5, with arrow=1](f), (d)--[scalar, with arrow=0,half left, out=180, in=45,relative=false,](a)};
\end{feynhand}
\end{tikzpicture}
 \begin{tikzpicture}[baseline=-\the\dimexpr\fontdimen22\textfont2\relax]
\begin{feynhand}
\vertex[squaredot] (a) at (-1,0){};
\vertex[squaredot] (a1) at (1,0){};
\vertex[squaredot](b) at (-1,-1){};
\vertex[squaredot](b1) at (1,-1){};
\vertex[squaredot] (c) at (-1.707-.2,0.707){};
\vertex[squaredot] (c1) at (1.707+.2,0.707){};
\vertex (d) at (0,.5){};
\vertex (e) at (0,-.5){};
\vertex (f) at (0,-1.5){};
\vertex (fp) at (-.12,-1.5){};
\vertex (g) at (0,1.5){};
\vertex (h) at (-1.707-1.2,0.707){};
\vertex (h1) at (1.707+1.2,0.707){};
\vertex (i) at (-1.707-.9,0.707+.8){$\mathbf{(b)}$};
\graph{(h)--[fermion](c) -- [scalar,half right,relative=false, out=60, in=180,looseness=1.2,with arrow=1](g), (c)--[chabos](a)--[fermion](b)-- [charged scalar,half right,relative=false, out=45, in=180,looseness=1.2, with arrow=1](e), (b)--[boson, half right,relative=false, out=-45, in=-45-90,looseness=1.1, with arrow=0.5](b1), (h1)--[anti fermion](c1) -- [scalar,half left,relative=false, out=120, in=0,looseness=1.2](g),(e)-- [scalar,half right,relative=false, out=0, in=45+89,looseness=1.2](b1)--[fermion](a1)--[chabos](c1), a1--[scalar, with arrow=0.5, half left, out=45+90, in=45,relative=false](a)};
\end{feynhand}
\end{tikzpicture}
 \begin{tikzpicture}[baseline=-\the\dimexpr\fontdimen22\textfont2\relax]
\begin{feynhand}
\vertex[squaredot] (a) at (-1,0){};
\vertex[squaredot] (a1) at (1,0){};
\vertex[squaredot](b) at (-1,-1){};
\vertex[squaredot](b1) at (1,-1){};
\vertex[squaredot] (c) at (-1.707-.2,0.707){};
\vertex[squaredot] (c1) at (1.707+.2,0.707){};
\vertex (d) at (0,.5){};
\vertex (e) at (0,-.5){};
\vertex (f) at (0,-1.5){};
\vertex (fp) at (-.12,-1.5){};
\vertex (g) at (0,1.5){};
\vertex (h) at (-1.707-1.2,0.707){};
\vertex (h1) at (1.707+1.2,0.707){};
\vertex (i) at (-1.707-.9,0.707+.8){$\mathbf{(c)}$};
\graph{(h)--[fermion](c) -- [scalar,half right,relative=false, out=60, in=90+45,looseness=1.2,with arrow=0.3](b1), (c)--[chabos](a)--[fermion](b)-- [scalar,half right,relative=false, out=45, in=90+30,looseness=1.2, with arrow=.7](c1), (b)--[boson, half right,relative=false, out=-45, in=-45-90,looseness=1.1, with arrow=0.5](b1), (h1)--[anti fermion](c1),(b1)--[fermion](a1)--[chabos](c1), a1--[scalar, with arrow=0.5, half left, out=45+90, in=45,relative=false](a)};
\end{feynhand}
\end{tikzpicture}
\end{align*}
    \caption{Diagrammatic representation of the decay of the excited state molecule. \textbf{(a)} Possible decay channel of an excited state molecule (solid line) into a ground state polaron (wavy line) and several bath particles and holes (dashed lines), which correspond to self-energy contributions \textbf{(b,c)} using the optical theorem. The decay channel in (a) allows for two distinct self-energy contributions, a crossed in \textbf{(c)} and a non-crossed in \textbf{(b)}. The square dots denote coupling vertices ${\sim} h$.}
    \label{fig:decaymoldiagrams}
\end{figure*}
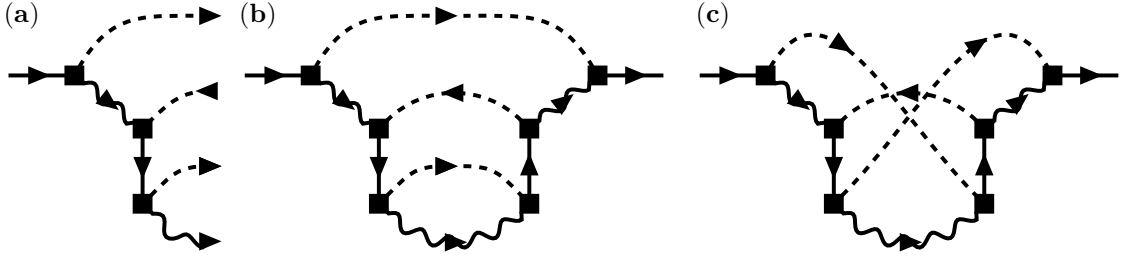

\subsection{Decay of the excited state molecule}\label{app_decay_mol}

To highlight the dependence of the decay width on the diagrammatic method used, in the following we discuss how a similar approach as used in Ref.~\cite{Bruun2010} may yield a different power law behaviour of the molecule decay width. As mentioned in the main text, the self-energy diagrammatics used in Ref.~\cite{Bruun2010} employ a $T$-matrix (containing no crossed diagrams) within a crossed diagram (see \cref{fig:decaymoldiagrams}(c)) to obtain a ${\sim} \Delta E^{9/2}$ dependence of the molecule decay width. We show that neglecting the crossed diagrams, as is done within conventional $T$-matrix approaches, and using a Fermi liquid theory approximation for these particles, one obtains a different power law dependence. 

The non-crossed self-energy contribution $\Sigma_t$ (see \cref{fig:decaymoldiagrams}(b)) corresponding to the decay process shown in \cref{fig:decaymoldiagrams}(a) is then proportional to 
    \begin{align}
    \Sigma_t(\omega, \zerov)\propto \int_{\kv,\kvp,\qv,\nu_1,\nu_2, \nu_3}
    \frac{G^R_{\phi}(i[\omega - \nu_1], -\kv )^2 G^R_{\phi}(i[\omega - \nu_1-\nu_2 + \nu_3], \qv- \kv -\kvp ) T^{R} (i[\omega -\nu_1+\nu_3],\qv-\kv)^2}{\left(- i \nu_1 + \kv^2- \epsilon_F\right)\left(- i \nu_2 + \kvp^2- \epsilon_F\right)\left(- i \nu_3 + \qv^2- \epsilon_F\right)} . \label{imagselftstart}
\end{align}
To proceed, we use a pole expansion for the retarded molecule propagator $T^R$ and the retarded impurity propagator $G_\phi^R$
\begin{align}
    G_{\phi}^R (z, \pv)&\propto \frac{Z_{\phi}}{- z + \frac{\pv^2}{ 2 m^{*}_\phi} }\label{polaronpropapprox} \\
    T^R (z, \pv)&\propto \frac{Z_{t}}{- z + \frac{\pv^2}{2 m^{*}_t} + \Delta E}, 
\end{align}
where $m^{*}_\phi$ and $m^{*}_t$ are the  effective masses of the attractive polaron and the molecule and $\Delta E$ denotes the energy difference between the attractive polaron and the molecule. Carrying out the frequency integrations and evaluating the self-energy near the pole of the molecule, we then obtain 
    \begin{align}
          \Im \Sigma^R_t(\Delta E + i0^+, \zerov)\propto   \int_{\kv,\kvp,\qv}
   \frac{Z_\phi^3 Z_t^2  \delta\left( \kv^2 + \kvp^2 - \qv^2 - \epsilon_F + \frac{(\qv-\kv -\kvp )^2}{2 m_\phi^{*}}-\Delta E \right) }{\left(\kv^2 - \qv^2 +\frac{(\qv- \kv)^2}{2 m_t^{*}} \right)^2\left(\kv^2 -\epsilon_F+ \frac{\kv^2}{2 m_\phi^*} -\Delta E \right)^2} . \label{imagselft}
    \end{align}
For $\Delta E \ll \epsilon_F$, the condition of the $\delta$-function in \cref{imagselft} is fulfilled when $\kv$, $\kvp$ and $\qv$ form an almost equilateral triangle at the Fermi surface with $|\kv|, |\kvp|, |\qv| \approx k_F$. Thus the two terms in the denominator of \cref{imagselft} approach $[k_F^2/(2 m_t^*)]^2$ and $[k_F^2/2 m_\phi^*]^2$, while in Ref.~\cite{Bruun2010} it was shown that the phase space integral scales as $ (m_\phi^*)^{3/2} \Delta E^{7/2}  $. Hence, within this approximation we obtain  that 
\begin{align}
     \Im \Sigma^R_t(\Delta E + i0^+, \zerov) \propto Z_\phi^3 Z_t^2 \Delta E^{7/2} (m_\phi^*)^{7/2} (m_t^{*})^2 .
\end{align}
Alternatively, one may disregard the dynamics of the propagators in \cref{imagselftstart}, as these propagators are not evaluated near their pole. Approximating these as constant instead, one obtains only the scaling due to the phase space integral given by ${\sim } Z_\phi  (m_\phi^*)^{3/2} \Delta E^{7/2}$ which is also shown in the main text.

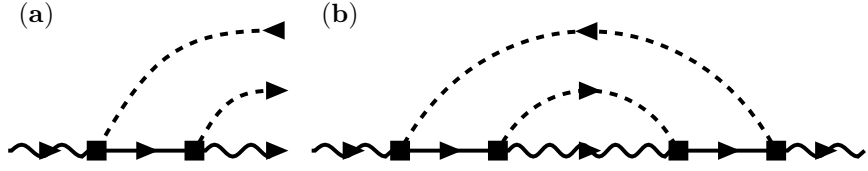
\begin{figure*}[ht]
    \normalsize
\begin{align*}
 \begin{tikzpicture}[baseline=-\the\dimexpr\fontdimen22\textfont2\relax]
\begin{feynhand}
\vertex (a) at (-3.8,0){};
\vertex[squaredot](b) at (-2.5,0){};
\vertex[squaredot](c) at (-1.2,0){};
\vertex (e) at (0,0){};
\vertex (f) at (0,0.8){};
\vertex (g) at (0,1.6){};
\vertex (i) at (-3.3,1.8){$\mathbf{(a)}$};
\graph{(a)--[charged boson](b)--[fermion](c)--[boson, with arrow=1](e), (c)--[scalar,looseness=1.2, half left, out=60, in=180,relative=false, with arrow=1](f), (g)--[scalar,looseness=1.2, half left, out=180, in=60,relative=false, with arrow=0](b)};
\end{feynhand}
\end{tikzpicture}
 \begin{tikzpicture}[baseline=-\the\dimexpr\fontdimen22\textfont2\relax]
\begin{feynhand}
\vertex (a) at (-3.8,0){};
\vertex[squaredot](b) at (-2.5,0){};
\vertex[squaredot](c) at (-1.2,0){};
\vertex (a1) at (3.8,0){};
\vertex[squaredot](b1) at (2.5,0){};
\vertex[squaredot](c1) at (1.2,0){};
\vertex (e) at (0,0){};
\vertex (f) at (0,0.8){};
\vertex (g) at (0,1.6){};
\vertex (i) at (-3.3,1.8){$\mathbf{(b)}$};
\graph{(a)--[charged boson](b)--[fermion](c)--[boson, with arrow=0.5](c1)--[fermion](b1)--[charged boson](a1), (c)--[scalar,looseness=1.2, half left, out=60, in=90+30,relative=false, with arrow=0.5](c1), (b1)--[scalar,looseness=1.2, half left, out=90+30, in=60,relative=false, with arrow=0.5](b)};
\end{feynhand}
\end{tikzpicture}
\end{align*}
    \caption{Diagrammatic representation of the decay of ground-state polarons at finite momentum into lower-lying polarons. \textbf{(a)} Possible decay channel of a polaron at finite momentum into a ground-state polaron at lower momentum and a particle and hole excitation, which corresponds to a self-energy contribution \textbf{(b)}. Unlike in \cref{fig:decaymoldiagrams}, the decay channel in \textbf{(a)} allows only for a single self-energy contribution \textbf{(b)}.}
    \label{fig:decaypoldiagrams}
\end{figure*}
\subsection{Decay of the ground state attractive polaron at finite momentum}\label{app_decay_pol}

At $1/k_F a \ll 1/k_F a_c$ the decay of the attractive polaron at small momentum  is only into attractive polaron states of a smaller momentum as the lowest-lying molecule state lies higher in energy. The simplest decay process representing this route is shown in \cref{fig:decaypoldiagrams}(a) and involves a particle-hole exchange with the bath particles. This process can easily be turned into a corresponding self-energy contribution, shown in \cref{fig:decaypoldiagrams}(b),  using the optical theorem such that the self-energy is proportional to 
\begin{align}
\Sigma_\phi( \omega, \pv) &\propto \int_{\kv, \qv, \nu_1, \nu_2} \frac{G^R_{\phi}(i[\omega+ \nu_1- \nu_2],\pv+ \qv -\kv) T(\omega+ \nu_1, \qv + \pv)^2}{(- i \nu_1 + \qv^2 - \epsilon_F)(- i \nu_2 + \kv^2 - \epsilon_F)} \nnl
& \propto \int_{\kv, \qv} G^R_{\phi}(i[\omega+ -i (\qv^2 -\kv^2)],\pv+ \qv -\kv) T(\omega-i (\qv^2 - \epsilon_F), \qv + \pv)^2\Theta(\epsilon_F- \qv^2)\Theta(\kv^2-\epsilon_F) ,
\end{align}
where $T$ denotes the $T$-matrix \cite{Combescot2007}. Furthermore, we have carried out the integration over $\nu_1$ and $\nu_2$ analytically by closing the contours in the right and in the left half of the complex plane, respectively. Considering the attractive polaron as a free particle, whose interactions with the bath have been taken into account via a modification of the quasiparticle gap (to zero, as the $\pv=0$ attractive polaron is the ground state), quasiparticle weight $Z_\phi$ and the effective mass $m^{*}_\phi$, we approximate the polaron propagator $G^R_{\phi}(z,\pv)$ using \cref{polaronpropapprox}.

\begin{figure}[bt]
	\begin{center}
	\includegraphics[width=0.4823\linewidth]{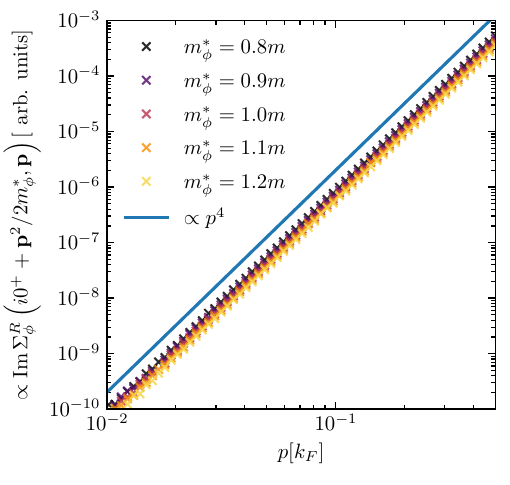}
	\end{center}
	\caption{Imaginary part of the polaron self-energy contribution in \cref{eq_fermi_liquid} for different effective polaron masses. The self-energy contribution $\Im \Sigma^R_\phi\left( \frac{\pv^2}{2 m_{\phi}^* } + i0^+, \pv\right)$ is shown in arbitrary units for different effective polaron masses ($m_\phi^* = 0.8m$(black), $0.9m$(purple), $m$(red), $1.1m$(orange), $1.2m$(yellow)) as a function of momentum $p$. The contributions follow a ${\sim}p^4$ scaling (blue line).  }
	\label{fig:fermiliquid}	
\end{figure}

Furthermore, as the decay of the attractive polaron is not into a molecule state, we approximate the scattering matrix $T\approx g$ via the bare coupling constant. Later we will investigate how the inclusion of $T$ changes the behavior of the decay width. Thus, evaluating the self-energy near the real axis at the location of the quasiparticle pole $\Omega=\pv^2/ 2 m_{\phi}^*$, we obtain that 
\begin{align}
    \Im \Sigma^R_\phi\left( \frac{\pv^2}{2 m_{\phi}^* } + i0^+, \pv\right) & \propto \int_{\kv> k_F, \qv< k_F}\! \! \! \delta \left(- \frac{\pv^2}{2 m_{\phi}^* } + \frac{(\pv+ \qv-\kv)^2}{2 m_\phi^{*}} - \qv^2 + \kv^2 \right) , \label{eq_fermi_liquid}
\end{align}    
where we have dropped the dependence on $Z_\phi$.

The imaginary part of the self-energy \cref{eq_fermi_liquid},  is shown in \cref{fig:fermiliquid} for different values of the effective mass $m_\phi^*$ and it can be seen that the imaginary part of the self-energy at the quasiparticle pole follows a $\propto p^4 $ scaling, as also seen for the full fRG model in the main text. 

Suppose now that the scattering $T$-matrix was not approximated by $g$, then along the real axis it is clear that for $i\omega\to \Omega+ i0^+= \pv^2/ 2 m_{\phi}^* + i 0^+$  and $\qv^2<\epsilon_F $ we have that 
\begin{align}
    \Im T^R\left(\frac{\pv^2}{ 2 m_{\phi}^*} + (\qv^2 - \epsilon_F)+i 0^+, \qv + \pv\right)=0 
\end{align}
because the lowest-lying molecule state lies higher in energy (see also the discussion \cref{analyticalstructure}). One thus arrives at 
\begin{align}
    \Im \Sigma^R_\phi\left( \frac{\pv^2}{2 m_{\phi}^* } + i0^+, \pv\right) & \propto \int_{\kv> k_F, \qv< k_F} \! \! \! \delta \left(- \frac{\pv^2}{2 m_{\phi}^* } + \frac{(\pv+ \qv-\kv)^2}{2 m_\phi^{*}} - \qv^2 + \kv^2 \right)  T^R\left(\frac{\pv^2}{2 m_{\phi}^*} + (\qv^2 - \epsilon_F)+ i0^+, \qv + \pv\right)^2 .
\end{align} 
Since the molecule is a higher-lying excited state by assumption, at small momentum $\pv$ the  $T$-matrix approaches a finite, constant value and thus the scaling of the imaginary part of the self-energy is solely determined by the phase space configuration scaling enforced by the $\delta$-function.

 \end{widetext}
\end{document}